\documentclass{article}

\PassOptionsToPackage{numbers}{natbib}
\usepackage[preprint]{neurips_2026}

\usepackage[utf8]{inputenc}
\usepackage[T1]{fontenc}
\usepackage[hidelinks]{hyperref}
\usepackage{url}
\usepackage{booktabs}
\usepackage{amsfonts}
\usepackage{amsmath}
\usepackage{amssymb}
\usepackage{nicefrac}
\usepackage{microtype}
\usepackage{xcolor}
\usepackage{graphicx}
\usepackage{float}
\usepackage{multirow}
\usepackage{array}
\usepackage{tabularx}
\usepackage{subcaption}
\usepackage{enumitem}
\usepackage[most]{tcolorbox}
\usepackage{placeins}    
\usepackage{bm}
\usepackage{bbm}

\usepackage{colortbl}
\usepackage{scrextend}
\usepackage{xurl}
\usepackage{wrapfig}

\definecolor{openaiDark}{HTML}{E31A1C}      
\definecolor{openaiLight}{HTML}{FB9A99}     
\definecolor{anthropicDark}{HTML}{FF7F00}   
\definecolor{anthropicLight}{HTML}{FDBF6F}  
\definecolor{googleDark}{HTML}{1F78B4}      
\definecolor{googleLight}{HTML}{A6CEE3}     
\definecolor{xaiDark}{HTML}{17BECF}         
\definecolor{xaiLight}{HTML}{9EDAE5}        
\definecolor{perplexityDark}{HTML}{6A3D9A}  
\definecolor{perplexityLight}{HTML}{CAB2D6} 
\definecolor{queryFg}{HTML}{4A7FC4}         
\definecolor{queryBg}{HTML}{E8F0FA}         
\definecolor{answerFg}{HTML}{C4883A}        
\definecolor{answerBg}{HTML}{FDF3E4}        
\definecolor{sourceFg}{HTML}{3A8A6E}        
\definecolor{sourceBg}{HTML}{E5F2ED}        
\definecolor{asOne}{HTML}{7A4A18}    
\definecolor{asTwo}{HTML}{A86F2A}    
\definecolor{asThree}{HTML}{C4883A}  
\definecolor{asFour}{HTML}{DEB57A}   
\definecolor{asFive}{HTML}{F0DDB8}   
\definecolor{textPrimary}{HTML}{333333}     
\definecolor{textSecondary}{HTML}{555555}   
\definecolor{textTertiary}{HTML}{666666}    
\definecolor{textFaint}{HTML}{999999}       
\definecolor{spineGray}{HTML}{AAAAAA}       
\definecolor{gridGray}{HTML}{E8E8E8}        
\providecommand{\srcbar}[1]{\textcolor{sourceFg}{\rule[0.1ex]{#1pt}{6pt}}}
\providecommand{\qrybar}[1]{\textcolor{queryFg}{\rule[0.1ex]{#1pt}{6pt}}}
\providecommand{\ansbar}[1]{\textcolor{answerFg}{\rule[0.1ex]{#1pt}{6pt}}}

\definecolor{tblBlueBg}{HTML}{EAF3FA}
\definecolor{tblBlueFg}{HTML}{1F5A8A}
\definecolor{tblRedBg}{HTML}{FBEAEA}
\definecolor{tblRedFg}{HTML}{A82020}

\providecommand{\redbar}[1]{\textcolor{tblRedFg}{\rule[0.1ex]{#1pt}{6pt}}}

\newcommand{\VM}{\textsc{Verified Misguidance}}
\newcommand{\VMshort}{\textsc{VM}}

\newcommand{\CiteTrace}{\textsc{CiteTrace}}

\definecolor{vmred}{RGB}{200, 50, 50}
\definecolor{vmblue}{RGB}{30, 80, 160}
\definecolor{successDk}{HTML}{0F6E56}
\definecolor{successLt}{HTML}{1D9E75}
\definecolor{successBg}{HTML}{E1F5EE}
\definecolor{failDk}{HTML}{A32D2D}
\definecolor{failLt}{HTML}{E24B4A}
\definecolor{failBg}{HTML}{FCEBEB}

\newcommand{\successbar}[1]{%
  \tikz[baseline=-0.3ex]{%
    \fill[successBg, rounded corners=1pt] (0,0) rectangle (2.4cm, 0.18cm);
    \fill[successLt, rounded corners=1pt] (0,0) rectangle (#1*0.024cm, 0.18cm);}}
\newcommand{\failbar}[1]{%
  \tikz[baseline=-0.3ex]{%
    \fill[failBg, rounded corners=1pt] (0,0) rectangle (2.4cm, 0.18cm);
    \fill[failLt, rounded corners=1pt] (0,0) rectangle (#1*0.024cm, 0.18cm);}}
\newcommand\blfootnote[1]{%
  \begingroup
  \renewcommand\thefootnote{}\footnote{#1}%
  \addtocounter{footnote}{-1}%
  \endgroup
}

\title{%
  Verified Misguidance: Measuring Structural Citation Failures~in~Search-Augmented~LLMs%
}
\author{%
  Yongsik Seo\textsuperscript{*\,1,5} \quad
  Wooseok Jeong\textsuperscript{*\,2} \quad
  Eunyoung Kim\textsuperscript{3} \quad
  Hyeonseo Jang\textsuperscript{4} \quad
  Dongha Lee\textsuperscript{\dag\,1,5} \\[0.8em]
  \textsuperscript{1}Department of Artificial Intelligence, Yonsei University \\
  \textsuperscript{2}Department of Computer Science and Engineering, Konkuk University \\
  \textsuperscript{3}Incheon International Airport Corporation \\
  \textsuperscript{4}Department of Computer Science and Engineering, Ewha Womans University \\
  \textsuperscript{5}ParamitaAI \\[0.4em]
  \texttt{ysseo@yonsei.ac.kr, jws010825@konkuk.ac.kr} \\
  \texttt{key@airport.kr, 0102jhshs@ewha.ac.kr, donalee@yonsei.ac.kr}
}

\begin{document}

\maketitle
\blfootnote{\textsuperscript{*}Equal contribution.}
\blfootnote{\textsuperscript{\dag}Corresponding author.}
\vspace{-1.5em}

\begin{abstract}
Users of search-augmented LLMs rely on citations as evidence that 
responses are grounded in real sources, and rarely verify the 
cited pages themselves.
Millions of queries per day now pass through these systems, 
making citation quality a silent determinant of whether users are 
informed or misled---yet existing benchmarks each address one 
facet in isolation, leaving the joint structure that determines 
citation trustworthiness unmeasured.
We construct \CiteTrace{}, a large-scale dataset that 
\emph{traces the full citation chain from user query through 
retrieved source to generated answer}: 11{,}200 real-world 
queries from 28 communities paired with 112{,}000 responses from 
ten models across five providers, yielding 761{,}495 evaluable 
citation pairs.
We design a three-dimension evaluation framework that scores each 
citation on intent--purpose alignment, source suitability, and 
answer--source fidelity, using expert-validated predefined 
matrices and a five-level fidelity rubric; the framework applies 
to any system that produces citation-bearing responses.
Applying this framework at scale, we identify a systematic 
pattern we call \VM{} (\VMshort{}): models cite real, accessible 
sources yet fail along one or more dimensions, producing a 
fidelity--suitability trade-off in which faithful models select 
inappropriate sources and vice versa.
Across our pool, 30.6\% of citations distort their sources and 
27.1\% originate from domain-inappropriate sources; at the 
response level, up to 96\% of users encounter at least one 
structurally misleading citation.
Provider-level differences explain 88--96\% of citation-quality 
variance, suggesting that source selection is governed more by factors beyond individual model capability than by the LLMs themselves. 
Together, \CiteTrace{} and its evaluation framework provide the first resource for diagnosing structural citation failures in deployed search-augmented systems.
\end{abstract}
\section{Introduction}
\label{sec:intro}

Users of search-augmented LLMs rely on citations as evidence that responses are grounded in real sources~\citep{ding2025citationstrustllmgenerated,aee2025}, and 
rarely verify the cited pages themselves~\citep{fogg2003prominence,liu2023evaluating}.
Millions of queries per day now pass through these systems~\citep{searcharena2026}, making citation quality a silent determinant of whether users are informed or misled.
Search-augmented LLMs must judge query intent, select domain-appropriate sources through a retrieval pipeline whose division of labor between search backend and model remains opaque, and faithfully ground the answer in those sources.
Each step in this multi-stage process can fail independently.
On the same medical query, one model may cite a government health agency while another cites a personal wellness blog; both answers may look identical, but only one rests on a trustworthy source.
The resulting failures are harder to detect than hallucination because the cited sources are real; we call this phenomenon \textbf{Verified Misguidance} (VM).

The tools available for auditing citations each address only one 
facet in isolation.
Citation verification checks whether a claim is supported by its 
source~\citep{alce,rashkin2023ais}; credibility rubrics such as 
the CRAAP test~\citep{craap} and Google's Your Money or Your 
Life (YMYL) 
classification~\citep{ymyl} rate source quality independently of 
the query; generative search audits examine system-level 
source-selection 
patterns~\citep{li2024generative,zhang2025source}.
A source can pass a fidelity check and a quality rubric 
independently yet still be the wrong type of source for the 
question at hand, and none of these tools would flag it.

Diagnosing VM requires linking each citing sentence to the 
crawled content of its source across diverse query 
domains~\citep{sourcebench,sourcecheckup}---a combination that 
existing benchmarks provide only partially or for single domains.
We construct \CiteTrace{} to close this gap, sourcing 
cross-domain queries from 28 Stack Exchange communities where 
expert knowledge and commercial incentives 
coexist~\citep{akerlof1970,arrow1963}, collecting 
search-augmented responses from ten LLMs across five providers 
under a neutral prompt that imposes no citation conventions, and 
crawling every cited URL to recover source content.
The resulting dataset comprises \textbf{761{,}495} evaluable 
citation pairs, each linking a user query to a model's citing 
sentence and the crawled content of its source.
This three-way linkage enables, for the first time, joint 
analysis of whether the right source was selected and whether it 
was faithfully used.

\begin{figure}[!t]
\centering
\includegraphics[width=0.99\linewidth]{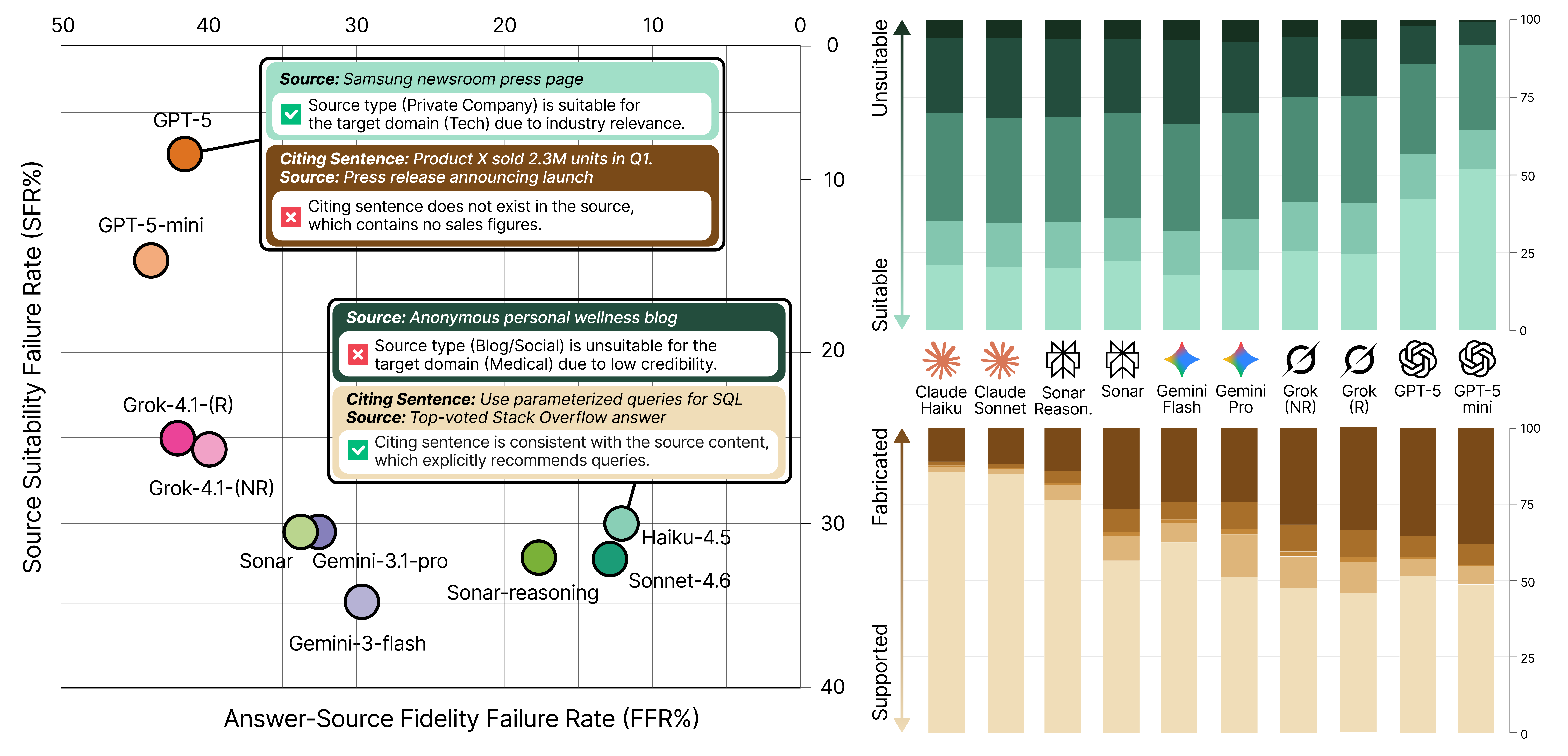}
\caption{The fidelity--suitability trade-off across ten 
search-augmented LLMs. 
\textbf{Left}: each model is plotted by its Fidelity Failure Rate 
(FFR, $x$-axis) and Suitability Failure Rate (SFR, $y$-axis). 
\texttt{gpt-5} selects domain-appropriate sources (SFR 8.0\%) but frequently fabricates their content (FFR 42.3\%), while \texttt{claude-haiku} faithfully reflects its sources (FFR 12.3\%) but draws heavily on unsuitable types (SFR 30.1\%); the ideal corner remains unoccupied. Within each provider, models show similar fidelity--suitability profiles, while between-provider spread is large.
\textbf{Right}: per-model distributions of source suitability (top, green) and answer--source fidelity (bottom, amber). 
Fidelity is sharply bipolar; most citations are either fully supported or fully fabricated, with few intermediate cases.}
\label{fig:teaser}
\end{figure}

To measure VM, we design a three-dimension evaluation framework: Intent--Purpose Alignment scores the fit between the query's information need and the source's communicative purpose; Source Suitability scores whether the source type is credible for the domain; and Answer--Source Fidelity scores whether the citing sentence faithfully reflects the source content. 
Each dimension captures a failure mode invisible to the other two: a source can be faithful yet unsuitable, suitable yet misaligned with the user's intent, or well-aligned yet unfaithful.
All classifications rely on a single LLM judge (\texttt{gpt-4o-mini}) validated against human annotators (Appendix~\ref{app:evaluating-vm}). 
The framework is not tied to \CiteTrace{}; it applies to any system that produces citation-bearing responses and can serve as a reusable instrument for citation quality evaluation beyond our setting.

Applying the framework to \CiteTrace{} reveals that VM is 
pervasive across all ten models.
Three patterns characterize the effect:
(1)~models that cite faithfully tend to select domain-inappropriate sources, and vice versa; 
a fidelity--suitability trade-off that no single model escapes;
(2)~88--96\% of citation-quality variance traces to the provider's search infrastructure rather than to model capability; 
and (3)~these failures compound at the response level, exposing up to 90\% of users to at least one structurally misleading citation (Figure~\ref{fig:teaser}).
These patterns suggest that improving generation alone is insufficient; source selection at the retrieval stage warrants 
equal attention.

We make three contributions to the study of citation quality in 
search-augmented systems:
\begin{itemize}[leftmargin=1.2em,itemsep=2pt,topsep=2pt]
  \item \textbf{Dataset.\ \ } \CiteTrace{} is the 
  first large-scale resource that links real user queries, 
  citation-bearing LLM responses, and crawled source content 
  across 28 domains and ten models, enabling researchers to study 
  citation quality jointly rather than one facet at a time 
  (\S\ref{sec:citetrace}).

  \item \textbf{Evaluation.\ \ } Our three-dimension evaluation framework scores each citation on intent--purpose alignment, source suitability, and answer--source fidelity, three facets that existing tools evaluate separately.
  The framework applies to any system that produces citation-bearing responses (\S\ref{sec:measuring-vm}).

  \item \textbf{Findings.\ \ } Joint evaluation reveals a fidelity--suitability trade-off governed at the provider level, exposing up to 90\% of users to at least one 
  structurally misleading citation.
  These findings suggest that dedicated attention is needed not only at the generation stage but also at the stage where sources are retrieved and selected 
  (\S\ref{sec:vm_effect}).
\end{itemize}

\CiteTrace{} is publicly released on HuggingFace at \url{https://huggingface.co/datasets/oseoko/citetrace-vm} with Croissant-compliant metadata~\citep{akhtar2024croissant}, and the evaluation code is available at \url{https://github.com/oseoko/verified-misguidance}; licensing and reproducibility scope are documented in Appendix~\ref{app:data-release}.

\section{\CiteTrace: A Large-Scale Dataset for Citation Quality Evaluation}
\label{sec:citetrace}

Measuring search-augmented LLM citation quality requires three 
components: real user queries, citation-bearing model responses, 
and the content of each cited source.
\CiteTrace{} is a large-scale citation evaluation dataset that 
combines cross-domain queries, search-augmented responses from ten 
models, and the crawled content of their cited sources.
We describe query sourcing 
(\S\ref{sec:citetrace-sourcing}), response collection 
(\S\ref{sec:citetrace-collection}), and source crawling 
(\S\ref{sec:citetrace-crawling}); full construction 
details appear in Appendix~\ref{app:constructing-citetrace}.

\begin{figure}[!t]
\centering
\includegraphics[width=0.99\linewidth]{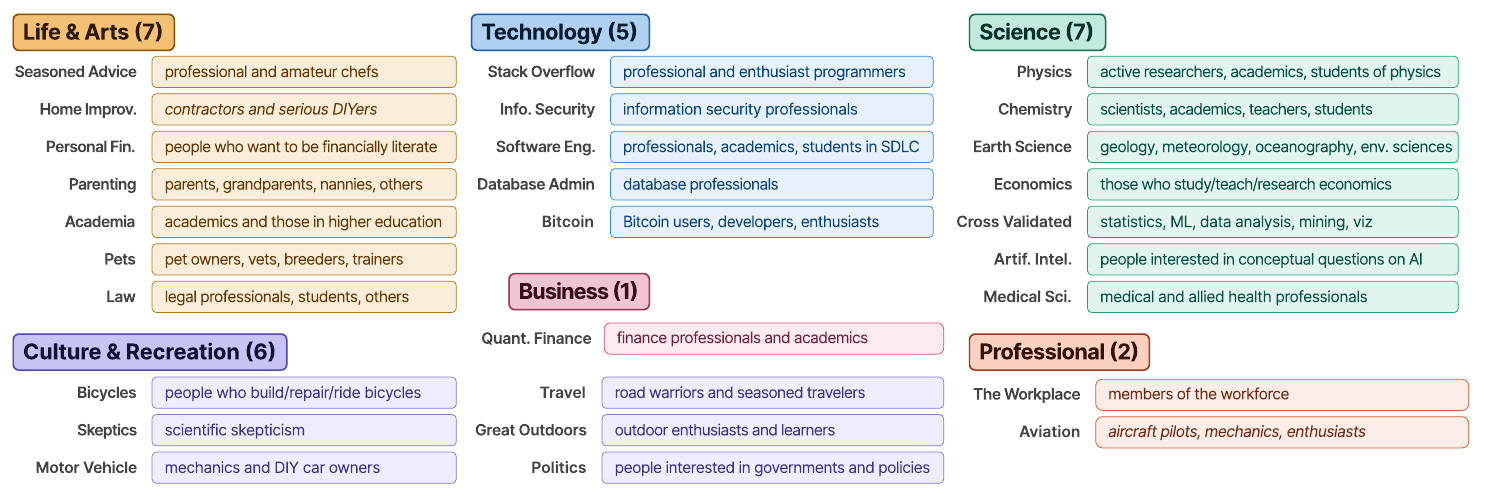}
\caption{Composition of the \CiteTrace{} query set. The 28 Stack 
Exchange communities are grouped by Stack Exchange's six official 
categories. Each community is annotated with its audience 
and domain.
Communities were selected to satisfy three conditions: expert knowledge is required, commercial actors are 
present, and queries involve substantive decisions about safety, 
cost, or design.}
\label{fig:dataset-overview}
\end{figure}

\subsection{Sourcing Queries}
\label{sec:citetrace-sourcing}

We source queries from Stack 
Exchange,\footnote{\href{https://archive.org/details/stackexchange_20251231}{https://archive.org/details/stackexchange\_20251231}} selecting 
28 communities that satisfy three conditions jointly:
expert knowledge is required to answer well, commercial actors are present and incentivized to shape perceptions, and the queries connect to substantive real-world decisions about safety, cost, or design. 
This combination is a key design choice;
it targets precisely the settings where citation misguidance carries the highest cost and is hardest for users to detect. 
General-interest, entertainment, and single-correct-answer communities were excluded because they lack this commercial-bias dynamic. 
The 28 sites span Stack Exchange's six official categories(Figure~\ref{fig:dataset-overview}), covering domains from medicine and law to programming and personal finance.

To ensure cross-domain comparability, we sample uniformly at 400 queries per site rather than proportionally, since proportional sampling would have been dominated by Stack Overflow alone. 
Combined with community-validation and recency filters, this yields a final dataset of 11,200 queries. Full selection criteria and per-site statistics appear in Appendix~\ref{app:sourcing-queries}.

\subsection{Collecting Responses and Citations}
\label{sec:citetrace-collection}

We query ten search-augmented LLMs from five providers (OpenAI, Anthropic, Google, xAI, and Perplexity), including standard and reasoning variant pairs where available. 
All models share a single neutral system prompt that specifies no citation format, count, or source preference, so that each model's citation behavior reflects its own defaults rather than prompt artifacts.
We issue all 11,200 queries to each model within the same 15-day window to keep the web results comparable across models, producing 112,000 responses and 1,271,046 citation pairs; 
8,069 responses (7.2\%) contain no citations. 
Model identifiers, API configurations, and the full system prompt appear in Appendix~\ref{app:collecting-responses}.

\subsection{Crawling and Verifying Sources}
\label{sec:citetrace-crawling}

Unlike prior work that evaluates citations against a pre-retrieved corpus, we crawl every unique URL cited across the ten models directly, since citation quality can only be assessed against the actual content of the cited source. 
We retrieve 231,105 of 396,670 unique URLs (58.3\%), covering 63.2\% of all citation pairs, under \texttt{robots.txt} compliance.

Crawl failures are not uniform: 
they concentrate on forum and Q\&A hosts due to bot blocking, which structurally under-represents community-based sources in the evaluable pool. 
Because these are precisely the source types most likely to be problematic, our reported failure rates are conservative lower bounds on the true population values. 
After removing citations whose text consists solely of code or tables, instances flagged as unevaluable by the LLM judge, and extremely short extractions~\citep{press2024citeme}, the final evaluable pool contains {761,495} citation pairs across the ten models. 
The full crawl pipeline, failure breakdown, host-tier bias, and per-model evaluability rates appear in Appendix~\ref{app:crawling-sources}.
\section{An Evaluation Framework for Diagnosing Verified Misguidance}
\label{sec:measuring-vm}

\begin{figure}[!t]
\centering
\includegraphics[width=0.99\linewidth]{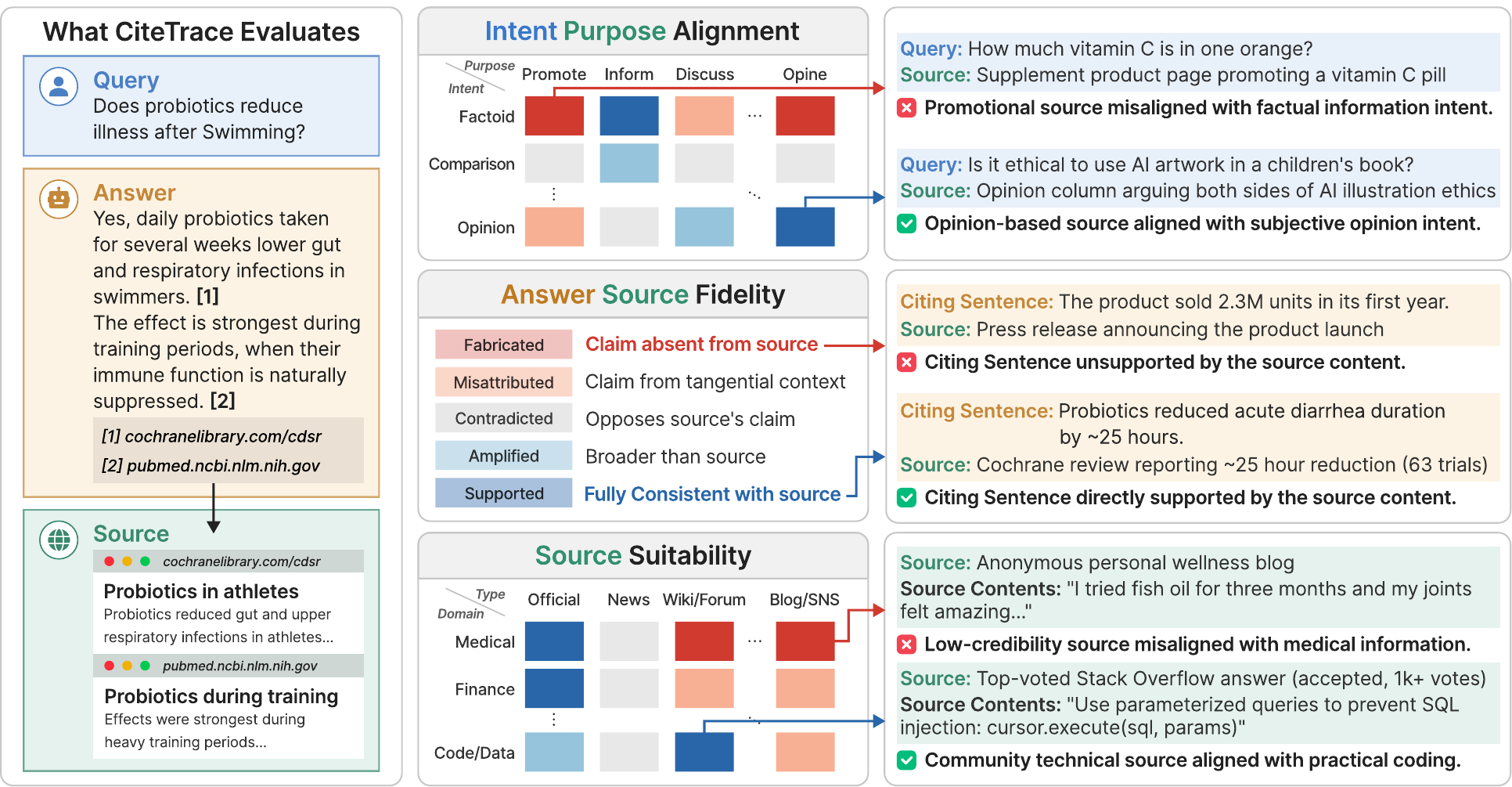}
\caption{Overview of the three-dimension evaluation framework for 
diagnosing Verified Misguidance.
\textbf{Left}: a user poses a query to a search-augmented LLM, which retrieves sources and generates a cited answer; the framework evaluates the resulting citation by jointly examining the query, the citing sentence, and the source content. 
\textbf{Center}: each citation is scored along three dimensions: Intent--Purpose Alignment (IPA) measures the fit between query intent and source purpose via a $5 \times 6$ matrix; Source Suitability (SS) measures whether the source type is sufficiently credible for the domain via a $10 \times 6$ matrix; Answer--Source Fidelity (ASF) measures how faithfully the citing sentence reflects the source content on a five-level rubric. Each dimension yields a 1--5 score; scores $\leq$2 define dimension-level failure.
\textbf{Right}: concrete pass/fail examples for each dimension: a factual query citing a promotional source (IPA failure), a fabricated claim absent from the source (ASF failure), and a medical query citing a personal blog (SS failure).}
\label{fig:qsa_framework}
\end{figure}

Having assembled the citation pairs, the next question is how to evaluate them. 
A cited source can mislead the user even when it is real and the answer looks correct: its purpose may conflict with the user's intent, the source type may be insufficiently credible for the domain, or the answer may distort what the source actually says.
To capture these three failure modes jointly, we propose a three-dimension evaluation framework (Figure~\ref{fig:qsa_framework}): 
the fit between query intent and source purpose (\S\ref{sec:query-source-alignment}), 
the  suitability of the source type for the domain (\S\ref{sec:source-suitability}), and the fidelity of the citing sentence to the source content (\S\ref{sec:answer-source-fidelity}).
The first two dimensions score each citation via expert-validated predefined matrices; the third uses a five-level fidelity rubric. 
All classifications rely on a single LLM judge (\texttt{gpt-4o-mini}) whose reliability we validate against human annotators in Appendix~\ref{app:evaluating-vm}.

\subsection{Alignment Between Query Intent and Source Purpose}
\label{sec:query-source-alignment}

A source can be factually accurate yet structurally misaligned with what the user needs;
a promotional page cited for a factual query, or an opinion piece cited for a request seeking causal explanation. 
Factual verification alone cannot detect this, because the mismatch lies in the source's communicative function rather than its content. 
We classify each query by its intent (e.g., factoid, explanation, instruction)~\citep{bolotova2022non} and each cited source by its communicative purpose (e.g., to promote, to inform, to discuss)~\citep{biber2015exploring,sharoff2018functional}, then score their alignment via a predefined 5$\times$6 Intent--Purpose Alignment (IPA) Matrix on a 1--5 scale.
Scores of 3--5 indicate functional alignment; scores of 1--2 flag structural conflict, where the source's incentive diverges from the user's need~\citep{akerlof1970}. 
We define citations with IPA score $\leq$2 as alignment failures and report their share as the Alignment Failure Rate (AFR).
Full taxonomy, matrix design, and expert validation appear in Appendix~\ref{app:axis1-detail}.

\subsection{Source Suitability Across Domains and Types}
\label{sec:source-suitability}

Even when a source's purpose aligns with the user's intent, the source type may be insufficiently credible for the domain in question. 
The same personal blog may be acceptable for an everyday cooking question but structurally inappropriate for a medical query;
yet existing source-quality frameworks evaluate sources independently of the domain they are cited for~\citep{craap}. 
We score each source against a predefined 10$\times$6 Source Suitability (SS) Matrix on a 1--5 scale that crosses substantive domain (e.g., medical, legal, finance) with publication type (e.g., official institution, paper/research, blog/social media).
The matrix is anchored in the CRAAP test and Google's YMYL classification~\citep{ymyl}, which define where source-type credibility requirements are highest. 
We define citations with SS score $\leq$2 as suitability failures and report their share as the Suitability Failure Rate (SFR). 
Full taxonomy, matrix design, and expert validation appear in Appendix~\ref{app:axis3-detail}.

\subsection{Fidelity of Answers to Their Sources}
\label{sec:answer-source-fidelity}

A source can be well-chosen yet still be misrepresented in the answer. 
Prior citation benchmarks reduce fidelity to a binary supported/not-attributable judgment~\citep{alce,rashkin2023ais}, but the characteristic distortions of Verified Misguidance live between those endpoints: 
a claim may exist in the source yet be stripped of a critical qualifier, reversed in direction, or attached to tangential context that changes its meaning. 
We compare each citing sentence against the crawled source content and assign an Answer--Source Fidelity (ASF) score on a five-level rubric~\citep{maynez2020faithfulness} that explicitly separates these intermediate failure modes. 
We define citations with ASF score $\leq$2 as fidelity failures and report their share as the Fidelity Failure Rate (FFR). 
The five-level rubric and human-agreement statistics appear in Appendix~\ref{app:axis2-detail}.

\subsection{Validating the LLM Judge}
All five classification tasks rely on a single LLM judge (\texttt{gpt-4o-mini}), which we validate against human annotators on 200 stratified samples per dimension with three independent annotators each. 
Cohen's $\kappa$ against majority-vote consensus ranges from 0.788 to 0.879 across the five dimensions, exceeding the substantial-agreement threshold of 0.667 throughout. 
Inter-annotator agreement among the three human annotators is Krippendorff's $\alpha \geq 0.811$, confirming that the taxonomy itself is stable and not merely an artifact of the judge. 
For the two matrix-based dimensions, expert panels of 10 domain specialists per domain independently rated each matrix cell; ICC(2,k) is 0.916 for the IPA Matrix and 0.958 (median) for the SS Matrix, with no cell deviating from expert consensus by two or more points. 
Together, these results support treating the judge's classifications as reliable inputs to the analyses in \S\ref{sec:vm_effect}. Full reliability statistics appear in Appendix~\ref{app:evaluating-vm}.
\section{Structural Citation Failures in Search-Augmented LLMs}
\label{sec:vm_effect}

Applying the framework to the 761,495 evaluable citation pairs in {}\CiteTrace, we ask three questions: 
(1)~how do failure rates differ across dimensions, and do they 
co-occur or trade off (\S\ref{sec:three-failures});
(2)~what explains the variation: model capability or systematic provider-level differences (\S\ref{sec:independent-failures}); and
(3)~how do citation-level failures compound into user-level exposure (\S\ref{sec:user-exposure}).
All reported rates are conservative lower bounds 
(Appendix~\ref{app:robustness-bounds}).

\begin{figure}[!t]
\centering
\includegraphics[width=0.99\linewidth]{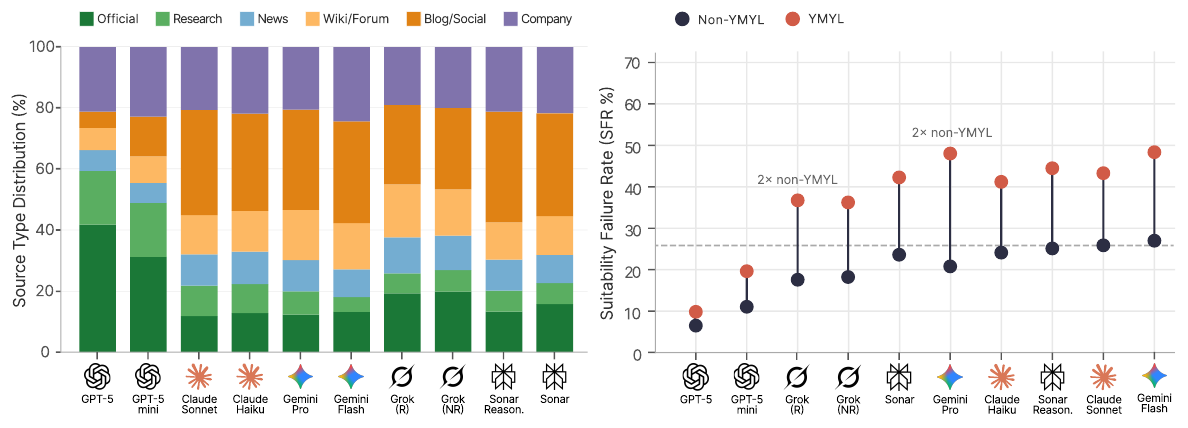}
\caption{Source-type profiles and YMYL amplification. 
\textbf{Left}: source-type distribution per model, grouped by 
provider. Models within the same provider share similar profiles; 
the main variation across providers concentrates on the 
Official--Blog axis, while Company sources remain near-constant 
at 21\%.
\textbf{Right}: Suitability Failure Rate in YMYL domains (filled) 
versus non-YMYL domains (hollow) for each model. In YMYL 
domains, SFR nearly doubles the non-YMYL baseline across most 
models, highlighting that the risks of domain-inappropriate sourcing are substantially amplified precisely in the high-stakes queries where source suitability matters most.}
\label{fig:source-ymyl}
\end{figure}

\subsection{The Three Dimensions Reveal Distinct and Independent Failure Patterns}
\label{sec:three-failures}

We begin by examining how the three dimensions fail individually and whether their failures align or diverge across models.
If citation quality were a single underlying construct, we would expect the three dimensions to co-vary;
models that fail on one would tend to fail on others.

\paragraph{The source pool is structurally skewed before any dimension-level failure occurs.}
Blog/Social (29.2\%) and Company (21.1\%) sources together account for over half of all citations, while Official Institution (17.5\%) and Research (8.8\%) sources combined make up just over a quarter (Table~\ref{tab:d_pool_composition}). 
This skew is not model-specific but consistent across all ten models, suggesting it reflects the composition of sources that search backends surface rather than deliberate model choices (Appendix~\ref{app:biased-pool}).

\paragraph{Fidelity and suitability failures are prevalent but no single factor explains both.}
The three dimensions differ not only in failure rates but in what drives them. 
Fidelity failures are the most common (FFR 30.6\%) and tend to be complete rather than partial: most citations are either fully supported or fully fabricated, with few intermediate distortions, suggesting that models make a near-binary choice to ground their claims in the source or ignore it entirely. 
The spread across models is large (12.3\%--44.9\%) and driven by model identity rather than query type (Appendix~\ref{app:failure-profile}). 
Source Suitability fails at a comparable rate (SFR 27.1\%) but with a partly inverted model ranking, already hinting that fidelity and suitability are governed by different factors. 
Intent--Purpose Alignment is an order of magnitude lower (AFR 5.1\%) and shaped by query composition rather than model choice, reflecting the dominance of inform-purpose sources in the citation pool.

\paragraph{Models that cite faithfully tend to select unsuitable 
sources.}
The two main dimensions do not fail together but against each other (Figure~\ref{fig:teaser}). 
\texttt{claude-haiku} achieves the lowest FFR (12.3\%) but ranks only 6th on suitability (SFR 30.1\%), drawing 31.8\% of its citations from Blog sources; \texttt{gpt-5} shows the reverse (FFR 42.3\%, SFR 8.0\%), drawing 41.8\% from Official sources. 
{Company sources account for a near-constant 21\% across all models; the variation concentrates on the Official--Blog axis (Figure~\ref{fig:source-ymyl}, Left).}
{The inversion sharpens} in YMYL domains, where SFR nearly doubles the non-YMYL baseline (Figure~\ref{fig:source-ymyl}, Right).
{It also recurs} along citation density: as citations per response grow, FFR drops (from 35\% to 28\%) while SFR rises (from 22\% to 31\%).
No model occupies the ideal corner of both low FFR and low SFR. 
Models sharing a search backend exhibit similar source-type profiles, suggesting the trade-off reflects retrieval infrastructure rather than generation.

\paragraph{The three dimensions fail independently, not jointly.}
If the three dimensions captured a single underlying quality factor, citations that fail on one dimension would tend to fail on others. To test this, we define a Critical VM (CritVM) instance as a citation that fails on all three dimensions simultaneously (i.e., IPA $\leq$2, ASF $\leq$2, and SS $\leq$2). 
Under the null hypothesis of independent failures, the expected CritVM rate is the product of the three marginal rates: $0.051 \times 0.306 \times 0.271 \approx 0.42$\%. 
The observed rate is 3,174 of 761,495 citations (0.42\%), matching the independent-failure expectation exactly. 
Single-dimension failures dwarf joint failures by an order of magnitude: ASF-only failures account for 22.2\% of all citations, SS-only for 19.8\%, and IPA-only for 2.2\%, while the three-dimension intersection is 0.42\%. 
This confirms that the three dimensions capture structurally distinct failure modes, and that any single-dimension evaluation would miss the majority of citation failures.

\subsection{Provider-Level Differences, Not Model Capability, Explain Most Quality Variance}
\label{sec:independent-failures}

The fidelity--suitability trade-off suggests that citation quality may be shaped more by provider-level factors than by individual model capability. 
We test this with five within-provider comparisons that hold the search backend constant while varying model scale and reasoning capability, isolating how much room the generator has to affect each dimension.

\paragraph{Scaling up does not improve citation quality.}
Within-provider model pairs that differ in scale but share a search backend show marginal quality differences (Table~\ref{tab:d_model_size_paired}). 
The Anthropic pair (\texttt{claude-sonnet} vs. \texttt{claude-haiku}) differs by less than 0.8~pp on FFR and 1.6~pp on SFR despite \texttt{claude-sonnet} issuing roughly twice as many citations per response.
The Google pair reproduces the fidelity--suitability trade-off in miniature: 
\texttt{gemini-pro} improves SFR by 3.4~pp but worsens FFR by 3.1~pp. 
The largest within-provider gap (\texttt{gpt-5} vs. \texttt{gpt-5-mini}: $-$2.6~pp FFR, $-$6.3~pp SFR) remains far smaller than the between-provider gaps observed above, suggesting that scaling model capacity within the same provider has little effect on which sources the search backend returns.

\paragraph{Reasoning helps fidelity but not source selection.}
Paired reasoning and non-reasoning models sharing a search backend isolate the generation-side effect (Table~\ref{tab:d_reasoning_paired}). 
The xAI pair shows differences of at most 2.1~pp on all three dimensions. 
The Perplexity pair shows a sharper pattern: the reasoning model issues half as many citations and cuts FFR by 16~pp (18.0\% vs. 34.1\%), but SFR and AFR change by less than 2~pp. 
When reasoning helps, it appears to affect how the model uses a source rather than which sources are retrieved. 
Fidelity is the dimension most responsive to generation-side improvements; source selection appears largely outside the generator's control.

\paragraph{Provider identity accounts for most quality variance.}
A variance decomposition confirms the pattern (Table~\ref{tab:d_variance}): provider effects account for 96\% of alignment and suitability variance, leaving under 4\% for within-provider model differences. 
Fidelity variance is also provider-dominated (88\%), but the remaining 12\% is attributable to model-level differences, consistent with the reasoning effect above. 
The asymmetry is structurally interpretable: alignment and suitability are determined by which sources the search backend retrieves, before the generator acts; fidelity additionally depends on how the generator uses those sources, leaving some room for model-level variation.

\subsection{Citation Failures Compound into Widespread User Exposure}
\label{sec:user-exposure}

Citation-level failure rates understate the user-facing impact because failures compound across the multiple citations in each response. 
Even a modest per-citation failure rate translates into near-certain exposure at the response level.

\begin{figure}[!t]
\centering
\includegraphics[width=0.99\linewidth]{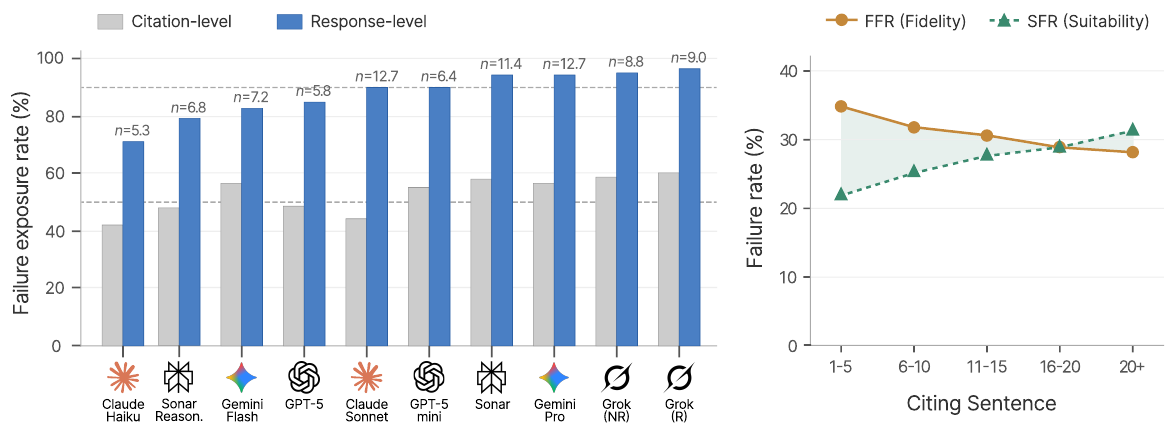}
\caption{Response-level amplification of citation failures. 
A response is counted as exposed if at least one of its citations scores $\leq 2$ on any of the three dimensions; the response-level rate is the share of such responses across all 
responses of that model.
\textbf{Left}: citation-level failure rate (gray) versus response-level exposure (blue) for each model, sorted by response-level exposure.
\textbf{Right}: fidelity and suitability failure rates by citation density bin. As citations per response grow, FFR decreases while SFR increases, reproducing the fidelity--suitability trade-off along the density axis.}
\label{fig:response-amplification}
\end{figure}

\paragraph{Most responses contain at least one flawed citation.}
\texttt{claude-sonnet}'s FFR is only 13.1\%, but with $\bar{n} = 12.7$ citations per response its response-level fidelity exposure reaches 62.9\%---a four-fold amplification. 
{Response-level exposure ranges from 71.3\% (\texttt{claude-haiku}, $\bar{n} = 5.3$) to 96.1\% (\texttt{grok-reasoning}, $\bar{n} = 9.0$); even the best-performing model leaves nearly three in four responses with at least one structurally flawed citation (Figure~\ref{fig:response-amplification}, Left).}
{The amplification tracks citation density: as $\bar{n}$ grows from the 1--5 bin to the 20{+} bin, FFR falls (33.7\% to 28.2\%) while SFR rises (22.4\% to 31.7\%), so denser responses shift the failure mode from fidelity to suitability without reducing overall exposure (Figure~\ref{fig:response-amplification}, Right).}
The fidelity--suitability trade-off of \S\ref{sec:three-failures} is not diluted but amplified at the response level: \texttt{claude-sonnet} achieves the second-lowest response-level FFR yet the highest response-level SFR, confirming that no model offers reliable protection across both dimensions.

\paragraph{Some citations are unreachable before any dimension-level failure occurs.}
Compounding this further, a non-trivial share of citations points to sources that are entirely inaccessible at the time of use due to link rot or domain decommissioning: phantom-citation rates range from 2.1\% to 15.5\% across models, with Google's two models exceeding 14\% (Table~\ref{tab:d_crawl_bias}). 
One illustrative case involves a query about U.S. climate change projections, where a model cited the Fifth National Climate Assessment hosted at a government domain that had been decommissioned months prior. 
Had the source been accessible, it would have scored SS = 5, an optimal source selection degraded entirely by infrastructure decay rather than model error. 
These phantom citations represent a failure mode distinct from the structural mismatches we measure: the source was once real and authoritative, but is no longer verifiable by the user (Appendix~\ref{app:crawl-bias}).

\paragraph{The same answer can rest on very different citation reliability.}
On a fish-oil vitamin-A query (YMYL Medical), all ten models reached the same conclusion, yet Anthropic cited NIH (FFR 0\%) while Perplexity and Google cited marketing blogs (FFR 80--100\%)---a gap invisible from the answer alone. 
On a VIX-futures contango query (YMYL Finance), provider-mean suitability scores diverged by a factor of two. 
Together, these cases illustrate that no surface-level signal distinguishes reliable from unreliable citations, and that improvement on one dimension does not transfer to another: the three dimensions are statistically independent (Appendix~\ref{app:robustness-bounds}).
Users cannot judge citation quality without examining each source individually, the very step that search-augmented responses are designed to spare them.
\section{Related Work}
\label{sec:related-work}

\paragraph{Citation faithfulness and attribution verification.}
Whether a model's generated claim is supported by its cited source is a central question in citation quality research.
Early work established the task through NLI-based and QA-based sentence-level adjudication~\citep{bohnet2022attributed,alce,press2024citeme,rashkin2023ais,aliice2024}, and verification was then scaled to full RAG pipelines through lightweight trained verifiers and automated evaluation frameworks~\citep{ragas,ares,minicheck}; citation-specific tools further classify citation intent and apply full-text evidence reasoning~\citep{haan2025semanticcite, scite2021}.
Multi-level faithfulness rubrics from abstractive summarization provide a finer-grained vocabulary for citation failures beyond binary support judgments~\citep{kryscinski2020factcc,laban2022summac,maynez2020faithfulness,min2023factscore,pagnoni2021frank,wang2020qags}.
Empirical audits confirm that failures are pervasive: correct answers can be unfaithful to their sources~\citep{generation_time_posthoc,wallat2025correctness}, up to 90\% of medical LLM responses lack full source support~\citep{sourcecheckup}, and fabricated or hallucinated citations add a further failure mode~\citep{rao2026detecting,xu2026ghostcite}.
All of these frameworks treat the A--S link as the sole criterion, leaving source appropriateness for the user's query out of scope.

\paragraph{Source quality and reliability assessment.}
A separate line of work evaluates the quality or reliability of sources independently of claim fidelity.
Credibility research shows that users judge source trustworthiness through prominence, authority, and contextual fit~\citep{fogg2003prominence,kakol2017understanding,rieh2007credibility,sun2019consumer}, and evaluation rubrics such as the CRAAP test~\citep{craap} and SourceBench~\citep{sourcebench} operationalize these criteria at the document level; Google's YMYL classification~\citep{ymyl} and RAG reliability estimation~\citep{hwang2025retrieval} extend this to high-stakes domains and retrieval-time signals.
Theoretical grounding comes from the adverse-selection model~\citep{akerlof1970} and information-asymmetry framework~\citep{arrow1963}, which explain how information quality degrades in markets for expert knowledge.
These frameworks share a structural limitation: source quality is treated as a fixed property of the page, yet the same source can be appropriate for one query intent or domain and inappropriate for another.

\paragraph{Generative search engine audits.}
A growing body of work examines how search-augmented LLMs select and present sources at the system level.
Source-selection audits show that generative engines disproportionately surface news and business sources~\citep{li2024generative}, exhibit outlet-name-driven political bias~\citep{bang2024measuring}, and differ from traditional search in coverage and citation patterns~\citep{zhang2025source}; content producers have begun optimizing for these dynamics~\citep{aggarwal2024geo}.
On the user side, citation count raises perceived credibility regardless of actual support~\citep{ding2025citationstrustllmgenerated,searcharena2026}, and system-level evaluations identify structural failure modes~\citep{aee2025}, find that only 51.5\% of cited statements are fully supported~\citep{liu2023evaluating}, and jointly assess credibility and groundedness~\citep{vykopal2026assessing}.
Each of these studies examines a single facet, including bias, preference, or verifiability, without connecting source selection to query intent or answer fidelity within a unified framework.

\paragraph{Positioning.}
Prior work measures A--S faithfulness~\citep{bohnet2022attributed,ragas,alce,minicheck,aliice2024}, absolute source quality~\citep{craap,hwang2025retrieval,sourcebench}, or system-level source patterns~\citep{li2024generative,searcharena2026,zhang2025source} in isolation; none model the conditional dependence between source type, query intent, and answer fidelity that drives the failure patterns we report.
This fragmentation means that a system can score well on any single dimension while failing structurally on another, a gap that single-dimension benchmarks cannot detect by design.
\CiteTrace{} is the first dataset to measure all three simultaneously, enabling joint analysis of the fidelity--suitability trade-off, provider-level variance, and response-level exposure that prior benchmarks leave unmeasured.
\section{Conclusion}
\label{sec:conclusion}

We {introduce} \CiteTrace{}, a large-scale dataset of 761{,}495 citation pairs from ten search-augmented LLMs across $11{,}200$ real-world queries, and a three-dimension evaluation framework for citation quality jointly across intent--purpose alignment, source suitability, and answer--source fidelity{,} aspects that existing benchmarks assess in isolation or not at all.
\VM{} (\VMshort{}) names the phenomenon in which search-augmented LLMs cite real, accessible sources that mislead through intent--purpose misalignment, domain-inappropriate sourcing, or distortion of source content.
Our analyses demonstrate that models that cite faithfully tend to select unsuitable sources, while models that select suitable sources distort their content, a trade-off that appears to stem from factors beyond any single model's generation capability, and remains invisible to any single-dimension evaluation.
{Our findings suggest that improving generation alone may be insufficient to resolve citation quality, and that dedicated attention to source selection, at either the retrieval or generation stage, may be equally warranted.}
{\CiteTrace{} and our evaluation framework are released as a reusable resource; we hope they catalyze future efforts to improve source selection and establish citation quality as a first-class evaluation criterion for search-augmented systems.}
As these systems mediate access to information for millions of users daily, the structural failures we document are not edge cases but routine events whose scale and invisibility make them a concern for researchers and developers.

\bibliography{references}

@inproceedings{searcharena2026,
  title     = {Search Arena: Analyzing Search-Augmented {LLM}s},
  author    = {Miroyan, Mihran and Wu, Tsung-Han and King, Logan and Li, Tianle and Pan, Jiayi and Hu, Xinyan and Chiang, Wei-Lin and Angelopoulos, Anastasios N. and Darrell, Trevor and Norouzi, Narges and Gonzalez, Joseph E.},
  booktitle = {Proceedings of the International Conference on Learning Representations ({ICLR})},
  year      = {2026},
  note      = {arXiv:2506.05334},
}

@inproceedings{wallat2025correctness,
title={Correctness is not Faithfulness in Retrieval Augmented Generation Attributions}, 
author={Wallat, Jonas and Heuss, Maria and Rijke, Maarten de and Anand, Avishek}, booktitle={Proceedings of the 2025 International ACM SIGIR Conference on Innovative Concepts and Theories in Information Retrieval (ICTIR)}, 
pages={22--32}, 
year={2025}
}

@inproceedings{ragas,
  title     = {{RAGAS}: Automated Evaluation of Retrieval Augmented Generation},
  author    = {Es, Shahul and James, Jithin and Anke, Luis Espinosa and Schockaert, Steven},
  booktitle = {Proceedings of the 18th Conference of the European Chapter of the Association for Computational Linguistics: System Demonstrations},
  pages     = {150--158},
  year      = {2024},
}

@inproceedings{ares,
  title     = {{ARES}: An Automated Evaluation Framework for Retrieval-Augmented Generation Systems},
  author    = {Saad-Falcon, Jon and Khattab, Omar and Potts, Christopher and Zaharia, Matei},
  booktitle = {Proceedings of the 2024 Conference of the North American Chapter of the Association for Computational Linguistics: Human Language Technologies (Volume 1: Long Papers)},
  pages     = {338--354},
  year      = {2024},
}

@inproceedings{alce,
  title     = {Enabling Large Language Models to Generate Text with Citations},
  author    = {Gao, Tianyu and Yen, Howard and Yu, Jiatong and Chen, Danqi},
  booktitle = {Proceedings of the 2023 Conference on Empirical Methods in Natural Language Processing ({EMNLP})},
  pages     = {6465--6488},
  year      = {2023},
}

@article{sourcebench,
  title   = {{SourceBench}: Can {AI} Answers Reference Quality Web Sources?},
  author  = {Jin, Hexi and Liu, Stephen and Li, Yuheng and Malik, Simran and Zhang, Yiying},
  journal = {arXiv preprint arXiv:2602.16942},
  year    = {2026},
}

@article{sourcecheckup,
  title   = {An automated framework for assessing how well {LLM}s cite relevant medical references},
  author  = {Wu, Kevin and Wu, Eric and Wei, Kevin and Zhang, Angela and Casasola, Allison and Nguyen, Teresa and Riantawan, Sith and Shi, Patricia and Ho, Daniel and Zou, James},
  journal = {Nature Communications},
  volume  = {16},
  number  = {1},
  pages   = {3615},
  year    = {2025},
  doi     = {10.1038/s41467-025-58551-6},
}

@inproceedings{aee2025,
  title     = {Search Engines in the {AI} Era: A Qualitative Understanding to the False Promise of Factual and Verifiable Source-Cited Responses in {LLM}-based Search},
  author    = {Narayanan Venkit, Pranav and Laban, Philippe and Zhou, Yilun and Mao, Yixin and Wu, Chien-Sheng},
  booktitle = {Proceedings of the 2025 ACM Conference on Fairness, Accountability, and Transparency ({FAccT})},
  pages     = {1325--1340},
  year      = {2025},
  doi       = {10.1145/3715275.3732089},
}

@article{xu2026ghostcite,
  title={GhostCite: A Large-Scale Analysis of Citation Validity in the Age of Large Language Models},
  author={Xu, Zuyao and Qiu, Yuqi and Sun, Lu and Miao, FaSheng and Wu, Fubin and Wang, Xinyi and Li, Xiang and Lu, Haozhe and Zhang, ZhengZe and Hu, Yuxin and others},
  journal={arXiv preprint arXiv:2602.06718},
  year={2026}
}

@article{rao2026detecting, 
  title={Detecting and Correcting Reference Hallucinations in Commercial LLMs and Deep Research Agents}, 
  author={Rao, Delip and Wong, Eric and Callison-Burch, Chris}, 
  journal={arXiv preprint arXiv:2604.03173}, 
  year={2026}
}

@inproceedings{akhtar2024croissant,
  title     = {Croissant: A Metadata Format for {ML}-Ready Datasets},
  author    = {Akhtar, Mubashara and Benjelloun, Omar and Conforti, Costanza
               and Foschini, Luca and Gijsbers, Pieter and Giner-Miguelez, Joan
               and Goswami, Sujata and Jain, Nitisha and Karamousadakis, Michalis
               and Krishna, Satyapriya and Kuchnik, Michael and Lesage, Sylvain
               and Lhoest, Quentin and Marcenac, Pierre and Maskey, Manil
               and Mattson, Peter and Oala, Luis and Oderinwale, Hamidah
               and Ruyssen, Pierre and Santos, Tim and Shinde, Rajat
               and Simperl, Elena and Suresh, Arjun and Thomas, Goeffry
               and Tykhonov, Slava and Vanschoren, Joaquin and Varma, Susheel
               and van der Velde, Jos and Vogler, Steffen and Wu, Carole-Jean
               and Zhang, Luyao},
  booktitle = {Advances in Neural Information Processing Systems ({NeurIPS})},
  volume    = {37},
  pages     = {82133--82148},
  publisher = {Curran Associates, Inc.},
  year      = {2024},
  note      = {Datasets and Benchmarks Track, Spotlight},
}

@article{akerlof1970,
  title   = {The Market for ``Lemons'': Quality Uncertainty and the Market Mechanism},
  author  = {Akerlof, George A.},
  journal = {The Quarterly Journal of Economics},
  volume  = {84},
  number  = {3},
  pages   = {488--500},
  year    = {1970},
}

@article{arrow1963,
  title   = {Uncertainty and the Welfare Economics of Medical Care},
  author  = {Arrow, Kenneth J.},
  journal = {The American Economic Review},
  volume  = {53},
  number  = {5},
  pages   = {941--973},
  year    = {1963},
}

@inproceedings{cqadupstack,
  title     = {{CQADupStack}: A Benchmark Data Set for Community Question-Answering Research},
  author    = {Hoogeveen, Doris and Verspoor, Karin M. and Baldwin, Timothy},
  booktitle = {Proceedings of the 20th Australasian Document Computing Symposium},
  pages     = {3:1--3:8},
  year      = {2015},
}

@inproceedings{prism,
  title     = {{PRISM}: A Participatory, Representative and Individualised Evaluation of Language Model Alignment},
  author    = {Kirk, Hannah Rose and Vidgen, Bertie and Röttger, Paul and Hale, Scott A.},
  booktitle = {Advances in Neural Information Processing Systems ({NeurIPS}) Datasets and Benchmarks Track},
  year      = {2024},
  note      = {Best Paper Award},
}

@article{koster2022rfc, 
  title={Rfc 9309 robots exclusion protocol}, author={Koster, Martijn and Illyes, Gary and Zeller, Henner and Sassman, Lizzi}, journal={Internet Engineering Task Force}, year={2022}
}

@article{craap,
  title   = {The {CRAAP} Test},
  author  = {Blakeslee, Sarah},
  journal = {LOEX Quarterly},
  volume  = {31},
  number  = {3},
  pages   = {4},
  year    = {2004},
}

@article{rieh2007credibility,
  title   = {Credibility: A Multidisciplinary Framework},
  author  = {Rieh, Soo Young and Danielson, David R.},
  journal = {Annual Review of Information Science and Technology},
  volume  = {41},
  number  = {1},
  pages   = {307--364},
  year    = {2007},
  doi     = {10.1002/aris.2007.1440410114},
}

@article{sun2019consumer,
  title   = {Consumer Evaluation of the Quality of Online Health Information: Systematic Literature Review of Relevant Criteria and Indicators},
  author  = {Sun, Yalin and Zhang, Yan and Gwizdka, Jacek and Trace, Ciaran B.},
  journal = {Journal of Medical Internet Research},
  volume  = {21},
  number  = {5},
  pages   = {e12522},
  year    = {2019},
  doi     = {10.2196/12522},
}

@article{tanzil2025stackoverflow,
  title   = {A Systematic Mapping Study of Crowd Knowledge Enhanced Software Engineering Research Using {Stack Overflow}},
  author  = {Tanzil, Minaoar Hossain and Chowdhury, Shaiful and Modaberi, Somayeh and Uddin, Gias and Hemmati, Hadi},
  journal = {Journal of Systems and Software},
  volume  = {226},
  pages   = {112405},
  year    = {2025},
  doi     = {10.1016/j.jss.2025.112405},
}

@inproceedings{pagnoni2021frank,
  title     = {Understanding Factuality in Abstractive Summarization with {FRANK}: A Benchmark for Factuality Metrics},
  author    = {Pagnoni, Artidoro and Balachandran, Vidhisha and Tsvetkov, Yulia},
  booktitle = {Proceedings of the 2021 Conference of the North American Chapter of the Association for Computational Linguistics: Human Language Technologies},
  pages     = {4812--4829},
  year      = {2021},
  doi       = {10.18653/v1/2021.naacl-main.383},
}

@inproceedings{maynez2020faithfulness,
  title     = {On Faithfulness and Factuality in Abstractive Summarization},
  author    = {Maynez, Joshua and Narayan, Shashi and Bohnet, Bernd and McDonald, Ryan},
  booktitle = {Proceedings of the 58th Annual Meeting of the Association for Computational Linguistics},
  pages     = {1906--1919},
  year      = {2020},
  doi       = {10.18653/v1/2020.acl-main.173},
  url       = {https://aclanthology.org/2020.acl-main.173/},
}

@inproceedings{kryscinski2020factcc,
  title     = {Evaluating the Factual Consistency of Abstractive Text Summarization},
  author    = {Kryscinski, Wojciech and McCann, Bryan and Xiong, Caiming and Socher, Richard},
  booktitle = {Proceedings of the 2020 Conference on Empirical Methods in Natural Language Processing},
  pages     = {9332--9346},
  year      = {2020},
  doi       = {10.18653/v1/2020.emnlp-main.750},
  url       = {https://aclanthology.org/2020.emnlp-main.750/},
}

@inproceedings{wang2020qags,
  title     = {Asking and Answering Questions to Evaluate the Factual Consistency of Summaries},
  author    = {Wang, Alex and Cho, Kyunghyun and Lewis, Mike},
  booktitle = {Proceedings of the 58th Annual Meeting of the Association for Computational Linguistics},
  pages     = {5008--5020},
  year      = {2020},
  doi       = {10.18653/v1/2020.acl-main.450},
  url       = {https://aclanthology.org/2020.acl-main.450/},
}

@article{laban2022summac,
  title   = {{SummaC}: Re-Visiting {NLI}-based Models for Inconsistency Detection in Summarization},
  author  = {Laban, Philippe and Schnabel, Tobias and Bennett, Paul N. and Hearst, Marti A.},
  journal = {Transactions of the Association for Computational Linguistics},
  volume  = {10},
  pages   = {163--177},
  year    = {2022},
  doi     = {10.1162/tacl_a_00453},
}

@inproceedings{min2023factscore,
  title     = {{FActScore}: Fine-grained Atomic Evaluation of Factual Precision in Long Form Text Generation},
  author    = {Min, Sewon and Krishna, Kalpesh and Lyu, Xinxi and Lewis, Mike and Yih, Wen-tau and Koh, Pang and Iyyer, Mohit and Zettlemoyer, Luke and Hajishirzi, Hannaneh},
  booktitle = {Proceedings of the 2023 Conference on Empirical Methods in Natural Language Processing},
  pages     = {12076--12100},
  year      = {2023},
  doi       = {10.18653/v1/2023.emnlp-main.741},
  url       = {https://aclanthology.org/2023.emnlp-main.741/},
}

@misc{ymyl,
  title        = {Search Quality Evaluator Guidelines},
  author       = {{Google}},
  year         = {2022},
  howpublished = {\url{https://guidelines.raterhub.com/searchqualityevaluatorguidelines.pdf}},
}

@article{haan2025semanticcite, 
  title={SemanticCite: Citation Verification with AI-Powered Full-Text Analysis and Evidence-Based Reasoning}, 
  author={Haan, Sebastian}, 
  journal={arXiv preprint arXiv:2511.16198}, 
  year={2025}
}

@inproceedings{vykopal2026assessing, 
  title={Assessing Web Search Credibility and Response Groundedness in Chat Assistants}, 
  author={Vykopal, Ivan and Pikuliak, Mat{\'u}{\v{s}} and Ostermann, Simon and Simko, Marian}, 
  booktitle={Proceedings of the 19th Conference of the European Chapter of the Association for Computational Linguistics (Volume 1: Long Papers)}, 
  pages={2539--2560}, year={2026}
}

@article{bohnet2022attributed, 
title={Attributed question answering: Evaluation and modeling for attributed large language models},
 author={Bohnet, Bernd and Tran, Vinh Q and Verga, Pat and Aharoni, Roee and Andor, Daniel and Soares, Livio Baldini and Ciaramita, Massimiliano and Eisenstein, Jacob and Ganchev, Kuzman and Herzig, Jonathan and others}, 
journal={arXiv preprint arXiv:2212.08037}, year={2022}
}

@inproceedings{minicheck,
  title     = {{MiniCheck}: Efficient Fact-Checking of {LLM}s on Grounding Documents},
  author={Tang, Liyan and Laban, Philippe and Durrett, Greg}, 
  booktitle = {Proceedings of the 2024 Conference on Empirical Methods in Natural Language Processing ({EMNLP})},
  year      = {2024},
}

@article{generation_time_posthoc,
title   = {Generation-Time vs.\ Post-hoc Citation: A Holistic Evaluation
of {LLM} Attribution},
author  = {Saxena, Yash and Bommireddy, Raviteja and Padia, Ankur
and Gaur, Manas},
journal = {arXiv preprint arXiv:2509.21557},
year    = {2025},
}

@article{li2024generative,
 title={Generative ai search engines as arbiters of public knowledge: An audit of bias and authority}, 
author={Li, Alice and Sinnamon, Luanne}, 
journal={Proceedings of the Association for Information Science and Technology},
 volume={61}, 
number={1}, 
pages={205--217}, 
year={2024}, 
publisher={Wiley Online Library}
}

@article{press2024citeme,
  title={CiteME: Can Language Models Accurately Cite Scientific Claims?},
  author={Press, Ori and Hochlehnert, Andreas and Prabhu, Ameya and Udandarao, Vishaal and Press, Ofir and Bethge, Matthias},
  journal={Advances in Neural Information Processing Systems},
  volume={37},
  pages={7847--7877},
  year={2024}
}

@inproceedings{hwang2025retrieval,
  title={Retrieval-Augmented Generation with Estimation of Source Reliability},
  author={Hwang, Jeongyeon and Park, Junyoung and Park, Hyejin and Kim, Dongwoo and Park, Sangdon and Ok, Jungseul},
  booktitle={Proceedings of the 2025 Conference on Empirical Methods in Natural Language Processing},
  pages={34279--34303},
  year={2025},
  publisher={Association for Computational Linguistics},
  url={https://aclanthology.org/2025.emnlp-main.1738/}
}

@article{zhang2025source,
  title={Source Coverage and Citation Bias in LLM-based vs. Traditional Search Engines},
  author={Zhang, Peixian and Ye, Qiming and Peng, Zifan and Garimella, Kiran and Tyson, Gareth},
  journal={arXiv preprint arXiv:2512.09483},
  year={2025}
}

@article{kakol2017understanding, 
  title={Understanding and predicting web content credibility using the content credibility corpus}, author={Kakol, Michal and Nielek, Radoslaw and Wierzbicki, Adam}, journal={Information Processing \& Management}, volume={53}, number={5}, pages={1043--1061}, year={2017}, publisher={Elsevier}
}

@article{cohen1960,
  title   = {A Coefficient of Agreement for Nominal Scales},
  author  = {Cohen, Jacob},
  journal = {Educational and Psychological Measurement},
  volume  = {20},
  number  = {1},
  pages   = {37--46},
  year    = {1960},
}

@techreport{krippendorff2011,
author      = {Krippendorff, Klaus},
title       = {Computing {K}rippendorff's Alpha-Reliability},
institution = {Annenberg School for Communication, University of Pennsylvania},
year        = {2011},
number      = {43},
url         = {https://repository.upenn.edu/asc_papers/43},
}

@article{ji2023hallucination,
  title   = {Survey of Hallucination in Natural Language Generation},
  author  = {Ji, Ziwei and Lee, Nayeon and Frieske, Rita and Yu, Tiezheng and Su, Dan and Xu, Yan and Ishii, Etsuko and Bang, Ye Jin and Madotto, Andrea and Fung, Pascale},
  journal = {ACM Computing Surveys},
  volume  = {55},
  number  = {12},
  pages   = {1--38},
  year    = {2023},
}

@inproceedings{broder2002,
  title     = {A Taxonomy of Web Search},
  author    = {Broder, Andrei},
  booktitle = {ACM SIGIR Forum},
  volume    = {36},
  number    = {2},
  pages     = {3--10},
  year      = {2002},
}

@book{swales1990,
  author    = {Swales, John M.},
  title     = {Genre Analysis: {E}nglish in Academic and 
               Research Settings},
  publisher = {Cambridge University Press},
  year      = {1990},
  address   = {Cambridge},
}

@article{taylor1968question,
  title   = {Question-Negotiation and Information Seeking in Libraries},
  author  = {Taylor, Robert S.},
  journal = {College \& Research Libraries},
  volume  = {29},
  number  = {3},
  pages   = {178--194},
  year    = {1968},
  doi     = {10.5860/crl_29_03_178},
}

@book{ingwersen2005turn,
  title     = {The Turn: Integration of Information Seeking and Retrieval in Context},
  author    = {Ingwersen, Peter and J{\"a}rvelin, Kalervo},
  publisher = {Springer},
  address   = {Dordrecht},
  series    = {The Information Retrieval Series},
  volume    = {18},
  year      = {2005},
  doi       = {10.1007/1-4020-3851-8},
  isbn      = {978-1-4020-3851-8},
}

@inproceedings{barbaresi2021,
  title     = {Trafilatura: A Web Scraping Library and Command-Line Tool for Text Discovery and Extraction},
  author    = {Barbaresi, Adrien},
  booktitle = {Proceedings of the Joint Conference of the 59th Annual Meeting of the Association for Computational Linguistics and the 11th International Joint Conference on Natural Language Processing: System Demonstrations},
  pages     = {122--131},
  year      = {2021},
}

@article{scite2021,
  title   = {Scite: A Smart Citation Index that Displays the Context of Citations and Classifies their Intent using Deep Learning},
  author  = {Nicholson, Josh M. and Mordaunt, Milo and Lopez, Patrice and Uppala, Ashish and Rosber, Domenic and Bber, Sean P. and Thaler, Christoph and Deng, Yuhao and Greene, Casey S. and Nishi, Satoshi},
  journal = {Quantitative Science Studies},
  volume  = {2},
  number  = {3},
  pages   = {882--898},
  year    = {2021},
}

@inproceedings{fogg2003prominence,
  title     = {How Do Users Evaluate the Credibility of Web Sites? A Study with Over 2,500 Participants},
  author    = {Fogg, B. J. and Soohoo, Cathy and Danielson, David R. and Marable, Leslie and Stanford, Julianne and Tauber, Ellen R.},
  booktitle = {Proceedings of the 2003 Conference on Designing for User Experiences ({DUX})},
  pages     = {1--15},
  year      = {2003},
}

@inproceedings{aggarwal2024geo, 
title={Geo: Generative engine optimization}, 
author={Aggarwal, Pranjal and Murahari, Vishvak and Rajpurohit, Tanmay and Kalyan, Ashwin and Narasimhan, Karthik and Deshpande, Ameet}, 
booktitle={Proceedings of the 30th ACM SIGKDD conference on knowledge discovery and data mining}, 
pages={5--16}, 
year={2024}
}

@article{liu2023evaluating,
  title   = {Evaluating Verifiability in Generative Search Engines},
  author  = {Liu, Nelson F. and Zhang, Tianyi and Liang, Percy},
  journal = {Findings of the Association for Computational Linguistics ({EMNLP})},
  pages   = {7001--7025},
  year    = {2023},
}

@misc{schemaorg,
  author       = {{Schema.org Community Group}},
  title        = {Schema.org: A Shared Vocabulary for Structured Data},
  howpublished = {\url{https://schema.org}},
  year         = {2011},
  note         = {Founded by Google, Microsoft, Yahoo, and Yandex},
}

@misc{iab_taxonomy,
  author       = {{IAB Tech Lab}},
  title        = {Content Taxonomy 3.0},
  howpublished = {IAB Tech Lab Standard},
  year         = {2022},
  url          = {https://iabtechlab.com/standards/content-taxonomy/},
}

@inproceedings{bang2024measuring, 

title={Measuring political bias in large language models: What is said and how it is said},
 author={Bang, Yejin and Chen, Delong and Lee, Nayeon and Fung, Pascale}, 
booktitle={Proceedings of the 62nd Annual Meeting of the Association for Computational Linguistics (Volume 1: Long Papers)}, 
pages={11142--11159}, year={2024}
}

@inproceedings{aliice2024,
  title     = {{ALiiCE}: Evaluating Positional Fine-Grained Citation Generation},
  author    = {Xu, Yilong and Niu, Jinhua and Xie, Guoxin},
  booktitle = {Proceedings of the 2024 Conference on Empirical Methods in Natural Language Processing ({EMNLP})},
  year      = {2024},
}

@article{koo2016guideline,
  title={A guideline of selecting and reporting intraclass correlation coefficients for reliability research},
  author={Koo, Terry K and Li, Mae Y},
  journal={Journal of chiropractic medicine},
  volume={15},
  number={2},
  pages={155--163},
  year={2016},
  publisher={National University of Health Sciences}
}

@article{landis1977measurement,
  title={The measurement of observer agreement for categorical data},
  author={Landis, J Richard and Koch, Gary G},
  journal={biometrics},
  pages={159--174},
  year={1977},
  publisher={JSTOR}
}

@inproceedings{rose2004understanding,
  title={Understanding user goals in web search},
  author={Rose, Daniel E and Levinson, Danny},
  booktitle={Proceedings of the 13th international conference on World Wide Web},
  pages={13--19},
  year={2004}
}

@article{jansen2008determining,
  title={Determining the informational, navigational, and transactional intent of Web queries},
  author={Jansen, Bernard J and Booth, Danielle L and Spink, Amanda},
  journal={Information Processing \& Management},
  volume={44},
  number={3},
  pages={1251--1266},
  year={2008},
  publisher={Elsevier}
}

@article{biber2015exploring,
  title={Exploring the composition of the searchable web: A corpus-based taxonomy of web registers},
  author={Biber, Douglas and Egbert, Jesse and Davies, Mark},
  journal={Corpora},
  volume={10},
  number={1},
  pages={11--45},
  year={2015},
  publisher={Edinburgh University Press 22 George Square, Edinburgh EH8 9LF UK}
}

@article{sharoff2018functional,
  title={Functional text dimensions for the annotation of web corpora},
  author={Sharoff, Serge},
  journal={Corpora},
  volume={13},
  number={1},
  pages={65--95},
  year={2018},
  publisher={Edinburgh University Press The Tun-Holyrood Road, 12 (2f) Jackson's Entry~…}
}

@article{rashkin2023ais,
    title = "Measuring Attribution in Natural Language Generation Models",
    author = "Rashkin, Hannah  and
      Nikolaev, Vitaly  and
      Lamm, Matthew  and
      Aroyo, Lora  and
      Collins, Michael  and
      Das, Dipanjan  and
      Petrov, Slav  and
      Tomar, Gaurav Singh  and
      Turc, Iulia  and
      Reitter, David",
    journal = "Computational Linguistics",
    volume = "49",
    number = "4",
    month = dec,
    year = "2023",
    address = "Cambridge, MA",
    publisher = "MIT Press",
    url = "https://aclanthology.org/2023.cl-4.2/",
    doi = "10.1162/coli_a_00486",
    pages = "777--840"
}

@inproceedings{bolotova2022non,
  title={A non-factoid question-answering taxonomy},
  author={Bolotova, Valeriia and Blinov, Vladislav and Scholer, Falk and Croft, W Bruce and Sanderson, Mark},
  booktitle={Proceedings of the 45th International ACM SIGIR Conference on Research and Development in Information Retrieval},
  pages={1196--1207},
  year={2022}
}

@misc{ding2025citationstrustllmgenerated,
      title={Citations and Trust in LLM Generated Responses}, 
      author={Yifan Ding and Matthew Facciani and Amrit Poudel and Ellen Joyce and Salvador Aguinaga and Balaji Veeramani and Sanmitra Bhattacharya and Tim Weninger},
      year={2025},
      eprint={2501.01303},
      archivePrefix={arXiv},
      primaryClass={cs.CL},
      url={https://arxiv.org/abs/2501.01303}, 
}
\bibliographystyle{plainnat}

\clearpage
\appendix

{\Large\bfseries Appendices}
\addcontentsline{toc}{section}{Appendices}
\vspace{1.0em}

\noindent
\textbf{A\quad Discussions: Scope, Limitations, and Broader Impact}%
\dotfill\textbf{\pageref{app:scope}}\par
\nopagebreak
\hspace*{1.5em}A.1\quad Scope and Operational Assumptions\dotfill\pageref{app:scope-assumptions}\par
\hspace*{1.5em}A.2\quad Limitations\dotfill\pageref{app:limitations}\par
\hspace*{1.5em}A.3\quad Future Directions\dotfill\pageref{app:future-directions}\par
\hspace*{1.5em}A.4\quad Broader Impact\dotfill\pageref{app:broader-impact}\par
\hspace*{1.5em}A.5\quad Comparison to Adjacent Failure Modes\dotfill\pageref{app:adjacent-modes}\par
\vspace{0.7em}

\noindent
\textbf{B\quad \CiteTrace{}: Dataset Construction}%
\dotfill\textbf{\pageref{app:constructing-citetrace}}\par
\nopagebreak
\hspace*{1.5em}B.1\quad Sourcing Real-World Queries from Stack Exchange Communities\dotfill\pageref{app:sourcing-queries}\par
\hspace*{1.5em}B.2\quad Collecting Search-Augmented Responses and Extracting Citations\dotfill\pageref{app:collecting-responses}\par
\hspace*{1.5em}B.3\quad Crawling and Verifying Source Content\dotfill\pageref{app:crawling-sources}\par
\vspace{0.7em}

\noindent
\textbf{C\quad Three-Dimension Evaluation Framework}%
\dotfill\textbf{\pageref{app:evaluating-vm}}\par
\nopagebreak
\hspace*{1.5em}C.1\quad Query--Source Alignment\dotfill\pageref{app:axis1-detail}\par
\hspace*{1.5em}C.2\quad Source Suitability\dotfill\pageref{app:axis3-detail}\par
\hspace*{1.5em}C.3\quad Answer--Source Fidelity\dotfill\pageref{app:axis2-detail}\par
\hspace*{1.5em}C.4\quad Cross-Dimension Integration and Robustness\dotfill\pageref{app:cross-axis-detail}\par
\vspace{0.7em}

\noindent
\textbf{D\quad The \VMshort{} Effect: Detailed Empirical Results}%
\dotfill\textbf{\pageref{app:results}}\par
\nopagebreak
\hspace*{1.5em}D.1\quad A Structurally Biased Source Pool\dotfill\pageref{app:biased-pool}\par
\hspace*{1.5em}D.2\quad Aggregate Failure Rates Across Three Dimensions\dotfill\pageref{app:failure-profile}\par
\hspace*{1.5em}D.3\quad The Fidelity--Suitability Trade-off Across Models\dotfill\pageref{app:three-failures}\par
\hspace*{1.5em}D.4\quad How Provider, Scale, and Reasoning Shape Citation Quality\dotfill\pageref{app:search-gen-independent}\par
\hspace*{1.5em}D.5\quad Response-Level Failure Exposure\dotfill\pageref{app:citations-to-users}\par
\hspace*{1.5em}D.6\quad Robustness and Lower Bounds\dotfill\pageref{app:robustness-bounds}\par
\hspace*{1.5em}D.7\quad Qualitative Failure Analysis\dotfill\pageref{app:cases}\par
\vspace{0.7em}

\noindent
\textbf{E\quad Data Release and Reproducibility}%
\dotfill\textbf{\pageref{app:data-release}}\par
\nopagebreak
\hspace*{1.5em}E.1\quad Data Access and Licensing\dotfill\pageref{app:data-access}\par
\hspace*{1.5em}E.2\quad Schema and Field Documentation\dotfill\pageref{app:schema}\par
\hspace*{1.5em}E.3\quad Reproducibility Notes\dotfill\pageref{app:reproducibility}\par
\vspace{0.7em}

\clearpage

\section{Discussion: Scope, Limitations, and Broader Impact}
\label{app:scope}

This appendix articulates the scope of evaluative claims our benchmark supports (\S\ref{app:scope-assumptions}), identifies five categories of limitations and their downstream effects (\S\ref{app:limitations}), proposes four research directions through which the framework can be extended (\S\ref{app:future-directions}), discusses societal and engineering implications (\S\ref{app:broader-impact}), and positions \VMshort{} relative to adjacent failure modes (\S\ref{app:adjacent-modes}).

\subsection{Scope and Operational Assumptions}
\label{app:scope-assumptions}

\paragraph{Construct.}
\VM{} (\VMshort{}) refers to the structural condition under which a citation-bearing answer may mislead users despite the cited source being real, accessible, and faithfully cited, a condition distinct from hallucination~\citep{ji2023hallucination}.
We operationalize this as the simultaneous failure of three structurally distinct axes: query intent--source purpose alignment, answer--source fidelity, and source-type suitability for the domain.
The three dimensions capture statistically independent failure modes; as Appendix~\ref{app:cross-axis-detail} shows, single-dimension failures dwarf joint failures by an order of magnitude, so a single-dimension evaluation framework would miss the structural patterns we report.

\paragraph{Supported claims.}
\CiteTrace{} supports three claim types: \emph{existence}, that citation pairs failing all three dimensions are measurable and non-degenerate within our setting (28 communities, 10 models, English, March--April~2026); \emph{comparison}, that ranking-level differences across models, providers, and domains are robust (Appendix~\ref{app:variance}, \ref{app:threshold}); and \emph{diagnosis}, identifying which axis fails individually and how the three co-fail (Appendix~\ref{app:axis1-results}--\ref{app:variance}).
We do \emph{not} support claims of permanent provider quality, generalization to non-English settings, or causal isolation of retrieval versus generation.

\paragraph{Measurement assumptions.}
Five assumptions support our measurements; each is paired with a corresponding limitation in \S\ref{app:limitations}.
\begin{itemize}[leftmargin=1.4em,itemsep=2pt,topsep=2pt]
\item[\textbf{A1}] \textbf{Citation-marker semantics.} We interpret each provider's citation marker as asserting source--answer grounding, consistent with all five providers' API documentation (Appendix~\ref{app:collecting-responses}).
\item[\textbf{A2}] \textbf{Predefined-matrix validity.} IPA and SS matrices are expert-designed instruments; rankings remain stable under cell-wise $\pm 1$ perturbations (Appendix~\ref{app:human-validation}, \ref{app:threshold}).
{\item[\textbf{A3}] \textbf{LLM-judge reliability.} GPT-4o-mini's per-dimension $\kappa$ ($0.788$--$0.879$) meets the Landis--Koch substantial-agreement threshold~\citep{landis1977measurement}; three-annotator human validation (Krippendorff $\alpha \geq 0.811$~\citep{krippendorff2011}) supports taxonomy stability (Appendix~\ref{app:judge_reliability})}.
\item[\textbf{A4}] \textbf{Query representativeness.} Stack Exchange Q\&A represents one dimension of real-world information needs; we make no claim about conversational or multi-turn search.
\item[\textbf{A5}] \textbf{Temporal snapshot.} Results reflect a 15-day window; we do not claim trends outside this window.
\end{itemize}

\subsection{Limitations}
\label{app:limitations}

We acknowledge five categories of limitations that bound the interpretation of our findings, pairing each with its downstream effect.

\paragraph{Crawl-failure coverage bias.}
36.8\% of cited URLs failed to crawl, with failures concentrated on Forum/Q\&A (67.9\%) and Social/Blog (55.0\%) hosts due to Cloudflare bot blocking (Appendix~\ref{app:crawl-bias}).
{The downstream metrics (AFR, SFR, FFR) are computed only over successfully crawled citations and therefore exclude these failures from the denominator;} we estimate that including failed pages would raise YMYL SFR by 7--9 percentage points beyond the reported 27.1\%.
Future work could supplement live crawling with archival snapshots (e.g., the Internet Archive) for bot-blocked URLs.

\paragraph{LLM-as-Judge dependence.}
All taxonomy classifications and fidelity adjudication share a single LLM dependency (GPT-4o-mini), which may propagate systematic biases into downstream metrics in ways not fully captured by aggregate $\kappa$.
This single-judge design follows the established methodology of comparable frameworks (ALCE~\citep{alce} validates against a single NLI model, RAGAS~\citep{ragas} prompts a single LLM, and ARES~\citep{ares} trains one judge per dimension), and our per-dimension $\kappa$ (0.788--0.879) substantially exceeds their reported agreement levels.
We further mitigate single-judge risk through an explicit \texttt{UNEVALUABLE} escape label that removes ambiguous cases from the evaluation pool (Appendix~\ref{app:crawling-sources}).
Nonetheless, absolute values may shift under judge replacement; we report \emph{relative} cross-model and cross-provider patterns as the primary claims and encourage replication with an alternative judge family.

\paragraph{Predefined-matrix subjectivity.}
The IPA and SS matrices are expert-designed instruments rather than data-derived scores.
ICC~\citep{koo2016guideline} and threshold-sensitivity analyses confirm rank stability under cell-wise $\pm 1$ perturbations (Kendall $\tau \geq 0.82$ in five of seven variants), but reasonable experts may assign different scores to specific cells.
A more ambitious extension would elicit cell values from a structured panel of 20--50 experts per domain, allowing publication of a probabilistic version of each matrix.

\paragraph{Linguistic and cultural scope.}
\CiteTrace{}'s 11{,}200 queries are drawn from English-language Stack Exchange.
Information-asymmetry dynamics may differ in non-English contexts where the commercial-source landscape, regulatory environment, and platform ecosystems differ; for example, in markets with prominent state-affiliated media or single-platform Q\&A communities (Naver Knowledge-iN, Zhihu), our SP/ST taxonomies may need refinement.
We view cross-cultural extension (\S\ref{app:future-directions}) as a high-priority direction.

\paragraph{Temporal snapshot and causal attribution.}
Each provider's search tool issues live web queries, so identical prompts may return different sources at different times.
We minimize variation through a 15-day collection window, with all axes within ±5 percentage points across the window's two halves (Appendix~\ref{app:threshold}), but full reproducibility is not attainable.
Our framework also characterizes \emph{where} \VMshort{} concentrates but does not isolate \emph{which pipeline stage} (retrieval, generation, or interaction) produces it.
Controlled experiments with fixed retrieval or fixed generation would enable sharper causal decomposition; this is currently blocked by the closed-source nature of commercial retrieval pipelines.

\subsection{Future Directions}
\label{app:future-directions}

The limitations above suggest four concrete research directions.

\paragraph{Multilingual and cross-cultural extension.}
Extending \CiteTrace{} to non-English Stack Exchange and to non-Stack-Exchange platforms (Zhihu, Naver Knowledge-iN, Reddit) would test whether \VMshort{} concentrates on the same domain--source-type combinations across cultures and whether the 88--96\% provider variance dominance holds when search backends are tuned to non-English markets.

\paragraph{Longitudinal observation.}
A quarterly variant (\CiteTrace{}-Longitudinal) re-issuing the same 11{,}200 queries would track whether absolute FFR/SFR/AFR levels are improving, whether cross-provider trade-offs persist, and whether YMYL-domain failures narrow under retrieval-policy updates.
The marginal collection cost is modest, and the result is a continuous monitoring signal that single-shot benchmarks cannot provide.

\paragraph{Causal decomposition through controlled retrieval.}
A controlled experiment with the retrieved-context set held fixed while the generator is varied (and vice versa) would isolate the contribution of each pipeline stage, moving from descriptive characterization (where does \VMshort{} concentrate?) to mechanistic understanding (which pipeline stage produces it?).

\paragraph{Mitigation-oriented benchmarking.}
Three concrete mitigation paths follow from our findings: \emph{intent-aware retrieval reranking} (inverting IPA matrix as a scoring function), \emph{domain-aware source filtering for YMYL} (given the 2.3-fold Fisher OR), and \emph{citation-quality nudging at generation time} (since reasoning models reduce FFR but not SFR).
We release the framework, matrices, and per-citation labels to support replication of these mitigation experiments.

\subsection{Broader Impact}
\label{app:broader-impact}

\paragraph{Societal implications: misguidance at scale.}
Search-augmented LLMs increasingly mediate access to information in high-stakes decision-making domains.
Hundreds of millions of users now consult these systems in lieu of traditional search engines~\citep{searcharena2026}, treating the presence of a citation as {a sufficient guarantee of accuracy~\citep{ding2025citationstrustllmgenerated}.}
Our work demonstrates empirically that this trust is not always warranted: 30.6\% of citations distort their sources, 27.1\% originate from domain-inappropriate sources, and at the response level up to 90\% of users encounter at least one such citation.
Unlike a hallucinated fact, a citation-bearing answer that is faithful to a structurally inappropriate source presents no surface signal that distinguishes it from a trustworthy one; at scale, even single-digit failure rates translate into millions of misguidance events per day in the YMYL domains where misguidance carries the highest cost.

\paragraph{Implications for AI evaluation and system design.}
Our cross-dimension analysis (Appendix~\ref{app:cross-axis-detail}) shows that a system can score well on any single citation-quality dimension while failing structurally on another:
{ALCE-style citation precision~\citep{alce} and RAGAS-style faithfulness~\citep{ragas} target answer-source fidelity, while CRAAP- and SourceBench-style source quality~\citep{craap, sourcebench} target source suitability, and neither captures their interaction.}
\CiteTrace{} is designed to support multi-dimension structural evaluation in which a system's citation behavior is reported as a profile across IPA, SS, and ASF rather than a scalar.
For practitioners, our finding that 88--96\% of citation-quality variance is provider-level (Appendix~\ref{app:variance}) implies that retrieval-backend choice matters more than generator choice for citation quality, and that fidelity (Answer--Source Fidelity) and source-selection (Query--Source Alignment and Source Suitability) require separate engineering efforts.

\paragraph{Responsibility in model comparison.}
Our work reports systematic differences across providers, including ranking reversals between fidelity and source suitability.
These differences reflect specific model versions during a specific collection period and are \emph{not} permanent properties of any provider; search-augmented systems evolve on weekly timescales, and the rankings may invert within a quarter.
We release model identifiers and collection dates alongside all results to discourage citation of our rankings as static characterizations.

\paragraph{Ethics, safeguards, and misuse risk.}
All queries derive from the Stack Exchange Data Dump (CC BY-SA 4.0), and we comply with attribution requirements.
Source crawling respects \texttt{robots.txt} and rate limits (Appendix~\ref{app:crawling-sources}); the public release contains URLs and metadata only, not source bodies (Appendix~\ref{app:redistribution-policy}); no human-subjects data are collected.
{In principle, content producers could reverse-engineer IPA matrix to boost their pages' search-result visibility. We assess this risk as low because IPA matrix is a diagnostic tool, not a retrieval signal used by any production system.}

\subsection{Comparison to Adjacent Failure Modes}
\label{app:adjacent-modes}

\VMshort{} occupies a position adjacent to several phenomena studied in the {LLM} evaluation literature, and distinguishing it from each is essential for interpreting our findings.

\paragraph{Hallucination.}
Hallucination~\citep{ji2023hallucination} concerns claims that are factually false or unsupported by any source.
\VMshort{} is structurally distinct: by construction, the cited source in a \VMshort{} instance is real and accessible, and the cited claim need not be false.
The structural failure lies in alignment of source purpose to query intent (Intent--Purpose Alignment) or in source-type suitability for the domain (Source Suitability), not in factual accuracy per se.
Empirically, ASF1 (Fabricated) cases (closest to classical hallucination) account for 24.5\% of citations, while structural failures on {IPA or SS} without fabrication add roughly another $20$ percentage points.

\paragraph{Citation-precision benchmarks.}
ALCE~\citep{alce}, AutoAIS~\citep{bohnet2022attributed}, and related work measure whether each cited claim is supported by the cited source.
This corresponds to a portion of our Answer--Source Fidelity (ASF), but treats the suitability of the cited source as exogenous: if the source supports the claim, the citation is counted as correct, regardless of whether the source is structurally suitable for the user's intent.
Our cross-dimension trade-off (Anthropic high-ASF/low-SS versus OpenAI low-ASF/high-SS) shows precisely why this is insufficient.

\paragraph{Source-quality benchmarks.}
SourceBench~\citep{sourcebench}, the CRAAP test~\citep{craap}, and related work evaluate the absolute quality of retrieved sources with a universal rubric, scoring each source independently of how its type interacts with its content domain.
Our SS matrix (Appendix~\ref{app:axis3-detail}) explicitly encodes this interaction: the same Wiki/Forum type is rated SS~$5$ when the source falls in Code/Data and SS~$1$ when it falls in Medical. 
A benchmark that ignores the domain--type interaction would conflate these cases and miss the YMYL-specific failure pattern that drives much of our reported SFR ($38.3\%$ in YMYL versus $21.1\%$ outside).

\paragraph{Information asymmetry and trust calibration.}
Our work inherits the information-asymmetry frame of \citet{akerlof1970,arrow1963} but operationalizes it for the search-augmented LLM setting, where the LLM rather than the user is the proximate consumer of search results, and the user receives a summary that hides the structural mismatch from view.
Recent trust-calibration research~\citep{searcharena2026} shows that citation count raises user trust regardless of whether citations support the claim; our work contributes the system-side measurement that complements this user-side finding, quantifying the gap between surface citation behavior and the three structural quality axes.

\section{\CiteTrace{}: Dataset Construction}
\label{app:constructing-citetrace}

This appendix details the dataset construction procedure summarized in Section~\ref{sec:citetrace}.
We describe how we source queries from Stack Exchange (\S\ref{app:sourcing-queries}), collect search-augmented responses and extract citations (\S\ref{app:collecting-responses}), and crawl and filter cited sources (\S\ref{app:crawling-sources}).

\subsection{Sourcing Real-World Queries from Stack Exchange Communities}
\label{app:sourcing-queries}

\paragraph{Site selection.}
Stack Exchange Network~\citep{tanzil2025stackoverflow}, based on its December 31, 2025 snapshot, hosts $182{+}$ communities.
We narrow this pool to $28$ sites through a four-stage selection process designed to retain communities where commercial information asymmetry is structurally possible.
\emph{Stage 0 (candidate pool, $182{+}$ down to $48$)} excludes communities that are unsuitable for studying commercial citation bias on substantive grounds: single-correct-answer domains (e.g., math, codegolf), fiction and entertainment sites lacking decision-making contexts (gaming, anime, movies), language-correction sites without commercial actors (english, ell), platform-specific tools dominated by a single ecosystem (emacs, tex, blender), religious-doctrine sites whose claims resist factual verification (christianity, islam), sites lacking a structural commercial-bias mechanism (lifehacks, history), and beta sites with fewer than ten thousand valid questions.
The full exclusion criteria appear in Table~\ref{tab:b1_exclusion}.
\emph{Stages 1--3 ($48$ down to $28$)} apply five sequential gates summarized in Table~\ref{tab:b1_gates}: requirement of expert knowledge (C1), presence of commercial actors with incentive to distort answers (C2), connection to substantive decisions (C3), distinctness of the commercial-bias mechanism across selected sites (C4), and data sufficiency after quality filtering (C5).
Criterion C1 is satisfied by all $48$ Stage~0 candidates by construction; C2--C3 remove three sites with absent or unverified commercial actors (interpersonal, philosophy, ux); C4 deduplicates eight sites that share an existing site's bias mechanism; and C5 removes nine sites whose post-filter pool falls below $400$ queries, following CQADupStack's per-site minimum-threshold convention~\citep{cqadupstack}.
The remaining $28$ sites map directly to Stack Exchange's official six-category taxonomy: Technology~$5$, Science~$7$, Life~\&~Arts~$7$, Culture~\&~Recreation~$6$, Professional~$2$, Business~$1$.
We follow PRISM's~\citep{prism} approach of adopting an external taxonomy (UN subregion in their case) to avoid imposing our own categorization.

\begin{table}[!htbp]
\centering
\caption{Stage~0 exclusion criteria for Stack Exchange site selection. Sites meeting any single criterion are removed from consideration before the substantive gates (C2--C5) are applied.}
\label{tab:b1_exclusion}
\small
\setlength{\tabcolsep}{6pt}
\renewcommand{\arraystretch}{1.15}
\begin{tabular}{p{0.2\textwidth}p{0.39\textwidth}p{0.27\textwidth}}
\toprule
\textbf{Exclusion criterion} & \textbf{Rationale} & \textbf{Excluded sites (examples)} \\
\midrule
Single-correct-answer domain & Commercial bias does not arise structurally & math, puzzles, codegolf \\
Fiction \& entertainment & No decision-making context & gaming, anime, scifi, movies \\
Language correction & Commercial actors absent & english, ell \\
Platform-specific tools & Single-tool ecosystem yields trivial bias & emacs, tex, blender, drupal \\
Religious doctrine & Belief claims resist verification & christianity, islam, judaism \\
Absent commercial structure & No commercial-bias mechanism & lifehacks, history \\
Insufficient scale ($< 10$K Qs) & Beta sites, inadequate sample size & various beta sites \\
\bottomrule
\end{tabular}
\end{table}

\begin{table}[!htbp]
\centering
\caption{Selection gates C1--C5 applied in Stages 1--3. C1 is automatically satisfied by all Stage~0 survivors; C2--C5 are the substantive gates that reduce the candidate pool from $48$ to $28$ sites.}
\label{tab:b1_gates}
\small
\setlength{\tabcolsep}{6pt}
\renewcommand{\arraystretch}{1.2}
\begin{tabular}{cp{0.27\textwidth}p{0.50\textwidth}}
\toprule
\textbf{\#} & \textbf{Criterion} & \textbf{Decision question} \\
\midrule
C1 & Expertise required & Does answering require domain expertise? \\
C2 & Commercial actor present & Are there commercial stakeholders incentivized to distort answers? \\
C3 & Decision-making context & Does the question relate to substantive decisions (safety, cost, design)? \\
C4 & Bias-mechanism uniqueness & Is the commercial-bias mechanism distinct from other selected sites? \\
C5 & Data sufficiency & Are at least $400$ queries available after quality filtering? \\
\bottomrule
\end{tabular}
\end{table}

\paragraph{Quality filtering and sampling.}
A five-step filter applied to all $28$ sites yields a final pool of $471{,}876$ eligible queries from $25{,}483{,}555$ raw posts, a $98.1\%$ reduction summarized stage-by-stage in Table~\ref{tab:b2_filter}.
The single largest reduction is the community-validation gate (Score~$\geq 5$, $90.96\%$ drop), which removes unanswered or low-quality posts; the recency gate ($\geq 2018{-}01{-}01$, $78.91\%$ drop) further restricts the pool to queries posed within the search-augmented LLM era.
We then apply uniform per-site sampling of $400$ queries to obtain the final $11{,}200$-query dataset ($28 \times 400$).
Uniform sampling is essential because Stack Overflow alone accounts for $81.8\%$ of the eligible pool ($385{,}770 / 471{,}876$); proportional sampling would make cross-site comparisons impossible.
This design choice follows PRISM's~\citep{prism} per-country uniform-sampling strategy for cross-cultural balance.

\begin{table}[!htbp]
\centering
\caption{Five-step quality filter funnel applied to all $28$ sites combined. The community-validation gate (Step~1) and recency gate (Step~4) account for $99\%$ of the total reduction; per-site uniform sampling (Step~6) produces the final $11{,}200$-query dataset.}
\label{tab:b2_filter}
\small
\setlength{\tabcolsep}{6pt}
\renewcommand{\arraystretch}{1.2}
\begin{tabular}{clrr}
\toprule
\textbf{Stage} & \textbf{Filter} & \textbf{Remaining} & \textbf{Stage drop \%} \\
\midrule
0 & Total ($28$ sites combined) & $25{,}483{,}555$ & -- \\
1 & Score $\geq 5$ & $2{,}303{,}787$ & $90.96$ \\
2 & Title $15$--$150$ chars & $2{,}303{,}628$ & $0.01$ \\
3 & Body $\geq 50$ chars (HTML stripped) & $2{,}303{,}487$ & $0.01$ \\
4 & Posted $\geq 2018{-}01{-}01$ & $485{,}816$ & $78.91$ \\
5 & Intent classified (drop \texttt{unclear}) & $471{,}876$ & $2.87$ \\
6 & Per-site uniform sample of $400$ & $\mathbf{11{,}200}$ & -- \\
\bottomrule
\end{tabular}
\end{table}

\paragraph{Per-site composition.}
The selected $28$ sites span Stack Exchange's six categories with the distribution Technology~$5$, Science~$7$, Life~\&~Arts~$7$, Culture~\&~Recreation~$6$, Professional~$2$, Business~$1$.
Table~\ref{tab:b3_sites} reports for each site the official category, audience description (verbatim from the Stack Exchange Sites API), pool size after Step~5 of the quality filter, sample size ($400$ uniform), and mean posting year.
The mean year across the full pool is $2020.27$, indicating that the dataset reflects the period when search-augmented LLMs entered mainstream use.

\begin{table}[!p]
\centering
\caption{Per-site statistics for the $28$ selected Stack Exchange communities. \textbf{Category} follows the official Stack Exchange Network six-category taxonomy; \textbf{Audience} is taken verbatim from the Sites API \texttt{audience} field (fetched $2026{-}04{-}24$); \textbf{Pool} reports queries remaining after Step~5 of the quality filter; \textbf{Sample} is uniform $400$ across all sites; \textbf{Year} is the mean posting year within the pool, rounded to the nearest integer.}
\label{tab:b3_sites}
\small
\setlength{\tabcolsep}{4pt}
\renewcommand{\arraystretch}{1.10}
\resizebox{\linewidth}{!}{
\begin{tabular}{p{0.13\textwidth}p{0.22\textwidth}p{0.37\textwidth}rrr}
\toprule
\textbf{Category} & \textbf{Site} & \textbf{Audience} & \textbf{Pool} & \textbf{Sample} & \textbf{Year} \\
\midrule
Technology & Stack Overflow & professional and enthusiast programmers & $385{,}770$ & $400$ & $2020$ \\
Technology & Information Security & information security professionals & $3{,}037$ & $400$ & $2020$ \\
Technology & Software Engineering & professionals, academics, students in SDLC & $2{,}231$ & $400$ & $2020$ \\
Technology & Database Administrators & database professionals & $2{,}692$ & $400$ & $2020$ \\
Technology & Bitcoin & Bitcoin users, developers, enthusiasts & $816$ & $400$ & $2020$ \\
\midrule
Science & Physics & active researchers, academics, students of physics & $11{,}326$ & $400$ & $2021$ \\
Science & Chemistry & scientists, academics, teachers, students & $2{,}981$ & $400$ & $2020$ \\
Science & Earth Science & geology, meteorology, oceanography, env.\ sciences & $960$ & $400$ & $2020$ \\
Science & Economics & those who study/teach/research economics & $622$ & $400$ & $2020$ \\
Science & Cross Validated & statistics, ML, data analysis, mining, viz & $9{,}952$ & $400$ & $2020$ \\
Science & Artificial Intelligence & people interested in conceptual questions on AI & $1{,}224$ & $400$ & $2020$ \\
Science & Medical Sciences & medical and allied health professionals & $422$ & $400$ & $2020$ \\
\midrule
Life~\&~Arts & Seasoned Advice & professional and amateur chefs & $1{,}966$ & $400$ & $2021$ \\
Life~\&~Arts & Home Improvement & contractors and serious DIYers & $3{,}358$ & $400$ & $2021$ \\
Life~\&~Arts & Personal Finance \& Money & people who want to be financially literate & $2{,}739$ & $400$ & $2020$ \\
Life~\&~Arts & Parenting & parents, grandparents, nannies, others & $554$ & $400$ & $2020$ \\
Life~\&~Arts & Academia & academics and those in higher education & $7{,}515$ & $400$ & $2021$ \\
Life~\&~Arts & Pets & pet owners, vets, breeders, trainers & $481$ & $400$ & $2019$ \\
Life~\&~Arts & Law & legal professionals, students, others & $3{,}667$ & $400$ & $2022$ \\
\midrule
Culture~\&~Rec. & Bicycles & people who build/repair/ride bicycles & $2{,}556$ & $400$ & $2021$ \\
Culture~\&~Rec. & Skeptics & scientific skepticism & $2{,}245$ & $400$ & $2020$ \\
Culture~\&~Rec. & Motor Vehicle Maint. & mechanics and DIY car owners & $473$ & $400$ & $2021$ \\
Culture~\&~Rec. & Travel & road warriors and seasoned travelers & $5{,}849$ & $400$ & $2021$ \\
Culture~\&~Rec. & The Great Outdoors & outdoor enthusiasts and learners & $1{,}171$ & $400$ & $2020$ \\
Culture~\&~Rec. & Politics & people interested in governments and policies & $6{,}308$ & $400$ & $2021$ \\
\midrule
Professional & The Workplace & members of the workforce & $5{,}148$ & $400$ & $2020$ \\
Professional & Aviation & aircraft pilots, mechanics, enthusiasts & $4{,}964$ & $400$ & $2020$ \\
\midrule
Business & Quantitative Finance & finance professionals and academics & $849$ & $400$ & $2020$ \\
\midrule
\textbf{Total} & & & $\mathbf{471{,}876}$ & $\mathbf{11{,}200}$ & $\mathbf{2020}$ \\
\bottomrule
\end{tabular}
}
\end{table}

\subsection{Collecting Search-Augmented Responses and Extracting Citations}
\label{app:collecting-responses}

\paragraph{Model selection.}
We select ten search-augmented LLMs from five major providers, including standard/reasoning variant pairs to enable controlled comparisons of reasoning effects on citation behavior. GPT-5 and GPT-5 mini (OpenAI), Claude Sonnet~4.6 and Claude Haiku~4.5 (Anthropic), Gemini~3 Flash and Gemini~3.1 Pro (Google), Grok~4.1 Fast-Non-Reasoning and Grok-4.1 Fast-Reasoning (xAI), and Sonar and Sonar Reasoning Pro (Perplexity).
Selection criteria are: (i) provider diversity to capture distinct retrieval-backend behaviors, (ii) built-in search tool support via the production API, (iii) non-deprecated availability throughout the collection window, and (iv) the existence of at least one standard/reasoning pair within each provider that supports such pairing.
Table~\ref{tab:b4_models} summarizes per-model API parameters, search tools, and collection windows.

\begin{table}[!htbp]
\centering
\caption{API configuration for the ten search-augmented {LLMs}. Search tool is the web-search tool exposed through the provider's API. Perplexity Sonar models perform retrieval natively and require no tool specification. Temperature is N/A for the GPT-5 family, where the parameter is not exposed in the Responses API.}
\label{tab:b4_models}
\small
\setlength{\tabcolsep}{5pt}
\renewcommand{\arraystretch}{1.15}
\begin{tabular}{lllccl}
\toprule
\textbf{Provider} & \textbf{Model} & \textbf{Search tool} & \textbf{Temp.} & \textbf{Max tokens} & \textbf{Window} \\
\midrule
OpenAI & gpt-5-2025-08-07 & web\_search & N/A & $8192$ & $03/26$--$04/09$ \\
OpenAI & gpt-5-mini-2025-08-07 & web\_search & N/A & $8192$ & $03/26$--$04/09$ \\
Anthropic & claude-sonnet-4-6 & web\_search\_20250305 & $0$ & $8192$ & $03/26$--$04/09$ \\
Anthropic & claude-haiku-4-5-20251001 & web\_search\_20250305 & $0$ & $8192$ & $03/26$--$04/09$ \\
Google & gemini-3-flash & google\_search & $0$ & $8192$ & $03/26$--$04/09$ \\
Google & gemini-3.1-pro & google\_search & $0$ & $8192$ & $03/26$--$04/09$ \\
xAI & grok-4-1-fast-non-reasoning & web\_search & $0$ & $8192$ & $03/26$--$04/09$ \\
xAI & grok-4-1-fast-reasoning & web\_search & $0$ & $8192$ & $03/26$--$04/09$ \\
Perplexity & sonar & built-in & $0$ & $8192$ & $03/26$--$04/09$ \\
Perplexity & sonar-reasoning-pro & built-in & $0$ & $8192$ & $03/26$--$04/09$ \\
\bottomrule
\end{tabular}
\end{table}

\begin{figure}[!t]
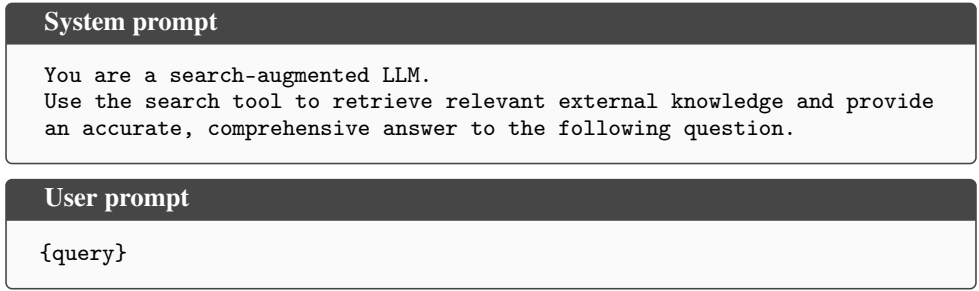

\centering
\begin{tcolorbox}[
  colback=gray!4,
  colframe=gray!55!black,
  title=\textbf{System prompt},
  width=0.92\textwidth,
  arc=2pt,
  boxrule=0.6pt,
  fontupper=\small\ttfamily
]
You are a search-augmented LLM. 

Use the search tool to retrieve relevant external knowledge and provide an accurate, comprehensive answer to the following question.
\end{tcolorbox}
\begin{tcolorbox}[
  colback=gray!4,
  colframe=gray!55!black,
  title=\textbf{User prompt},
  width=0.92\textwidth,
  arc=2pt,
  boxrule=0.6pt,
  fontupper=\small\ttfamily
]
\{query\}
\end{tcolorbox}
\caption{Prompt for Response Generation}
\label{fig:b4_system_prompt}
\end{figure}

\paragraph{Prompt design.}
We use a single system prompt across all ten models to isolate citation behavior specific to each model from variance introduced by the prompt.
The prompt is designed to intervene as little as possible, providing no specification of citation format, count, or preference for particular source types so that each model's natural citation behavior can surface.
At the same time, it preserves a realistic query format by passing only the Stack Exchange post title as the user query, mimicking the short questions users typically ask rather than inputs that have been artificially expanded.
The full system prompt appears in Figure~\ref{fig:b4_system_prompt}.

\paragraph{Response collection.}
We dispatch each of the $11{,}200$ queries to each of the 10 models, yielding $11{,}200 \times 10 = 112{,}000$ responses.
All collection occurs within the same $15$ days to limit temporal drift in the live web index that each provider's search tool consults.
Rate limits on the provider side, transient errors, and timeout retries are handled via exponential backoff with a maximum of three retries per call.
Calls that fail all three retries are logged with an empty response body and excluded from downstream analysis.

\paragraph{Citation extraction and normalization.}
The five providers return citations in five different API formats, {summarized in Table~\ref{tab:b5_extraction}.}
We parse each format and normalize all citations to a common schema $\langle \texttt{cited\_sentence}, \texttt{source\_url} \rangle$, where \texttt{cited\_sentence} is the portion of the answer that relies on the source and \texttt{source\_url} is its URL.
For providers that use inline markers (OpenAI, xAI, Perplexity), we extract a context window of two sentences around each marker, which aligns with the citation blocks of $1$ to $2$ sentences that Anthropic and Google return directly.
After normalization we obtain $1{,}271{,}046$ citation pairs across the full $112{,}000$ responses.
$8{,}069$ responses contain no citations and are excluded from evaluation on individual dimensions but retained in the public release.

\paragraph{Extraction reliability.}
For providers that return citations as structured blocks (Anthropic, Google), extraction is deterministic because the API returns JSON fields (\texttt{cited\_text}, \texttt{source\_url}) that require no heuristic parsing.
For providers that use inline markers (OpenAI, xAI, Perplexity), the context window of two sentences captures more than the minimum grounding context, so fidelity judgments (Answer-Source Fidelity, \S\ref{app:axis2-detail}) remain conservative. A wider window may include text that is not grounded but will not miss the grounded claim.
Residual parsing failures such as truncated sentences or malformed markers produce defective \texttt{cited\_sentence} entries.
These entries are caught by the evaluability filters in \S\ref{app:crawling-sources}, which remove code or table content, overly short extractions, and entries under five words before evaluation on individual dimensions.

\begin{table}[!htbp]
\centering
\caption{Citation extraction and normalization by provider.}
\label{tab:b5_extraction}
\small
\renewcommand{\arraystretch}{1.2}
\resizebox{\linewidth}{!}{
\begin{tabular}{lllll}
\toprule
\textbf{Provider} & \textbf{Type} & \textbf{Citation unit} & \textbf{URL source} & \textbf{Sentences} \\
\midrule
OpenAI    & Marker & \texttt{annotation indices}     & \texttt{annotations[].url}        & Last 2 before marker \\
xAI       & Marker & \texttt{annotation indices}     & \texttt{annotations[].url}        & Last 2 before marker \\
Perplexity& Marker & \texttt{[N]} marker     & \texttt{citations[]} array        & Last 2 before marker \\
\midrule
Anthropic & Block  & \texttt{cited\_text}     & \texttt{citations[].url}          & Last 2 from block \\
Google    & Block  & \texttt{grounding\_supports[]}            & \texttt{grounding\_chunks[].web.uri}  & Last 2 from block \\
\bottomrule
\end{tabular}
}
\end{table}

\subsection{Crawling and Verifying Source Content}
\label{app:crawling-sources}

\paragraph{Pipeline.}
Each unique URL extracted from citations is processed through a pipeline of four stages.
First, we resolve redirects (e.g., Gemini proxy URLs) and strip tracking parameters to obtain the canonical URL.
Second, we render the page using a headless browser (Playwright Chromium), falling back to an async HTTP client when rendering fails.
Third, we convert the retrieved HTML to plain text, applying \texttt{trafilatura}~\citep{barbaresi2021}, \texttt{readability-lxml},, and raw \texttt{innerText} as successive fallbacks to maximize extraction coverage.
Finally, we detect error pages by checking whether the extracted text is too short or matches known patterns for bot blocking.
All crawling respects \texttt{robots.txt} and enforces rate limiting per domain\citep{koster2022rfc}.
Table~\ref{tab:crawl_params} lists the specific thresholds and timeout values used at each stage.

\begin{table}[!htbp]
\centering
\caption{Crawling pipeline parameters.}
\label{tab:crawl_params}
\small
\setlength{\tabcolsep}{6pt}
\renewcommand{\arraystretch}{1.2}
\begin{tabular}{lll}
\toprule
\textbf{Stage} & \textbf{Parameter} & \textbf{Value} \\
\midrule
URL resolution   & Timeout              & 10\,s \\
Page collection  & Timeout per URL      & 15\,s \\
                 & Global concurrency   & 5 \\
                 & Content cap          & $50{,}000$ chars \\
Error detection  & Fail if length below & 50 chars \\
                 & Bot blocking patterns & 16 signatures \\
Rate limiting    & Delay per domain     & 2\,s \\
\bottomrule
\end{tabular}
\end{table}

\begin{table}[!htbp]
\centering
\caption{Crawl pipeline outcome by unique URL and citation pair.
Inline bars visualize Success and Failed shares for each unit.}
\label{tab:b6_outcome}
\small
\setlength{\tabcolsep}{6pt}
\renewcommand{\arraystretch}{1.30}
\begin{tabular}{lrrp{2.4cm}rrp{2.4cm}}
\toprule
\textbf{Status} & \textbf{URLs} & \textbf{\%} & \textbf{Share}
                & \textbf{Citations} & \textbf{\%} & \textbf{Share} \\
\midrule
{\textbf{Success}}
  & 231{,}105 & \textbf{58.3} & \successbar{58.3}
  & 802{,}945 & \textbf{63.2} & \successbar{63.2} \\
{\textbf{Failed}}
  & 165{,}565 & \textbf{41.7} & \failbar{41.7}
  & 468{,}101 & \textbf{36.8} & \failbar{36.8} \\
\midrule
\textbf{Total}
  & \textbf{396{,}670} & \textbf{100.0} &
  & \textbf{1{,}271{,}046} & \textbf{100.0} & \\
\bottomrule
\end{tabular}
\end{table}

\begin{table}[!htbp]
\centering
\caption{Failure category breakdown for crawl pipeline. Highlighted cells mark the two largest categories (JS rendering / bot blocking and unsupported file formats), which together account for $74.7\%$ of all crawl failures.}
\label{tab:b6_failures}
\small
\setlength{\tabcolsep}{4pt}
\renewcommand{\arraystretch}{1.3}
\resizebox{\linewidth}{!}{
\begin{tabular}{p{4.2cm}
                >{\raggedleft\arraybackslash}p{1.4cm}
                >{\raggedleft\arraybackslash}p{0.8cm} p{2.3cm}
                >{\raggedleft\arraybackslash}p{1.4cm}
                >{\raggedleft\arraybackslash}p{0.8cm} p{2.3cm}}
\toprule
\textbf{Failure category} & \textbf{URLs} & \textbf{\%} & \textbf{Share}
                          & \textbf{Citations} & \textbf{\%} & \textbf{Share} \\
\midrule
\textbf{JS rendering / bot blocking}
       & \cellcolor{tblRedBg}80{,}517 & \cellcolor{tblRedBg}48.6 & \redbar{58.3}
       & \cellcolor{tblRedBg}248{,}430 & \cellcolor{tblRedBg}53.1 & \redbar{63.7} \\
\textbf{File format (PDF / Office)}
       & \cellcolor{tblRedBg}43{,}165 & \cellcolor{tblRedBg}26.1 & \redbar{31.3}
       & \cellcolor{tblRedBg}119{,}592 & \cellcolor{tblRedBg}25.6 & \redbar{30.7} \\
Empty response                 & 29{,}290 & 17.7 & \redbar{21.2} & 68{,}787 & 14.7 & \redbar{17.6} \\
Server error / access denied   & 6{,}662  & 4.0  & \redbar{4.8}  & 16{,}837 & 3.6  & \redbar{4.3} \\
Timeout                        & 3{,}125  & 1.9  & \redbar{2.3}  & 7{,}316  & 1.6  & \redbar{1.9} \\
Other (menu / login pages)     & 2{,}018  & 1.2  & \redbar{1.4}  & 5{,}130  & 1.1  & \redbar{1.3} \\
DNS resolution / domain expiry & 788      & 0.5  & \redbar{0.6}  & 2{,}009  & 0.4  & \redbar{0.5} \\
\midrule
\textbf{Total}                 & \textbf{165{,}565} & \textbf{100.0} & & \textbf{468{,}101} & \textbf{100.0} & \\
\bottomrule
\end{tabular}
}
\end{table}

\begin{table}[!htbp]
\centering
\caption{Crawl failure rates by host tier (top) and query category (bottom). Forum/Q\&A and Social/Blog tiers fail at substantially higher rates due to bot blocking, leading to under-representation of community-based and personal-author sources in the evaluable pool.}
\label{tab:b6_bias}
\small
\setlength{\tabcolsep}{4pt}
\renewcommand{\arraystretch}{1.3}
\begin{tabular}{p{3.4cm}
                >{\raggedleft\arraybackslash}p{1.8cm}
                >{\raggedleft\arraybackslash}p{1.8cm}
                >{\raggedleft\arraybackslash}p{0.9cm} p{3.0cm}}
\toprule
\textbf{Host tier} & \textbf{Total} & \textbf{Failed} & \textbf{\%} & \textbf{Share} \\
\midrule
\textbf{Forum / Q\&A}  & \cellcolor{tblRedBg}72{,}435  & \cellcolor{tblRedBg}49{,}152 & \cellcolor{tblRedBg}67.9 & \redbar{81.5} \\
\textbf{Social / Blog} & \cellcolor{tblRedBg}28{,}146  & \cellcolor{tblRedBg}15{,}490 & \cellcolor{tblRedBg}55.0 & \redbar{66.0} \\
Academic               & 52{,}657  & 19{,}489 & 37.0 & \redbar{44.4} \\
Gov / Org              & 28{,}704  & 9{,}545  & 33.3 & \redbar{40.0} \\
Commercial / Other     & 246{,}331 & 70{,}487 & 28.6 & \redbar{34.3} \\
News                   & 7{,}138   & 1{,}402  & 19.6 & \redbar{23.5} \\
\midrule
\textbf{Total}         & \textbf{435{,}411} & \textbf{165{,}565} & \textbf{38.0} & \\
\bottomrule
\end{tabular}

\vspace{2pt}
\centerline{\small (a) By host tier (URL level)}
\vspace{8pt}

\begin{tabular}{p{3.4cm}
                >{\raggedleft\arraybackslash}p{1.8cm}
                >{\raggedleft\arraybackslash}p{1.8cm}
                >{\raggedleft\arraybackslash}p{0.9cm} p{3.0cm}}
\toprule
\textbf{Query category} & \textbf{Total} & \textbf{Failed} & \textbf{\%} & \textbf{Share} \\
\midrule
\textbf{Business}     & \cellcolor{tblRedBg}43{,}975 & \cellcolor{tblRedBg}21{,}002  & \cellcolor{tblRedBg}47.8 & \redbar{57.4} \\
Culture \& Recreation & 275{,}576     & 111{,}815 & 40.6 & \redbar{48.7} \\
Professional          & 78{,}567      & 31{,}896  & 40.6 & \redbar{48.7} \\
Science               & 320{,}204     & 127{,}185 & 39.7 & \redbar{47.6} \\
Life \& Arts          & 349{,}622     & 124{,}075 & 35.5 & \redbar{42.6} \\
Technology            & 203{,}102     & 52{,}128  & 25.7 & \redbar{30.8} \\
\midrule
\textbf{Total}        & \textbf{1{,}271{,}046} & \textbf{468{,}101} & \textbf{36.8} & \\
\bottomrule
\end{tabular}

\vspace{2pt}
\centerline{\small (b) By query category (citation level)}
\end{table}
\paragraph{Crawl results.}
The pipeline successfully retrieves $231{,}105$ of $396{,}670$ unique URLs ($58.3\%$), covering $63.2\%$ of all citation pairs (Table~\ref{tab:b6_outcome}).
The most common reasons for failure are JavaScript rendering or bot blocking and documents in formats other than HTML such as PDF (Table~\ref{tab:b6_failures}).
Failures are not uniform across host types.
Forum and Q\&A hosts fail most often, primarily due to Cloudflare bot blocking, while news domains have the highest success rate (Table~\ref{tab:b6_bias}).
As a result, community and personal sources are underrepresented in the evaluable pool.
For example, the share of forum and Q\&A sources drops from $16.6\%$ of all sources to $8.6\%$ after crawling.
We therefore treat our SFR estimate as a conservative lower bound, as discussed further in Appendix~\ref{app:robustness-bounds} and Section~\ref{sec:vm_effect}.

\begin{table}[!htbp]
\centering
\caption{Evaluability filtering stages applied to crawled citations.}
\label{tab:b7_outcome}
\small
\setlength{\tabcolsep}{4pt}
\renewcommand{\arraystretch}{1.3}
\begin{tabular}{p{4.5cm}
                >{\raggedleft\arraybackslash}p{1.6cm}
                >{\raggedleft\arraybackslash}p{0.9cm} p{0.9cm}
                >{\raggedleft\arraybackslash}p{1.8cm}}
\toprule
\textbf{Filter} & \textbf{Removed} & \textbf{\%} & \textbf{Share} & \textbf{Remaining} \\
\midrule
--                  & --         & --     &              & 802{,}945 \\
\textbf{Code or table content}
                         & \cellcolor{tblRedBg}28{,}530 & \cellcolor{tblRedBg}3.55 & \redbar{4.3} & 774{,}415 \\
Judge unevaluable        & 6{,}878  & 0.86 & \redbar{1.0} & 767{,}537 \\
Too short ($<$ 20 chars) & 4{,}774  & 0.62 & \redbar{0.7} & 762{,}763 \\
Under 5 words            & 1{,}268  & 0.17 & \redbar{0.2} & 761{,}495 \\
\midrule
\textbf{Total removed}            & \textbf{41{,}450} & \textbf{5.16} &              & \textbf{761{,}495} \\
\bottomrule
\end{tabular}
\end{table}

\begin{table}[!htbp]
\centering
\caption{Per-model evaluability statistics, sorted by Eval~\%. \textbf{Eval \%} reports each model's evaluable share among its crawled citations; values range from $85.2\%$ (gemini-3-flash) to $99.3\%$ (claude-sonnet). Differences in \texttt{cited\_code\_table} flag rates account for most of the spread. The top performer (claude-sonnet) is highlighted; the bottom row (gemini-3-flash) is shown in bold. \emph{Inline bars use an anchored scale (length~$=$~Eval\%~$-$~80) rather than the $1\% = 1.2$pt scale of preceding tables, because the narrow $85$--$99\%$ range would otherwise make all bars visually identical.}}
\label{tab:b7_per_model}
\small
\setlength{\tabcolsep}{3pt}
\renewcommand{\arraystretch}{1.2}
\begin{tabular}{lrrrrrrl}
\toprule
\textbf{Model} & \textbf{Evaluable} & \textbf{code\_table} & \textbf{judge\_un}
               & \textbf{too\_short} & \textbf{lt5\_words}
               & \textbf{Eval \%} & \textbf{Share} \\
\midrule
\textbf{claude-sonnet-4-6}  & \cellcolor{sourceBg}129{,}936 & 223 & 399 & 263 & 58 & \cellcolor{sourceBg}99.3 & \srcbar{19.3} \\
claude-haiku-4-5            & 37{,}142  & 4       & 147    & 176    & 44  & 99.0 & \srcbar{19.0} \\
gpt-5-mini                  & 54{,}928  & 50      & 373    & 104    & 36  & 99.0 & \srcbar{19.0} \\
sonar-reasoning-pro         & 69{,}411  & 208     & 319    & 176    & 45  & 98.9 & \srcbar{18.9} \\
gpt-5                       & 38{,}322  & 33      & 294    & 146    & 53  & 98.7 & \srcbar{18.7} \\
gemini-3.1-pro              & 77{,}999  & 191     & 559    & 832    & 86  & 97.9 & \srcbar{17.9} \\
grok-4-1-fast-reasoning     & 92{,}355  & 5{,}494 & 1{,}080 & 601   & 170 & 92.6 & \srcbar{12.6} \\
grok-4-1-fast-non-reasoning & 90{,}916  & 5{,}843 & 1{,}181 & 623   & 229 & 92.0 & \srcbar{12.0} \\
sonar                       & 122{,}157 & 12{,}042 & 942   & 434    & 106 & 90.0 & \srcbar{10.0} \\
gemini-3-flash              & 48{,}329  & 4{,}452 & 2{,}073 & 1{,}420 & 442 & 85.2 & \srcbar{5.2} \\
\midrule
\textbf{Total}              & \textbf{761{,}495} & \textbf{28{,}540} & \textbf{7{,}367}
                            & \textbf{4{,}775} & \textbf{1{,}269}
                            & \textbf{94.8} & \srcbar{14.8} \\
\bottomrule
\end{tabular}
\end{table}

\paragraph{Evaluability filtering.}
Among the $802{,}945$ successfully crawled citations, we apply four filters to exclude pairs whose \texttt{cited\_sentence} is defective.
Following the staged attrition protocol of CiteME~\citep{press2024citeme}, we report each filter's effect separately rather than as a single aggregate (Table~\ref{tab:b7_outcome}).
The filters remove sentences that consist solely of code or table fragments, sentences flagged as unevaluable by the LLM judge, and sentences that are too short in character length or word count, accounting for $5.22\%$ of the crawled pool.
The final evaluable set contains $\mathbf{761{,}495}$ citation pairs across the ten models.
Evaluable rates per model range from $85.2\%$ (Gemini 3 Flash) to $99.3\%$ (Claude Sonnet), with variation largely due to differences in how often each model produces code in its citations (Table~\ref{tab:b7_per_model}).
All metrics in Appendix~\ref{app:results} are computed on this filtered pool.

\section{Three-Dimension Evaluation Framework}
\label{app:evaluating-vm}
\label{app:judge_reliability}
\label{app:taxonomies}

This appendix expands the evaluation framework summarized in Section~\ref{sec:measuring-vm}.
Following the structure of the main text, we describe Intent--Purpose Alignment in \S\ref{app:axis1-detail}, Source Suitability in \S\ref{app:axis3-detail}, and Answer--Source Fidelity in \S\ref{app:axis2-detail}.
For each dimension we present the taxonomy, the scoring rubric, the aggregation method at the response level, and the supporting validation evidence.
{Integration across dimensions and robustness checks are covered in \S\ref{app:cross-axis-detail}.
Each dimension is built from one or two underlying classification tasks: Query Intent (QI) and Source Purpose (SP) for alignment, Source Domain (SD) and Source Type (ST) for suitability, and Answer--Source Fidelity (ASF) for fidelity.}

\paragraph{Judge configuration.}
All five classification tasks use a single LLM judge (gpt-4o-mini-2024-07-18), called with temperature 0, strict JSON-mode structured output, and a $4{,}096$-token output limit.
{The full prompt for each task appears in Figures~\ref{fig:prompt_qi} 
through~\ref{fig:prompt_st}.}


\subsection{Alignment Between Query Intent and Source Purpose}
\label{app:axis1-detail}
\label{app:coi_matrix}

\paragraph{Query Intent (QI) taxonomy.}
Query Intent classifies each query by its \emph{primary completion condition}~\citep{taylor1968question,broder2002,ingwersen2005turn}, defined as what the user must obtain to consider the query resolved.
We define five labels (Table~\ref{tab:qi_taxonomy}): QI1 Factoid (single verifiable datum), QI2 Explanation (causal mechanism or principle), QI3 Instruction (procedural steps or troubleshooting), QI4 Comparison (evaluation of alternatives), and QI5 Opinion (subjective or value-laden judgment), {extending NF-CATS~\citep{bolotova2022non,rose2004understanding,jansen2008determining} with a dedicated comparison category} (QI4) that prior work conflates with instruction.

\begin{table}[!htbp]
\centering
\caption{Query Intent (QI) taxonomy: five labels classifying each query by its primary completion condition.}
\label{tab:qi_taxonomy}
\small
\setlength{\tabcolsep}{5pt}
\renewcommand{\arraystretch}{1.25}
\begin{tabular}{clp{0.68\textwidth}}
\toprule
\textbf{Label} & \textbf{Name} & \textbf{Definition} \\
\midrule
QI1 & Factoid      & A query seeking a specific, verifiable datum that is fully resolved by a single lookup with a context-independent answer. \\
QI2 & Explanation  & A query seeking the causal mechanism or underlying principle behind an observed phenomenon. \\
QI3 & Instruction  & A query seeking procedural steps or methods to perform a task or resolve a discrepancy, including troubleshooting. \\
QI4 & Comparison   & A query seeking evaluation of alternatives against explicit, intersubjectively shareable criteria to support a decision or recommendation. \\
QI5 & Opinion      & A query seeking a subjective, ethical, or value-laden judgment on a socially contested question. \\
\bottomrule
\end{tabular}
\end{table}
 
\begin{table}[!htbp]
\centering
\caption{Source Purpose (SP) taxonomy: six labels classifying each source by its communicative function.}
\label{tab:sp_taxonomy}
\small
\setlength{\tabcolsep}{5pt}
\renewcommand{\arraystretch}{1.25}
\begin{tabular}{clp{0.68\textwidth}}
\toprule
\textbf{Label} & \textbf{Name} & \textbf{Definition} \\
\midrule
SP1 & To Promote  & A source that advocates for a specific product, service, or commercial entity to drive purchase or favorable perception. \\
SP2 & To Inform   & A source that transmits factual or conceptual knowledge with no commercial incentive. Encyclopedic entries, official documentation. \\
SP3 & To Instruct & A source that provides procedural guidance: tutorials, manuals, how-to guides. \\
SP4 & To Report   & A source that reports an event, observation, or finding: news articles, study reports. \\
SP5 & To Discuss  & A source hosting community discussion or deliberation: forum threads, Q\&A sites. \\
SP6 & To Opine    & A source presenting an individual perspective or opinion: op-eds, personal blogs, commentary. \\
\bottomrule
\end{tabular}
\end{table}

\paragraph{Source Purpose (SP) taxonomy.}
Source Purpose classifies each cited source by its \emph{communicative function}~\citep{biber2015exploring,sharoff2018functional,swales1990}, defined as the goal that the source's author pursues regardless of factual content.
We define six labels (Table~\ref{tab:sp_taxonomy}): SP1 To Promote (commercial advocacy), SP2 To Inform (neutral knowledge transmission), SP3 To Instruct (procedural guidance), SP4 To Report (event reporting), SP5 To Discuss (community deliberation), and SP6 To Opine (individual perspective).
This distinction matters because the purpose of a source shapes its structural incentive: a promotional source may present accurate facts yet frame them to favor a product, creating a mismatch when paired with informational queries.
\textcolor{blue}{}

\definecolor{s1}{RGB}{215,48,39}    
\definecolor{s2}{RGB}{252,174,145}  
\definecolor{s3}{RGB}{245,245,245}  
\definecolor{s4}{RGB}{146,197,222}  
\definecolor{s5}{RGB}{33,102,172}

\begin{table}[!htbp]
\centering
\caption{Intent--Purpose Alignment (IPA) Matrix. Each cell scores alignment between Query Intent (rows) and Source Purpose (columns) from $1$ (structural conflict) to $5$ (direct match). Red cells ($\leq 2$) define alignment failures contributing to the AFR metric. Matrix values are validated by $10$ IR/LIS researchers ($\mathrm{ICC}(2,k)=0.916$; Table~\ref{tab:c_validation}).}
\label{tab:ipa_matrix}
\small
\setlength{\tabcolsep}{8pt}
\renewcommand{\arraystretch}{1.3}
\resizebox{\linewidth}{!}{
\begin{tabular}{lcccccc}
\toprule
 & \textbf{SP1} & \textbf{SP2} & \textbf{SP3} & \textbf{SP4} & \textbf{SP5} & \textbf{SP6} \\
 & To Promote & To Inform & To Instruct & To Report & To Discuss & To Opine \\
\midrule
\textbf{QI1} Factoid     & \cellcolor{s1}\color{white}$1$ & \cellcolor{s5}\color{white}$5$ & \cellcolor{s3}$3$ & \cellcolor{s4}$4$ & \cellcolor{s2}\color{white}$2$ & \cellcolor{s1}\color{white}$1$ \\
\textbf{QI2} Explanation & \cellcolor{s1}\color{white}$1$ & \cellcolor{s5}\color{white}$5$ & \cellcolor{s4}$4$ & \cellcolor{s3}$3$ & \cellcolor{s3}$3$ & \cellcolor{s2}\color{white}$2$ \\
\textbf{QI3} Instruction & \cellcolor{s2}\color{white}$2$ & \cellcolor{s3}$3$ & \cellcolor{s5}\color{white}$5$ & \cellcolor{s2}\color{white}$2$ & \cellcolor{s4}$4$ & \cellcolor{s2}\color{white}$2$ \\
\textbf{QI4} Comparison  & \cellcolor{s3}$3$ & \cellcolor{s4}$4$ & \cellcolor{s2}\color{white}$2$ & \cellcolor{s3}$3$ & \cellcolor{s3}$3$ & \cellcolor{s3}$3$ \\
\textbf{QI5} Opinion     & \cellcolor{s2}\color{white}$2$ & \cellcolor{s3}$3$ & \cellcolor{s2}\color{white}$2$ & \cellcolor{s3}$3$ & \cellcolor{s4}$4$ & \cellcolor{s5}\color{white}$5$ \\
\bottomrule
\end{tabular}
}
\end{table}

\paragraph{IPA matrix design.}
The Intent--Purpose Alignment(IPA) Matrix is a $5 \times 6$ matrix that scores each (QI, SP) combination on a $1$--$5$ scale (Table~\ref{tab:ipa_matrix}).
For a citation pair $c = (q, a_c, s_c)$ classified as $\mathrm{QI}(q)$ and $\mathrm{SP}(s_c)$, the alignment score is
\begin{equation}
\mathrm{IPA}(c) = \mathbf{M}_{IP}[\mathrm{QI}(q),\, \mathrm{SP}(s_c)] \in \{1, 2, 3, 4, 5\}.
\label{eq:ipa}
\end{equation}
Scores of $3$--$5$ indicate functional alignment, where the source's communicative function contributes to the query's completion condition ($5$ = direct match, $3$ = partial relevance).
Scores of $1$--$2$ indicate structural misalignment, where the source's incentive conflicts with the query's information need ($1$ = structural conflict, $2$ = weak fit).
We define citations with $\mathrm{IPA}(c) \leq 2$ as alignment failures.
This threshold corresponds to the boundary between the two regimes in the rubric, and we verify in Appendix~\ref{app:robustness-bounds} that shifting the threshold by $\pm 1$ preserves relative model rankings.
Per-cell justifications appear in Table\ref{tab:ipa_justifications_1}--\ref{tab:ipa_justifications_2}.

\begin{table}[!p]
\centering
\caption{Full cell-level justifications for the IPA matrix (Part~1: QI1--QI3). Each row explains why the (QI, SP) pair receives its assigned score on the $1$--$5$ scale.}
\label{tab:ipa_justifications_1}
\setlength{\tabcolsep}{4pt}
\renewcommand{\arraystretch}{1.25}
\resizebox{\linewidth}{!}{
\begin{tabular}{ccp{0.82\textwidth}}
\toprule
\textbf{Cell} & \textbf{Score} & \textbf{Rationale} \\
\midrule
\multicolumn{3}{l}{\textbf{QI1 Factoid}} \\
QI1$\times$SP1 & $1$ & Factoid queries require a context-independent verifiable fact; a promotional source's incentive to selectively present favorable information introduces adverse selection even when the fact itself is accurate. \\
QI1$\times$SP2 & $5$ & Neutral knowledge-transmission sources (encyclopedias, documentation) directly resolve factoid queries without incentive distortion. \\
QI1$\times$SP3 & $3$ & Instructional sources may contain the target fact within procedural context, but it is embedded rather than foregrounded; partial relevance. \\
QI1$\times$SP4 & $4$ & News reports frequently contain verifiable facts with editorial accountability, providing substantial support for factoid resolution. \\
QI1$\times$SP5 & $2$ & Community discussions may surface correct facts but lack editorial accountability; the user cannot distinguish verified from anecdotal claims. \\
QI1$\times$SP6 & $1$ & Opinion sources foreground subjective judgment; citing an opinion piece to resolve a factoid query structurally misrepresents the evidentiary basis. \\
\midrule
\multicolumn{3}{l}{\textbf{QI2 Explanation}} \\
QI2$\times$SP1 & $1$ & Explanation queries seek causal understanding; promotional sources selectively frame mechanisms to favor their product, distorting the explanatory account. \\
QI2$\times$SP2 & $5$ & Informational sources are designed to transmit causal mechanisms and principles, directly matching the explanation completion condition. \\
QI2$\times$SP3 & $4$ & Instructional sources often embed explanatory content (why a step works), providing substantial though not primary explanatory value. \\
QI2$\times$SP4 & $3$ & News reports may describe causes of events but typically lack the depth required for mechanistic understanding; partial relevance. \\
QI2$\times$SP5 & $3$ & Community discussions may contain expert-level explanations but are mixed with speculative or incomplete accounts; partial relevance. \\
QI2$\times$SP6 & $2$ & Opinion pieces may offer interpretive frameworks but foreground the author's perspective over neutral mechanism; weak fit for explanation. \\
\midrule
\multicolumn{3}{l}{\textbf{QI3 Instruction}} \\
QI3$\times$SP1 & $2$ & Promotional sources may include product-specific procedures, but the commercial incentive to steer users toward a particular solution weakens procedural neutrality. \\
QI3$\times$SP2 & $3$ & Informational sources provide background knowledge but rarely offer executable step-by-step procedures; the user must infer the procedure from conceptual content. \\
QI3$\times$SP3 & $5$ & Instructional sources (tutorials, how-to guides) directly match the procedural completion condition of instruction queries. \\
QI3$\times$SP4 & $2$ & News reports describe events rather than actionable procedures; citing a report to resolve an instructional query mismatches the source function. \\
QI3$\times$SP5 & $4$ & Community discussions (e.g., Stack Overflow answers) frequently provide tested, peer-reviewed procedures; substantial procedural support. \\
QI3$\times$SP6 & $2$ & Opinion sources express preferences rather than executable procedures; weak fit for resolving a procedural query. \\
\bottomrule
\end{tabular}
}
\end{table}
\begin{table}[!p]
\centering
\caption{Full cell-level justifications for the IPA matrix (Part~2: QI4--QI5). Each row explains why the (QI, SP) pair receives its assigned score on the $1$--$5$ scale.}
\label{tab:ipa_justifications_2}
\setlength{\tabcolsep}{4pt}
\renewcommand{\arraystretch}{1.25}
\resizebox{\linewidth}{!}{
\begin{tabular}{ccp{0.82\textwidth}}
\toprule
\textbf{Cell} & \textbf{Score} & \textbf{Rationale} \\
\midrule
\multicolumn{3}{l}{\textbf{QI4 Comparison}} \\
QI4$\times$SP1 & $3$ & Promotional sources may contain comparative claims, but the commercial incentive biases the comparison; partially relevant if the user seeks feature lists. \\
QI4$\times$SP2 & $4$ & Informational sources provide neutral feature descriptions that support structured comparison, though they may not directly rank alternatives. \\
QI4$\times$SP3 & $2$ & Instructional sources focus on how to use a single option rather than comparing alternatives; weak fit for evaluative comparison. \\
QI4$\times$SP4 & $3$ & News reports may cover product launches or policy alternatives, offering partial comparative context without structured evaluation. \\
QI4$\times$SP5 & $3$ & Community discussions may contain experiential comparisons, but the lack of systematic criteria limits their evaluative reliability; partial relevance. \\
QI4$\times$SP6 & $3$ & Opinion sources may express ranked preferences, but subjectivity and lack of sharable criteria place them at partial relevance for structured comparison. \\
\midrule
\multicolumn{3}{l}{\textbf{QI5 Opinion}} \\
QI5$\times$SP1 & $2$ & Promotional sources may contain implicit value positions, but their commercial incentive makes them unreliable representatives of genuine opinion diversity. \\
QI5$\times$SP2 & $3$ & Informational sources may present multiple perspectives neutrally but do not themselves take a position; partial relevance for opinion-seeking queries. \\
QI5$\times$SP3 & $2$ & Instructional sources address procedural tasks, not value judgments; citing a how-to guide for an opinion query mismatches the source function. \\
QI5$\times$SP4 & $3$ & News reports may frame contested issues and present stakeholder positions, offering partial context for forming opinions. \\
QI5$\times$SP5 & $4$ & Community discussions surface diverse perspectives from experienced participants, providing substantial support for opinion formation. \\
QI5$\times$SP6 & $5$ & Opinion sources directly match the completion condition of opinion queries: the user seeks a subjective, value-laden judgment, which is precisely what opine sources provide. \\
\bottomrule
\end{tabular}
}
\end{table}

\paragraph{Response-level aggregation.}
For a response $r$ with citation set $C_r$, we define two metrics:
\begin{equation}
\mathrm{R\text{-}IPA}(r) = \frac{1}{|C_r|}\sum_{c \in C_r} \mathrm{IPA}(c),
\qquad
\mathrm{AFR}(r) = \frac{|\{c \in C_r : \mathrm{IPA}(c) \leq 2\}|}{|C_r|},
\label{eq:ripam}
\end{equation}
\begin{equation}
\mathrm{R\text{-}AFR}(r) = \mathbbm{1}\bigl[\exists\, c \in C_r : \mathrm{IPA}(c) \leq 2\bigr].
\label{eq:rafr}
\end{equation}
R-IPA is the mean alignment score per response (higher is better),
AFR (Alignment Failure Rate) is the share of citations in the conflict
region (lower is better), {and R-AFR (Response-level Alignment Failure Rate) is a binary indicator of whether the user encounters at least one
misaligned citation in the response.}
Responses with zero citations ($10.3\%$ of the corpus) are excluded from aggregation but retained in the public release.

\paragraph{Validation.}
A panel of $10$ {IR/LIS researchers independently} rated all $30$ IPA Matrix cells on a $1$--$5$ scale.
Inter-rater reliability is excellent ($\mathrm{ICC}(2,k) = 0.916$~\citep{koo2016guideline}, $95\%$ CI $= [0.86, 0.95]$), and the correlation between expert consensus and design values is strong ($r = 0.871$, $\mathrm{CCC} = 0.821$).
No cell deviates from expert consensus by $2$ or more points, and the four cells with $1$-point deviation are all in the conservative direction.
Full validation statistics appear in Table~\ref{tab:c_validation}.
{On the QI ($\kappa = 0.862$) and SP ($\kappa = 0.862$) label-assignment 
tasks, the LLM judge reaches substantial agreement against three human 
annotators (Table~\ref{tab:c_judge_kappa}).}

\subsection{Source Suitability Across Domains and Types}
\label{app:axis3-detail}
 
\begin{table}[!htbp]
\centering
\caption{Source Domain taxonomy: ten labels classifying the topical domain of each cited source.}
\label{tab:sd_taxonomy}
\small
\setlength{\tabcolsep}{8pt}
\renewcommand{\arraystretch}{1.25}
\begin{tabular}{clp{0.60\textwidth}}
\toprule
\textbf{Label} & \textbf{Domain} & \textbf{Definition} \\
\midrule
SD1  & Medical/Health       & Diseases, treatments, medications, mental health, nutrition \\
SD2  & Legal                & Laws, regulations, court decisions, legal rights, compliance \\
SD3  & Finance              & Personal finance, investing, economics, taxation, banking \\
SD4  & Education            & Education, curriculum, university, degree, scholarship \\
SD5  & Science              & Natural sciences, mathematics, physics, chemistry, biology \\
SD6  & Code/Data            & Programming, software, data analysis, machine learning, AI \\
SD7  & Technical            & IT systems, infrastructure, cloud services, mechanics \\
SD8  & Social/Professional  & Society, relationships, workplace, career, parenting \\
SD9  & Shopping/Travel      & Shopping, product reviews, travel, accommodation \\
SD10 & Everyday             & Daily life, DIY, hobby, lifestyle, sports, pets, cooking \\
\bottomrule
\end{tabular}
\end{table}
\begin{table}[!htbp]
\centering
\caption{Source Type taxonomy: six labels classifying the institutional type of each cited source.}
\label{tab:st_taxonomy}
\small
\setlength{\tabcolsep}{6pt}
\renewcommand{\arraystretch}{1.25}
\begin{tabular}{clp{0.60\textwidth}}
\toprule
\textbf{Label} & \textbf{Name} & \textbf{Definition} \\
\midrule
ST1 & Official Institution & Government, regulatory agencies, nonprofits, academic institutions \\
ST2 & Paper/Research       & Peer-reviewed academic paper with author, abstract, references \\
ST3 & News/Magazine        & News article with byline and publication date \\
ST4 & Wiki/Forum           & Community-created content: wikis, Q\&A, forums \\
ST5 & Blog/Social             & Individual-authored content: blogs, social media, personal pages \\
ST6 & Private Company      & Company-published content: product pages, docs, corporate blogs \\
\bottomrule
\end{tabular}
\end{table}

\paragraph{Source Domain (SD) taxonomy.}
SD classifies each cited source by the substantive domain of its content into ten labels~\citep{iab_taxonomy} (Table~\ref{tab:sd_taxonomy}): SD1 Medical (diseases, treatments, mental health), SD2 Legal (law, regulation), SD3 Finance (banking, taxation), SD4 Education (curriculum, university, scholarship), SD5 Science (natural science, research), SD6 Code/Data (software, datasets), SD7 Technical (engineering, specifications), SD8 Social/Professional (workplace, career), SD9 Shopping/Travel (commerce, lifestyle purchases), and SD10 Everyday (recipes, hobbies, routine information).
Labels SD1--SD3 correspond to the YMYL domains identified by Google's Search Quality Rater Guidelines~\citep{ymyl}; SD4--SD7 capture knowledge-intensive domains; SD8--SD10 capture lifestyle and routine information.
\textcolor{blue}{}

\paragraph{Source Type (ST) taxonomy.}
ST classifies each source by its publication infrastructure and editorial accountability~\citep{rieh2007credibility,sun2019consumer} into six labels (Table~\ref{tab:st_taxonomy}): ST1 Official (government, regulatory agencies, nonprofits), ST2 Research (peer-reviewed academic publication), ST3 News (mainstream journalism), ST4 Wiki/Forum (community-edited or community-discussion content), ST5 Blog/Social (individual-author or social-media content), and ST6 Company (corporate or commercial content).
This is distinct from SP, which captures communicative intent rather than institutional structure.
\textcolor{blue}{}

\paragraph{SS matrix design.}

The Source Suitability (SS) Matrix is a $10 \times 6$ matrix scoring each (SD, ST) combination on a $1$--$5$ scale (Table~\ref{tab:ss_matrix}).
For citation pair $c = (q, a_c, s_c)$ with source $s_c$, the suitability score is
\begin{equation}
\mathrm{SS}(c) = \mathbf{M}_{SS}[\mathrm{SD}(s_c),\, \mathrm{ST}(s_c)] \in \{1, 2, 3, 4, 5\}.
\label{eq:ssm}
\end{equation}
Cell values follow the CRAAP test~\citep{craap} for source evaluation and the YMYL classification~\citep{ymyl}, governed by two principles.
First, in YMYL domains (SD1--SD3), only Official and Research sources receive $5$, while Wiki/Forum and Blog/Social receive $1$ in Medical and Legal domains due to the risk of unverified information in high-stakes contexts.
Second, in Code/Data (SD6), Wiki/Forum sources (ST4) receive $5$ because developer communities such as Stack Overflow hold de facto authority in software engineering; this reflects the empirical authority structure of the domain rather than a deviation from the CRAAP framework.
In non-YMYL domains (SD9 Shopping/Travel, SD10 Everyday), the minimum cell value is $3$ rather than $1$, reflecting the lower stakes of misalignment in routine contexts.
We define citations with $\mathrm{SS}(c) \leq 2$ as suitability failures, and verify in Appendix~\ref{app:robustness-bounds} that shifting the threshold by $\pm 1$ preserves relative model rankings.
Per-cell justifications appear in Tables~\ref{tab:ssm_justifications_1}--\ref{tab:ssm_justifications_3}.

\begin{table}[!htbp]
\centering
\caption{Source Suitability Matrix {(SS Matrix).} Scores range from $1$ (structural mismatch) to $5$ (strong match). Red cells ($\leq 2$) indicate suitability failures.}
\label{tab:ss_matrix}
\small
\setlength{\tabcolsep}{8pt}
\renewcommand{\arraystretch}{1.3}
\begin{tabular}{lcccccc}
\toprule
 & \textbf{ST1} & \textbf{ST2} & \textbf{ST3} & \textbf{ST4} & \textbf{ST5} & \textbf{ST6} \\
 & Official & Research & News & Wiki/Forum & Blog/Social & Company \\
\midrule
\textbf{SD1} Medical   & \cellcolor{s5}\color{white}$5$ & \cellcolor{s5}\color{white}$5$ & \cellcolor{s3}$3$ & \cellcolor{s1}\color{white}$1$ & \cellcolor{s1}\color{white}$1$ & \cellcolor{s2}$2$ \\
\textbf{SD2} Legal     & \cellcolor{s5}\color{white}$5$ & \cellcolor{s5}\color{white}$5$ & \cellcolor{s3}$3$ & \cellcolor{s1}\color{white}$1$ & \cellcolor{s1}\color{white}$1$ & \cellcolor{s2}$2$ \\
\textbf{SD3} Finance   & \cellcolor{s5}\color{white}$5$ & \cellcolor{s5}\color{white}$5$ & \cellcolor{s3}$3$ & \cellcolor{s2}$2$ & \cellcolor{s2}$2$ & \cellcolor{s3}$3$ \\
\textbf{SD4} Education & \cellcolor{s5}\color{white}$5$ & \cellcolor{s5}\color{white}$5$ & \cellcolor{s3}$3$ & \cellcolor{s3}$3$ & \cellcolor{s2}$2$ & \cellcolor{s3}$3$ \\
\textbf{SD5} Science   & \cellcolor{s5}\color{white}$5$ & \cellcolor{s5}\color{white}$5$ & \cellcolor{s3}$3$ & \cellcolor{s3}$3$ & \cellcolor{s2}$2$ & \cellcolor{s2}$2$ \\
\textbf{SD6} Code/Data & \cellcolor{s4}$4$ & \cellcolor{s4}$4$ & \cellcolor{s3}$3$ & \cellcolor{s5}\color{white}$5$ & \cellcolor{s2}$2$ & \cellcolor{s3}$3$ \\
\textbf{SD7} Technical & \cellcolor{s5}\color{white}$5$ & \cellcolor{s4}$4$ & \cellcolor{s3}$3$ & \cellcolor{s3}$3$ & \cellcolor{s2}$2$ & \cellcolor{s3}$3$ \\
\textbf{SD8} Social    & \cellcolor{s4}$4$ & \cellcolor{s4}$4$ & \cellcolor{s4}$4$ & \cellcolor{s3}$3$ & \cellcolor{s3}$3$ & \cellcolor{s3}$3$ \\
\textbf{SD9} Shopping  & \cellcolor{s4}$4$ & \cellcolor{s3}$3$ & \cellcolor{s4}$4$ & \cellcolor{s4}$4$ & \cellcolor{s3}$3$ & \cellcolor{s3}$3$ \\
\textbf{SD10} Everyday & \cellcolor{s4}$4$ & \cellcolor{s3}$3$ & \cellcolor{s4}$4$ & \cellcolor{s4}$4$ & \cellcolor{s4}$4$ & \cellcolor{s3}$3$ \\
\bottomrule
\end{tabular}
\end{table}

\begin{table}[!p]
\centering
\caption{Full cell-level justifications for the SS matrix (Part~1: YMYL domains SD1--SD3). Each row explains why the (SD, ST) pair receives its assigned score.}
\label{tab:ssm_justifications_1}
\setlength{\tabcolsep}{4pt}
\renewcommand{\arraystretch}{1.25}
\resizebox{\linewidth}{!}{
\begin{tabular}{ccp{0.82\textwidth}}
\toprule
\textbf{Cell} & \textbf{Score} & \textbf{Rationale} \\
\midrule
\multicolumn{3}{l}{\textbf{SD1 Medical (YMYL)}} \\
SD1$\times$ST1 & $5$ & Government health agencies (CDC, WHO, NHS) undergo peer review and regulatory oversight; highest reliability for medical claims. \\
SD1$\times$ST2 & $5$ & Peer-reviewed medical research is the gold standard for clinical evidence; directly appropriate for medical queries. \\
SD1$\times$ST3 & $3$ & Mainstream health journalism may simplify or sensationalize findings but operates under editorial accountability; moderate reliability. \\
SD1$\times$ST4 & $1$ & Wiki/forum content in medical domains carries high misinformation risk; anonymous contributors lack clinical accountability. \\
SD1$\times$ST5 & $1$ & Individual health blogs routinely propagate unverified claims with no editorial or clinical oversight. \\
SD1$\times$ST6 & $2$ & Pharmaceutical and health-product companies have regulatory disclosure obligations but also commercial incentives that bias health claims. \\
\midrule
\multicolumn{3}{l}{\textbf{SD2 Legal (YMYL)}} \\
SD2$\times$ST1 & $5$ & Official legal sources (statutes, court opinions, regulatory guidance) are authoritative by definition in legal domains. \\
SD2$\times$ST2 & $5$ & Legal scholarship in peer-reviewed journals provides reliable doctrinal analysis and case interpretation. \\
SD2$\times$ST3 & $3$ & Legal journalism provides accessible summaries but may lack jurisdictional precision; moderate reliability. \\
SD2$\times$ST4 & $1$ & Forum-based legal advice from non-lawyers carries high risk of jurisdictional error and misapplied precedent. \\
SD2$\times$ST5 & $1$ & Personal legal blogs may offer opinions without bar admission or malpractice accountability; structurally inappropriate for legal guidance. \\
SD2$\times$ST6 & $2$ & Corporate legal content has professional accountability but also client-acquisition incentives. \\
\midrule
\multicolumn{3}{l}{\textbf{SD3 Finance (YMYL)}} \\
SD3$\times$ST1 & $5$ & Financial regulators (SEC, central banks) provide authoritative data and guidance with statutory accountability. \\
SD3$\times$ST2 & $5$ & Peer-reviewed finance research offers rigorous, methodologically transparent analysis of financial phenomena. \\
SD3$\times$ST3 & $3$ & Financial journalism provides timely market coverage under editorial standards but may amplify short-term sentiment. \\
SD3$\times$ST4 & $2$ & Finance forums mix informed analysis with speculative advice; mostly inappropriate for financial decisions. \\
SD3$\times$ST5 & $2$ & Personal finance blogs vary widely in quality; absence of fiduciary duty makes them mostly inappropriate for financial guidance. \\
SD3$\times$ST6 & $3$ & Financial institutions' content is subject to regulatory disclosure ($10$-K, prospectuses) but carries sales incentives; moderate reliability. \\
\bottomrule
\end{tabular}
}
\end{table}

\begin{table}[!p]
\centering
\caption{Full cell-level justifications for the SS matrix (Part~2: knowledge-intensive domains SD4--SD7). Each row explains why the (SD, ST) pair receives its assigned score.}
\label{tab:ssm_justifications_2}
\setlength{\tabcolsep}{4pt}
\renewcommand{\arraystretch}{1.25}
\resizebox{\linewidth}{!}{
\begin{tabular}{ccp{0.82\textwidth}}
\toprule
\textbf{Cell} & \textbf{Score} & \textbf{Rationale} \\
\midrule
\multicolumn{3}{l}{\textbf{SD4 Education}} \\
SD4$\times$ST1 & $5$ & Official educational bodies (ministries, accreditation agencies) set authoritative curricular and policy standards. \\
SD4$\times$ST2 & $5$ & Peer-reviewed educational research provides evidence-based pedagogical guidance. \\
SD4$\times$ST3 & $3$ & Education journalism covers policy and trends but may lack pedagogical depth; moderate relevance. \\
SD4$\times$ST4 & $3$ & Educational wikis and forums provide useful crowd-sourced study resources but lack systematic quality control. \\
SD4$\times$ST5 & $2$ & Individual education blogs may share teaching tips but lack institutional backing or peer review; mostly inappropriate. \\
SD4$\times$ST6 & $3$ & EdTech companies provide product-linked educational content with commercial incentives but often useful pedagogical material. \\
\midrule
\multicolumn{3}{l}{\textbf{SD5 Science}} \\
SD5$\times$ST1 & $5$ & Government science agencies (NASA, NOAA, NIH) provide authoritative scientific data and reports. \\
SD5$\times$ST2 & $5$ & Peer-reviewed scientific publications are the primary knowledge-creation mechanism in science. \\
SD5$\times$ST3 & $3$ & Science journalism translates findings for lay audiences but may oversimplify or misrepresent effect sizes. \\
SD5$\times$ST4 & $3$ & Science wikis and forums (e.g., Physics Stack Exchange) contain expert-level content but lack formal peer review. \\
SD5$\times$ST5 & $2$ & Science blogs may distort findings through selective reporting or lack of methodological training; mostly inappropriate. \\
SD5$\times$ST6 & $2$ & Corporate science content carries commercial incentives that may bias interpretation. \\
\midrule
\multicolumn{3}{l}{\textbf{SD6 Code/Data}} \\
SD6$\times$ST1 & $4$ & Official documentation (language specs, RFC standards) is authoritative but may lag behind practical usage patterns. \\
SD6$\times$ST2 & $4$ & Academic CS publications provide algorithmic foundations but may not address practical implementation details. \\
SD6$\times$ST3 & $3$ & Tech journalism covers releases and trends but rarely provides actionable code-level guidance. \\
SD6$\times$ST4 & $5$ & Developer communities (Stack Overflow, GitHub Issues) are the de facto authority for code solutions; peer-tested and version-specific. \\
SD6$\times$ST5 & $2$ & Individual coding blogs vary widely; outdated or untested code snippets may introduce bugs; mostly inappropriate as primary sources. \\
SD6$\times$ST6 & $3$ & Company developer docs and SDKs are authoritative for their own products but may not generalize; moderate reliability. \\
\midrule
\multicolumn{3}{l}{\textbf{SD7 Technical}} \\
SD7$\times$ST1 & $5$ & Official standards bodies (ISO, IEEE, building codes) provide authoritative technical specifications. \\
SD7$\times$ST2 & $4$ & Engineering research publications provide validated methods but may not cover practical implementation constraints. \\
SD7$\times$ST3 & $3$ & Technical journalism provides accessible overviews but lacks specification-level precision. \\
SD7$\times$ST4 & $3$ & Technical forums contain practitioner knowledge but lack formal verification; moderate reliability. \\
SD7$\times$ST5 & $2$ & Individual technical blogs may share useful experience but lack institutional backing or peer review. \\
SD7$\times$ST6 & $3$ & Manufacturer documentation is authoritative for specific products but may omit cross-product comparisons. \\
\bottomrule
\end{tabular}
}
\end{table}

\begin{table}[!p]
\centering
\caption{Full cell-level justifications for the SS matrix (Part~3: lifestyle and routine domains SD8--SD10). Non-YMYL domains use a minimum score of $3$, reflecting lower stakes of source-type misalignment.}
\label{tab:ssm_justifications_3}
\setlength{\tabcolsep}{4pt}
\renewcommand{\arraystretch}{1.25}
\resizebox{\linewidth}{!}{
\begin{tabular}{ccp{0.82\textwidth}}
\toprule
\textbf{Cell} & \textbf{Score} & \textbf{Rationale} \\
\midrule
\multicolumn{3}{l}{\textbf{SD8 Social/Professional}} \\
SD8$\times$ST1 & $4$ & Government labor agencies provide authoritative employment data and workplace regulations. \\
SD8$\times$ST2 & $4$ & Organizational-behavior and HR research provides evidence-based workplace guidance. \\
SD8$\times$ST3 & $4$ & Workplace and career journalism provides timely, editorially reviewed advice on professional norms. \\
SD8$\times$ST4 & $3$ & Workplace forums offer experiential advice but norms vary by culture and industry. \\
SD8$\times$ST5 & $3$ & Career blogs share individual experiences that may not generalize but carry lower risk than YMYL domains. \\
SD8$\times$ST6 & $3$ & Corporate HR content is informative but self-promotional. \\
\midrule
\multicolumn{3}{l}{\textbf{SD9 Shopping/Travel}} \\
SD9$\times$ST1 & $4$ & Government tourism and consumer-protection agencies provide reliable destination and product-safety information. \\
SD9$\times$ST2 & $3$ & Academic tourism or consumer research is reliable but rarely addresses practical purchasing or travel decisions. \\
SD9$\times$ST3 & $4$ & Travel and product journalism provides timely, editorially reviewed recommendations with moderate commercial influence. \\
SD9$\times$ST4 & $4$ & Community review sites and travel forums provide firsthand, peer-validated consumer experiences. \\
SD9$\times$ST5 & $3$ & Individual travel and shopping blogs provide personal experience; commercial sponsorship is common but stakes are low. \\
SD9$\times$ST6 & $3$ & E-commerce and travel-company sites provide product details with obvious sales incentives; useful but biased. \\
\midrule
\multicolumn{3}{l}{\textbf{SD10 Everyday}} \\
SD10$\times$ST1 & $4$ & Government agencies provide authoritative guidance on everyday matters (food safety, recycling rules). \\
SD10$\times$ST2 & $3$ & Academic research on everyday topics is reliable but often overly technical for casual queries. \\
SD10$\times$ST3 & $4$ & Lifestyle journalism provides accessible, editorially reviewed content on everyday topics. \\
SD10$\times$ST4 & $4$ & Community wikis and forums are well-suited for everyday advice where collective experience is the primary authority. \\
SD10$\times$ST5 & $4$ & Personal lifestyle blogs are a natural fit for casual everyday queries where no formal authority is required. \\
SD10$\times$ST6 & $3$ & Corporate lifestyle content is useful but shaped by commercial incentives. \\
\bottomrule
\end{tabular}
}
\end{table}

\paragraph{Response-level aggregation.}
For a response $r$ with citation set $C_r$, we define two metrics:
\begin{equation}
\mathrm{R\text{-}SS}(r) = \frac{1}{|C_r|} \sum_{c \in C_r} \mathrm{SS}(c),
\qquad
\mathrm{SFR}(r) = \frac{|\{c \in C_r : \mathrm{SS}(c) \leq 2\}|}{|C_r|}.
\label{eq:safr}
\end{equation}
\begin{equation}
\mathrm{R\text{-}SFR}(r) = \mathbbm{1}\bigl[\exists\, c \in C_r : \mathrm{SS}(c) \leq 2\bigr].
\label{eq:rsfr}
\end{equation}
R-SS is the mean suitability score per response (higher is better), SFR (Suitability Failure Rate) is the share of citations from unsuitable source types for the relevant domain (lower is better), and {R-SFR (Response-level Suitability Failure Rate) is a binary indicator of whether the user encounters at least one unsuitable source in the response.}
Responses with zero citations are excluded from aggregation but retained in the public release.

\paragraph{Validation.}
A panel of $100$ domain experts ($10$ per domain) independently rated the $6$ cells of their assigned domain on a $1$--$5$ scale.
SD1--SD7 were rated by domain professionals or researchers, and SD8--SD10 by graduate-level researchers.
Inter-rater reliability is excellent across domains ($9$ of $10$ domains with $\mathrm{ICC}(2,k) \geq 0.90$~\citep{koo2016guideline}, median $= 0.958$), and the correlation between expert consensus and design values is strong ($r = 0.943$, $\mathrm{CCC} = 0.934$).
No cell exhibits Tier~$1$ disagreement ($\geq 2$ points from expert consensus).
Full per-domain statistics appear in Table~\ref{tab:c_validation}.

\begin{table}[!htbp]
\centering
\caption{Expert validation of IPA and SS matrices. \textbf{ICC$(2, k)$} is 
two-way mixed-effects intraclass correlation for absolute agreement of 
average measurements; \textbf{Pearson $r$} is the correlation between 
expert consensus and design values; \textbf{MAD} is mean absolute deviation 
on the $1$-to-$5$ scale. The IPA aggregate row reports validation by $n=10$ 
experts on all $30$ cells; SS rows report per-domain validation by $n=10$ 
experts on the $6$ cells of each domain, and the SS aggregate row pools 
all $60$ cells. The \textbf{YMYL} column marks the three high-stakes 
Source Domains (Medical, Legal, and Finance) with 
\textcolor{sourceFg}{$\checkmark$}, and their rows are highlighted accordingly.}
\label{tab:c_validation}
\small
\setlength{\tabcolsep}{6pt}
\renewcommand{\arraystretch}{1.15}
\begin{tabular}{llcccccc}
\toprule
\textbf{Matrix} & \textbf{Row} & \textbf{YMYL} & \textbf{N} & \textbf{ICC${(2,k)}$} & \textbf{95\% CI} & \textbf{Pearson $r$} & \textbf{MAD} \\
\midrule
\textbf{IPA} & Aggregate (30 cells) & & 10 & 0.916 & $[0.86, 0.95]$ & 0.871 & 0.513 \\
\midrule
\multirow{11}{*}{\textbf{SS}}
 & \cellcolor{sourceBg}SD1 Medical
 & \cellcolor{sourceBg}\textcolor{sourceFg}{$\checkmark$}
 & \cellcolor{sourceBg}10
 & \cellcolor{sourceBg}0.980
 & \cellcolor{sourceBg}$[0.94, 1.00]$
 & \cellcolor{sourceBg}0.990
 & \cellcolor{sourceBg}0.350 \\
 & \cellcolor{sourceBg}SD2 Legal
 & \cellcolor{sourceBg}\textcolor{sourceFg}{$\checkmark$}
 & \cellcolor{sourceBg}10
 & \cellcolor{sourceBg}0.983
 & \cellcolor{sourceBg}$[0.95, 1.00]$
 & \cellcolor{sourceBg}0.990
 & \cellcolor{sourceBg}0.350 \\
 & \cellcolor{sourceBg}SD3 Finance
 & \cellcolor{sourceBg}\textcolor{sourceFg}{$\checkmark$}
 & \cellcolor{sourceBg}10
 & \cellcolor{sourceBg}0.973
 & \cellcolor{sourceBg}$[0.92, 1.00]$
 & \cellcolor{sourceBg}0.969
 & \cellcolor{sourceBg}0.267 \\
 & SD4 Education  & & 10 & 0.951 & $[0.86, 0.99]$ & 0.973 & 0.250 \\
 & SD5 Science    & & 10 & 0.964 & $[0.90, 0.99]$ & 0.988 & 0.250 \\
 & SD6 Code/Data  & & 10 & 0.908 & $[0.74, 0.98]$ & 0.913 & 0.350 \\
 & SD7 Technical  & & 10 & 0.931 & $[0.81, 0.99]$ & 0.937 & 0.383 \\
 & SD8 Social     & & 10 & 0.972 & $[0.92, 1.00]$ & 0.842 & 0.500 \\
 & SD9 Shopping   & & 10 & 0.945 & $[0.85, 0.99]$ & 0.753 & 0.300 \\
 & SD10 Everyday  & & 10 & 0.895 & $[0.71, 0.98]$ & 0.778 & 0.267 \\
\cmidrule(lr){2-8}
 & \textbf{Aggregate (60 cells)} & & \textbf{100} & \textbf{0.962} & $\mathbf{[0.95, 0.97]}$ & \textbf{0.943} & \textbf{0.327} \\
\bottomrule
\end{tabular}
\end{table}

\subsection{Fidelity of Answers to Their Sources}
\label{app:axis2-detail}
\label{app:human-validation}

\begin{table}[!htbp]
\centering
\caption{Answer--Source Fidelity (ASF) taxonomy: five labels adjudicating the fidelity of each cited claim to its source.}
\label{tab:asf_taxonomy}
\small
\setlength{\tabcolsep}{5pt}
\renewcommand{\arraystretch}{1.25}
\begin{tabular}{clp{0.68\textwidth}}
\toprule
\textbf{Label} & \textbf{Name} & \textbf{Definition} \\
\midrule
ASF1 & Fabricated    & The cited claim does not exist in the source content at all. \\
ASF2 & Misattributed & The claim comes from a tangential part or is attributed to a context the source does not address. \\
ASF3 & Contradicted  & The source concludes or argues the opposite of what the cited sentence presents. \\
ASF4 & Amplified     & The claim exists but is presented with materially greater certainty, scope, or generality. \\
ASF5 & Supported     & The claim is present in and consistent with the source content. \\
\bottomrule
\end{tabular}
\end{table}

\paragraph{Answer--Source Fidelity (ASF) taxonomy.}
The ASF taxonomy classifies the relationship between a citing sentence and the crawled source content on a $1$--$5$ scale (Table~\ref{tab:asf_taxonomy}): ASF5 Supported (every claim attributable to the source), ASF4 Amplified (claims attributable but a critical qualifier omitted), ASF3 Contradicted (a hedged claim presented as unconditional or reversed), ASF2 Misattributed (the source provides only tangential evidence), and ASF1 Fabricated (no claim supported by the source).
The taxonomy extends the binary supported/not-attributable judgment of ALCE~\citep{alce} and AIS~\citep{rashkin2023ais,maynez2020faithfulness,pagnoni2021frank,kryscinski2020factcc,wang2020qags,laban2022summac}; the three intermediate levels (ASF4--ASF2) capture the directional distortions characteristic of Verified Misguidance.

\paragraph{Citing sentence and source content.}
Each citation pair consists of a two-sentence cited unit extracted from the model's response and the corresponding crawled source content.
The extraction procedure for both marker and block providers is described in Appendix~\ref{app:crawling-sources}.
The ASF judge receives both the cited unit and the source content, and assigns one of the five verdicts (ASF1--ASF5).

\paragraph{Response-level aggregation.}
For a response $r$ with citation set $C_r$, we define two metrics:
\begin{equation}
\mathrm{FFR}(r) = \frac{|\{c \in C_r : \mathrm{ASF}(c) \leq 2\}|}{|C_r|},
\qquad
\mathrm{R\text{-}FFR}(r) = \mathbbm{1}\bigl[\exists\, c \in C_r : \mathrm{ASF}(c) \leq 2\bigr].
\label{eq:ffr}
\end{equation}
FFR (Fidelity Failure Rate) is the share of citations that fail (lower is better), 
and R-FFR (Response-level Fidelity Failure Rate) is a binary indicator of whether the user encounters at least one failed citation in the response.
The threshold $\mathrm{ASF}(c) \leq 2$ groups Fabricated (ASF1) and Misattributed (ASF2), both of which share the property that a user examining the source cannot independently verify the cited claim, distinguishing them from Contradicted (ASF3) or Amplified (ASF4) cases where the source contains the claim in some form.
Responses with zero citations are excluded from aggregation but retained in the public release.

\paragraph{Validation.}
The LLM judge's ASF classification was validated against three human annotators on $200$ stratified samples ($140$ from a base split with $10$ Supported and $4$ Fabricated cases per model, plus $60$ samples covering the three intermediate subtypes at $20$ each) to ensure adequate coverage of rare failure categories.
Cohen's $\kappa$~\citep{cohen1960} against majority vote consensus is $0.858$ ($95\%$ CI $= [0.789, 0.914]$), indicating excellent agreement, with balanced accuracy of $0.918$.
Agreement among the three human annotators is Krippendorff's $\alpha = 0.811$~\citep{krippendorff2011} (all pairwise $\kappa \geq 0.802$), exceeding the substantial agreement threshold~\citep{landis1977measurement} ($\alpha \geq 0.667$).
Full reliability statistics for all five dimensions appear in Table~\ref{tab:c_judge_kappa}.

\begin{table}[!htbp]
\centering
\caption{LLM judge reliability against human consensus across the five classification dimensions. Validation uses $n = 200$ stratified samples per dimension with three human annotators, oversampling rare categories. $\kappa_{\mathrm{cons}}$ is Cohen's $\kappa$ against majority vote gold standard, $\kappa_{\mathrm{pair}}$ is the average of pairwise $\kappa$ between the judge and each annotator, and Bal.\ Acc is balanced accuracy. All five dimensions exceed $\kappa \geq 0.78$.}
\label{tab:c_judge_kappa}
\small
\setlength{\tabcolsep}{6pt}
\renewcommand{\arraystretch}{1.15}
\begin{tabular}{lcccccc}
\toprule
\textbf{Dimension} & \textbf{N} & \textbf{Raw Agr} & \bm{$\kappa_{\mathrm{cons}}$} & \textbf{95\% CI} & \bm{$\kappa_{\mathrm{pair}}$} & \textbf{Bal.\ Acc} \\
\midrule
QI  & 200 & 0.890 & \textbf{0.862} & $[0.806, 0.918]$ & 0.817 & 0.900 \\
SP  & 200 & 0.885 & \textbf{0.862} & $[0.807, 0.910]$ & 0.824 & 0.896 \\
SD  & 200 & 0.835 & \textbf{0.817} & $[0.755, 0.872]$ & 0.783 & 0.872 \\
ST  & 200 & 0.830 & \textbf{0.796} & $[0.731, 0.856]$ & 0.772 & 0.860 \\
ASF & 200 & 0.905 & \textbf{0.858} & $[0.789, 0.914]$ & 0.839 & 0.918 \\
\bottomrule
\end{tabular}
\end{table}

\subsection{Cross-Dimension Integration and Robustness}
\label{app:cross-axis-detail}

\paragraph{Critical VM definition.}
We define a citation pair as a Critical VM instance when all three dimensions fail simultaneously:
\begin{equation}
\mathrm{CritVM}(c) = \mathbbm{1}\bigl[\mathrm{IPA}(c) \leq 2 \;\wedge\; \mathrm{ASF}(c) \leq 2 \;\wedge\; \mathrm{SS}(c) \leq 2\bigr].
\label{eq:critvm}
\end{equation}
Each individual dimension condition corresponds to an established evaluation framework: 
Intent--Purpose Alignment to query intent taxonomies such as 
NF-CATS~\citep{bolotova2022non},
Source Suitability to source quality frameworks such as CRAAP~\citep{craap}, and Answer--Source Fidelity to citation verification benchmarks such as ALCE~\citep{alce} and CiteME~\citep{press2024citeme}.
The simultaneous failure across all three is not captured by any single existing framework, and this composition across dimensions is the central evaluative contribution of \CiteTrace{}.

\paragraph{Statistical independence test.}
Under the null hypothesis that the three dimensions fail independently, the expected joint failure rate is the product of the marginal rates:
\begin{equation}
\mathbb{E}_{\text{indep}}[\mathrm{CritVM}] = \mathrm{AFR} \cdot \mathrm{FFR} \cdot \mathrm{SFR} = 0.051 \cdot 0.306 \cdot 0.271 \approx 0.0042 = 0.42\%.
\label{eq:critvm-expected}
\end{equation}
The observed CritVM rate is $3{,}174 / 761{,}495 = 0.42\%$, matching the independent failure expectation.
A formal $\chi^2$ test against full pairwise dependence is reported in Appendix~\ref{app:robustness-bounds}.

\paragraph{Single judge dependence.}
All five classification dimensions use a single LLM judge (GPT-4o-mini).
This is consistent with the evaluation methodology of comparable frameworks: ALCE~\citep{alce} validates citation quality against a single NLI model ($\kappa = 0.525$), RAGAS~\citep{ragas} prompts a single LLM, and ARES~\citep{ares} trains a single tailored judge per dimension.
Our $\kappa$ values ($0.796$--$0.862$, Table~\ref{tab:c_judge_kappa}) substantially exceed these baselines, and the stratified validation protocol (\S\ref{app:axis2-detail}) with three independent human annotators (Krippendorff $\alpha \geq 0.811$) provides strong evidence that the judge's classifications are reliable.
Nonetheless, absolute metric values may shift under judge replacement; we discuss this dependence in Appendix~\ref{app:scope} and encourage replication with alternative judge families.

\paragraph{Threshold sensitivity protocol.}
The failure threshold $\leq 2$ for all three dimensions is anchored in the conflict and failure regime built into the rubrics (\S\ref{app:axis1-detail}, \S\ref{app:axis3-detail}, \S\ref{app:axis2-detail}).
To quantify sensitivity to this choice, we evaluate seven $\pm 1$ perturbation variants across the three dimensions: $\{<2, \leq 2, \leq 3\}$ for each dimension independently, plus four mixed perturbations.
Rankings preserve Kendall $\tau \geq 0.82$ in five of the seven variants, with the two exceptions involving the $\leq 3$ perturbation of Answer--Source Fidelity (FFR), which compresses the spread across models but retains the overall provider ordering.
Full ranking tables for each variant appear in Appendix~\ref{app:robustness-bounds}.

\begin{figure}[!p]
\centering
\begin{tcolorbox}[
  colback=gray!4, colframe=gray!55!black,
  title=\textbf{System prompt},
  fonttitle=\bfseries, width=0.92\textwidth,
  arc=2pt, boxrule=0.6pt, fontupper=\footnotesize\ttfamily
]
You are a query intent classifier. Given a user query, you assign exactly one intent label.\\[3pt]
Use the full query as evidence for your classification --- not surface wording alone.\\
Classify the query by the user's primary completion condition: what they mainly need in order to consider the query resolved.\\
If multiple subparts are present, choose the single label that best captures the user's primary need.\\[5pt]
\#\# Intent Definitions\\[3pt]
\textbf{QI1 Factoid.} The user wants a specific, verifiable fact such as a name, number, date, definition, rule, status, or context-independent yes/no answer. A short factual answer fully resolves the query.\\[3pt]
\textbf{QI2 Explanation.} The user wants understanding of a cause, mechanism, principle, reason, relationship, or underlying process. The query is resolved when the user understands why or how something works.\\[3pt]
\textbf{QI3 Instruction.} The user wants steps, procedures, methods, or diagnostic actions to perform a task, fix a problem, or close a gap between expected and actual state. The query is resolved when the user can act on the answer. The subject of the task must be a concrete, executable procedure --- not an open-ended personal or interpersonal situation.\\[3pt]
\textbf{QI4 Comparison.} The user wants alternatives evaluated in order to choose, decide, compare, rank, or receive a recommendation. The query is resolved by a grounded judgment using identifiable, sharable criteria. This includes queries asking whether something is good, appropriate, sufficient, or worth doing --- where the answer requires evaluation against criteria rather than factual lookup or causal explanation.\\[3pt]
\textbf{QI5 Opinion.} The user wants subjective perspective-sharing, lived experience, ethical reflection, interpersonal interpretation, or open-ended discussion. The query is resolved through viewpoint exchange rather than factual lookup, explanation, procedural guidance, or criteria-based comparison.\\[5pt]
\#\# Output (strict JSON)\\[3pt]
\{ ``intent\_reasoning'': ``\textit{step-by-step reasoning}'',\\
\hphantom{xx}``intent'': ``QI1'' $\vert$ ``QI2'' $\vert$ ``QI3'' $\vert$ ``QI4'' $\vert$ ``QI5'' \}
\end{tcolorbox}
\begin{tcolorbox}[
  colback=gray!4, colframe=gray!55!black,
  title=\textbf{User prompt},
  fonttitle=\bfseries, width=0.92\textwidth,
  arc=2pt, boxrule=0.6pt, fontupper=\footnotesize\ttfamily
]
\texttt{<query>}\\
\texttt{\{query\}}\\
\texttt{</query>}
\end{tcolorbox}
\caption{Full classification prompt for Query Intent (QI1--QI5). Issued via \texttt{gpt-4o-mini-2024-07-18} with temperature~$0$ and structured-output JSON schema in strict mode. The system prompt provides the full taxonomy codebook; the user prompt contains the Stack Exchange post title.}
\label{fig:prompt_qi}
\end{figure}

\begin{figure}[!p]
\centering
\begin{tcolorbox}[
  colback=gray!4, colframe=gray!55!black,
  title=\textbf{System prompt},
  fonttitle=\bfseries, width=0.92\textwidth,
  arc=2pt, boxrule=0.6pt, fontupper=\footnotesize\ttfamily
]
You are a web source purpose classifier. Given crawled web page content and its source URL, you assign exactly one purpose label.\\[3pt]
Use both the source content and the full URL as evidence for your classification --- not isolated signals alone.\\
Reason step-by-step based on all available information and choose the label that best captures its primary reader-facing purpose.\\[5pt]
\#\# Purpose Definitions\\[3pt]
\textbf{SP1 To Promote.} The page primarily exists to market, sell, position, or represent a company, product, service, or brand. Its main function is commercial promotion, business representation, or conversion, even if it also contains factual or explanatory material.\\[3pt]
\textbf{SP2 To Inform.} The page primarily exists to present factual, descriptive, explanatory, or reference-style information about a topic, concept, entity, or subject. Its main function is to help the reader understand or look up something. The reader's role is to comprehend, not to execute.\\[3pt]
\textbf{SP3 To Instruct.} The page primarily exists to help the reader do something by providing steps, procedures, methods, walkthroughs, or other executable guidance. Its main function is to support task completion or skill execution. The reader's role is to follow along and act, not merely to understand.\\[3pt]
\textbf{SP4 To Report.} The page primarily exists to report events, developments, announcements, or other time-linked occurrences. Its main function is to tell the reader what happened, what changed, or what was announced. Includes news articles with a byline and publication date, press releases, government announcements, and research press coverage.\\[3pt]
\textbf{SP5 To Discuss.} The page primarily exists as a space for exchange among multiple contributors, such as questions, answers, replies, comments, or community problem-solving. Its main function depends on multi-party participation rather than a single authored voice.\\[3pt]
\textbf{SP6 To Opine.} The page primarily exists to express a viewpoint, judgment, interpretation, review, editorial stance, or advocacy position. Its main function is subjective evaluation or perspective-sharing rather than neutral information, procedural guidance, event reporting, or multi-party discussion.\\[5pt]
\#\# Output (strict JSON)\\[3pt]
\{ ``purpose\_reasoning'': ``\textit{step-by-step reasoning}'', ``purpose'': ``SP1''..``SP6'' \}
\end{tcolorbox}
\begin{tcolorbox}[
  colback=gray!4, colframe=gray!55!black,
  title=\textbf{User prompt},
  fonttitle=\bfseries, width=0.92\textwidth,
  arc=2pt, boxrule=0.6pt, fontupper=\footnotesize\ttfamily
]
\texttt{<source\_url>}\\
\texttt{\{source\_url\}}\\
\texttt{</source\_url>}\\[3pt]
\texttt{<source\_content>}\\
\texttt{\{source\_content\}}\\
\texttt{</source\_content>}
\end{tcolorbox}
\caption{Full classification prompt for Source Purpose (SP1--SP6).}
\label{fig:prompt_sp}
\end{figure}

\begin{figure}[!p]
\centering
\begin{tcolorbox}[
  colback=gray!4, colframe=gray!55!black,
  title=\textbf{System prompt},
  fonttitle=\bfseries, width=0.92\textwidth,
  arc=2pt, boxrule=0.6pt, fontupper=\footnotesize\ttfamily
]
You are a web source domain classifier. Given crawled web page content and its source URL, you assign exactly one subject area label.\\[3pt]
Use both the crawled content and the full URL structure (domain, hostname, path) as evidence for your classification --- not only the example signals listed below.\\
Reason step-by-step based on all available information.\\[5pt]
\#\# Domain Definitions\\[3pt]
\textbf{SD1 Medical/Health.} Diseases, treatments, medications, mental health, nutrition.\\[3pt]
\textbf{SD2 Legal.} Laws, regulations, court decisions, legal rights, compliance.\\[3pt]
\textbf{SD3 Finance.} Personal finance, investing, economics, taxation, banking, insurance, real estate.\\[3pt]
\textbf{SD4 Education.} Education, curriculum, student, degree, scholarship, exam, learning, tuition, admission.\\[3pt]
\textbf{SD5 Science.} Natural sciences, mathematics, physics, chemistry, biology, astronomy.\\[3pt]
\textbf{SD6 Code/Data.} Programming, software, data analysis, machine learning, AI.\\[3pt]
\textbf{SD7 Technical.} IT systems, infrastructure, cloud services, mechanics, electronics.\\[3pt]
\textbf{SD8 Social/Professional.} Society, relationships, workplace, career, parenting, job search.\\[3pt]
\textbf{SD9 Shopping/Travel.} Shopping, product reviews, travel, accommodation.\\[3pt]
\textbf{SD10 Everyday.} Daily life, culture, DIY, hobbies, home, lifestyle, sports, entertainment, pets, food, cooking.\\[5pt]
\#\# Output (strict JSON)\\[3pt]
\{ ``domain\_reasoning'': ``\textit{step-by-step reasoning}'', ``domain'': ``SD1''..``SD10'' \}
\end{tcolorbox}
\begin{tcolorbox}[
  colback=gray!4, colframe=gray!55!black,
  title=\textbf{User prompt},
  fonttitle=\bfseries, width=0.92\textwidth,
  arc=2pt, boxrule=0.6pt, fontupper=\footnotesize\ttfamily
]
\texttt{<source\_url>}\\
\texttt{\{source\_url\}}\\
\texttt{</source\_url>}\\[3pt]
\texttt{<source\_content>}\\
\texttt{\{source\_content\}}\\
\texttt{</source\_content>}
\end{tcolorbox}
\caption{Full classification prompt for Source Domain (SD1--SD10). YMYL labels (SD1--SD3) follow Google's Search Quality Rater Guidelines~\citep{ymyl}.}
\label{fig:prompt_sd}
\end{figure}

\begin{figure}[!p]
\centering
\begin{tcolorbox}[
  colback=gray!4, colframe=gray!55!black,
  title=\textbf{System prompt},
  fonttitle=\bfseries, width=0.92\textwidth,
  arc=2pt, boxrule=0.6pt, fontupper=\footnotesize\ttfamily
]
You are a web source type classifier. Given crawled web page content and its source URL, you assign exactly one structural type label.\\[3pt]
Use both the crawled content and the source URL as evidence for your classification --- not only the example signals listed below.\\
Evaluate ALL type definitions before making a final decision.\\
Reason step-by-step based on all available information.\\[5pt]
\#\# Type Definitions\\[3pt]
\textbf{ST1 Official Institution.} Content formally issued under the name of: government bodies, legislative institutions, public regulatory agencies; intergovernmental organizations (e.g., UN, WHO, EU, IMF); accredited nonprofits, professional associations, or academic institutions. URL signal examples: \texttt{.gov}, \texttt{.go.**}, \texttt{.int}, \texttt{.ac.**}, \texttt{.edu}.\\[3pt]
\textbf{ST2 Paper/Research.} Peer-reviewed academic paper published in a journal or conference proceedings. Must have author name, affiliation, abstract, and references. Excludes theses, preprints, and working papers.\\[3pt]
\textbf{ST3 News/Magazine.} News or magazine article published by a media outlet. Must have BOTH a named INDIVIDUAL author (byline) identifiable in the content AND a publication date.\\[3pt]
\textbf{ST4 Wiki/Forum.} Content collectively created and maintained by a community. Includes wikis, Q\&A platforms, forums, and discussion boards. URL signal examples: wikipedia.org, reddit.com, stackoverflow.com, quora.com.\\[3pt]
\textbf{ST5 Blog/Social.} Content created and published by an INDIVIDUAL. Includes blogs, social media posts, and personal channel pages. URL signal examples: twitter.com, x.com, youtube.com, instagram.com, tiktok.com, facebook.com, medium.com, substack.com.\\[3pt]
\textbf{ST6 Private Company.} Content published by a private company or non-accredited organization as the publisher. Includes product pages, documentation, and corporate blog posts.\\[5pt]
\#\# Output (strict JSON)\\[3pt]
\{ ``type\_reasoning'': ``\textit{step-by-step reasoning}'', ``source\_type'': ``ST1''..``ST6'' \}
\end{tcolorbox}
\begin{tcolorbox}[
  colback=gray!4, colframe=gray!55!black,
  title=\textbf{User prompt},
  fonttitle=\bfseries, width=0.92\textwidth,
  arc=2pt, boxrule=0.6pt, fontupper=\footnotesize\ttfamily
]
\texttt{<source\_url>}\\
\texttt{\{source\_url\}}\\
\texttt{</source\_url>}\\[3pt]
\texttt{<source\_content>}\\
\texttt{\{source\_content\}}\\
\texttt{</source\_content>}
\end{tcolorbox}
\caption{Full classification prompt for Source Type (ST1--ST6). }
\label{fig:prompt_st}
\end{figure}

\begin{figure}[!p]
\centering
\begin{tcolorbox}[
  colback=gray!4, colframe=gray!55!black,
  title=\textbf{System prompt},
  fonttitle=\bfseries, width=0.92\textwidth,
  arc=2pt, boxrule=0.6pt, fontupper=\scriptsize\ttfamily
]
You are an answer-source fidelity evaluator. Given a cited sentence from an LLM response and the full text of the cited source, you assign exactly one fidelity label.\\[3pt]
Use the cited\_sentence as the primary evaluation target. Read the ENTIRE source\_content before making a judgment.\\
Identify the cited\_sentence's main claim --- the central assertion the sentence is built around --- and use that as the basis for your judgment. Treat incidental details (extra adjectives, illustrative terminology, side phrases) as secondary; they do not by themselves disqualify a label.\\
Reason step-by-step based on all available information.\\[5pt]
\#\# Decision Procedure\\[3pt]
Decide the final label in two steps.\\[3pt]
\textbf{Step 1 --- Macro verdict.} Choose one of:\\[3pt]
\textbf{SUPPORTED.} The cited\_sentence's main claim is present in and consistent with the source content. Citing one valid aspect of a multi-faceted source is acceptable. Natural summarization that omits details, or restates the source in different words while preserving its meaning, scope, and certainty, is SUPPORTED. Minor extra wording (e.g., an illustrative term, an everyday example, a side phrase) that does not change the main claim's truth value does NOT by itself disqualify SUPPORTED.\\[3pt]
\textbf{DISTORTED.} The source discusses the SAME topic as the cited\_sentence AND contains a specific passage that the cited\_sentence is based on, BUT the cited\_sentence's version materially differs from the source in a way that changes the claim's meaning, scope, certainty, or attribution. Use DISTORTED when you can point to the specific source passage being altered. If the alteration is only cosmetic and the main claim still holds, prefer SUPPORTED.\\[3pt]
\textbf{FABRICATED.} The cited\_sentence's main claim cannot be located in the source at all. This applies when the source covers a different topic entirely, or discusses the same broad topic but never makes --- or even gestures at --- the specific claim asserted. Do NOT use FABRICATED merely because the cited\_sentence adds incidental details on top of a claim that IS supported; that case is SUPPORTED (if the extras are minor) or DISTORTED (if the extras materially change the claim).\\[5pt]
\textbf{Step 2 --- Final label.}\\[3pt]
If SUPPORTED $\to$ output \textbf{ASF5}. \quad If FABRICATED $\to$ output \textbf{ASF1}.\\
If DISTORTED $\to$ choose exactly one distortion mechanism below and output the corresponding ASF label.\\[5pt]
\#\# Distortion Mechanisms (used only when Step 1 = DISTORTED)\\[3pt]
\textbf{ASF4 Amplified.} The main claim exists in the source but the cited\_sentence presents it with materially greater certainty, scope, or generality. The source uses explicit qualifiers (e.g., ``may,'' ``suggests,'' ``in some cases,'' ``preliminary'') that the cited\_sentence strips away, OR the cited\_sentence extends a narrowly-scoped finding into a broader claim, OR the cited\_sentence adds a substantive procedural step / specific numerical figure / branded slogan that materially changes what the source actually says.\\[3pt]
\textbf{ASF3 Contradicted.} The source concludes or argues the opposite of what the cited\_sentence presents.\\[3pt]
\textbf{ASF2 Misattributed.} The cited\_sentence's main claim is plausible AND the source contains recognizably related content, BUT the claim comes from a different part of the source's discussion or is attributed to a context the source does not actually address.\\[5pt]
\#\# Tie-breakers\\[3pt]
$\bullet$ If the main claim is supported but extras are present: prefer SUPPORTED for minor incidental extras; prefer ASF4 (Amplified) only when the extras materially change the claim.\\
$\bullet$ If you are torn between SUPPORTED and FABRICATED, ask whether the source contains the main claim at all. If yes $\to$ SUPPORTED. If no $\to$ consider DISTORTED first, then FABRICATED.\\[5pt]
\#\# Output (strict JSON)\\[3pt]
\{ ``asf\_reasoning'': ``\textit{step-by-step reasoning}'', ``verdict'': ``ASF1''..``ASF5'' \}
\end{tcolorbox}
\begin{tcolorbox}[
  colback=gray!4, colframe=gray!55!black,
  title=\textbf{User prompt},
  fonttitle=\bfseries, width=0.92\textwidth,
  arc=2pt, boxrule=0.6pt, fontupper=\footnotesize\ttfamily
]
\texttt{<cited\_sentence>}\\
\texttt{\{cited\_sentence\}}\\
\texttt{</cited\_sentence>}\\[3pt]
\texttt{<source\_content>}\\
\texttt{\{source\_content\}}\\
\texttt{</source\_content>}
\end{tcolorbox}
\caption{Full adjudication prompt for Answer--Source Fidelity (ASF1--ASF5).}
\label{fig:prompt_as}
\end{figure}
\clearpage


\begin{figure}[!p]
\centering
\includegraphics[width=\textwidth]{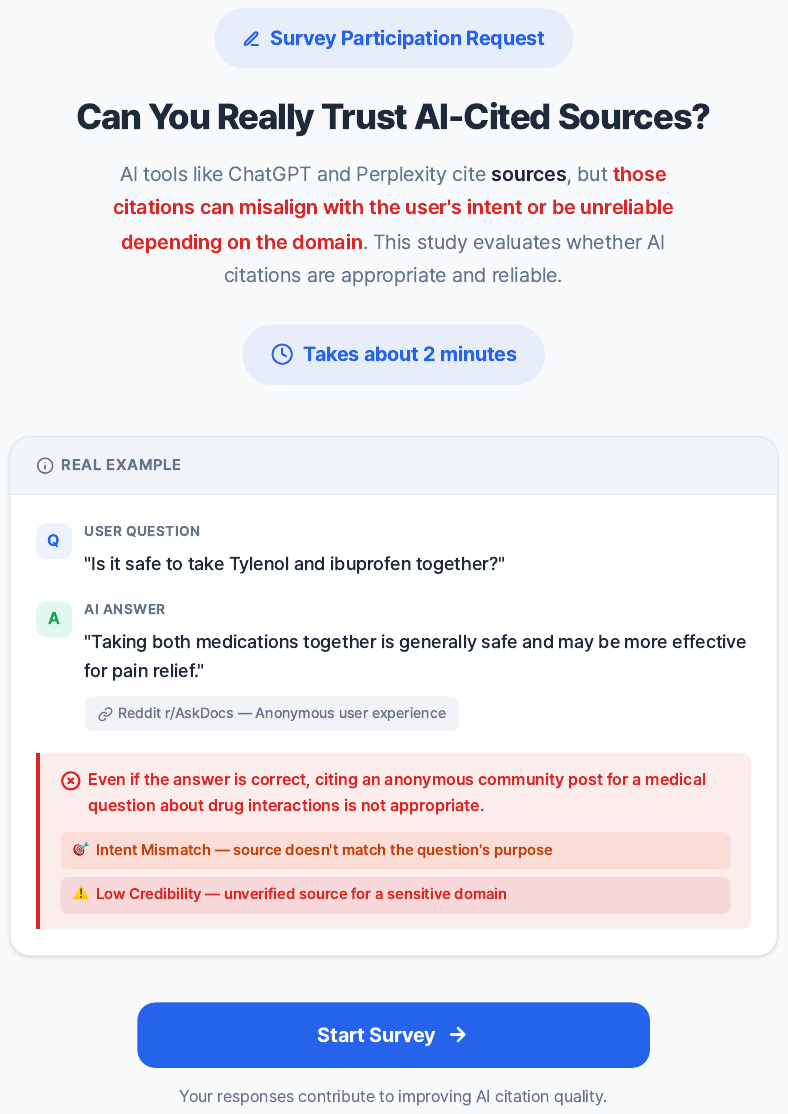}
\caption{Screenshot of the expert-panel survey landing page used for IPA Matrix validation (\S\ref{app:axis1-detail}) and SS Matrix validation (\S\ref{app:axis3-detail}). The page introduces the rating task and presents a real-world example before participants begin.}
\label{fig:ui_request}
\end{figure}
\clearpage

\begin{figure}[!p]
\centering
\includegraphics[width=\textwidth]{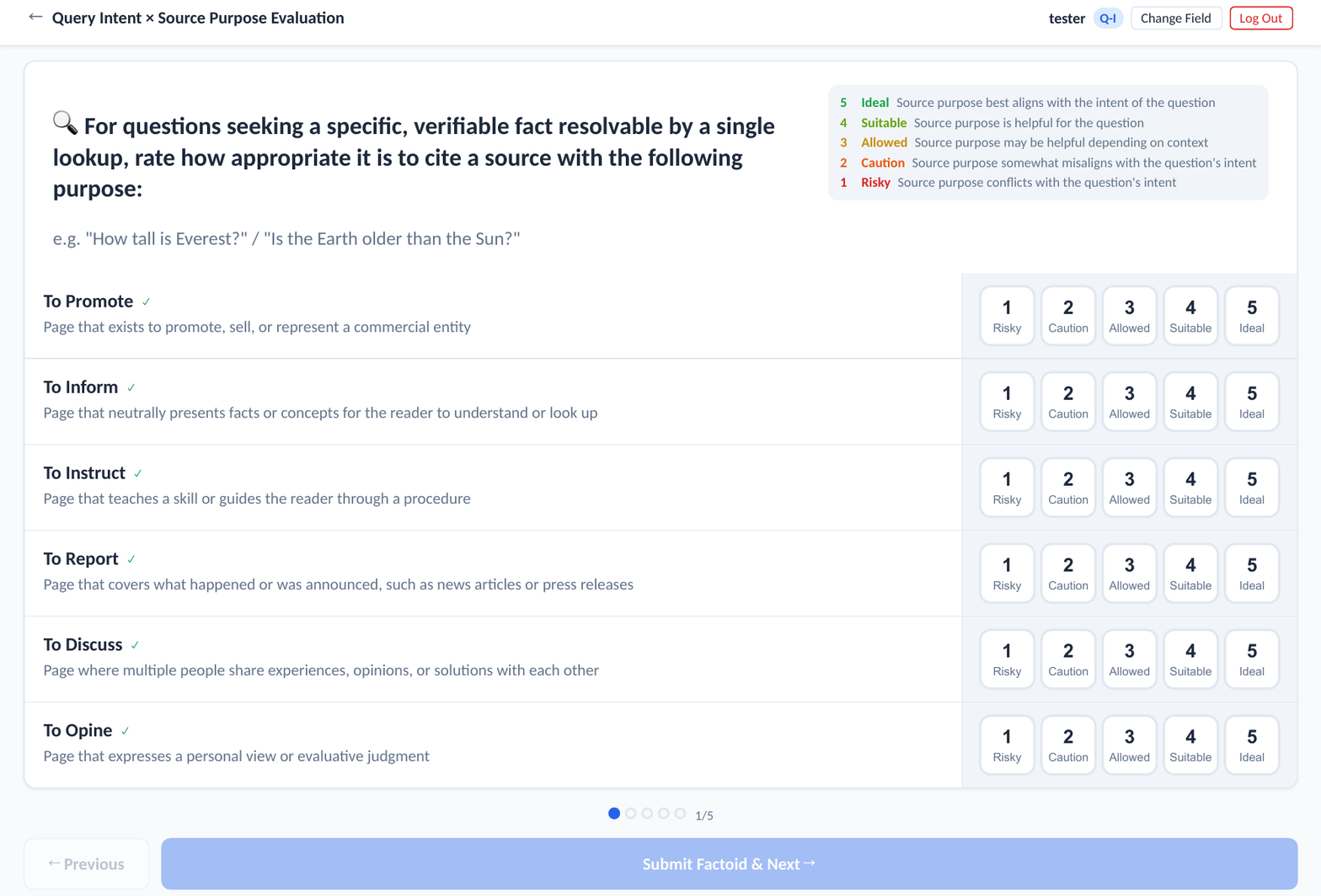}
\caption{Screenshot of the IPA matrix expert-validation interface (\S\ref{app:axis1-detail}). On a single page, validators rate the appropriateness of each of the six source purposes (SP1--SP6) for one query intent (here QI1 Factoid) on the $1$--$5$ scale. Validators complete five such pages, one per intent.}
\label{fig:ui_qisp}
\end{figure}
\clearpage

\begin{figure}[!p]
\centering
\includegraphics[width=\textwidth]{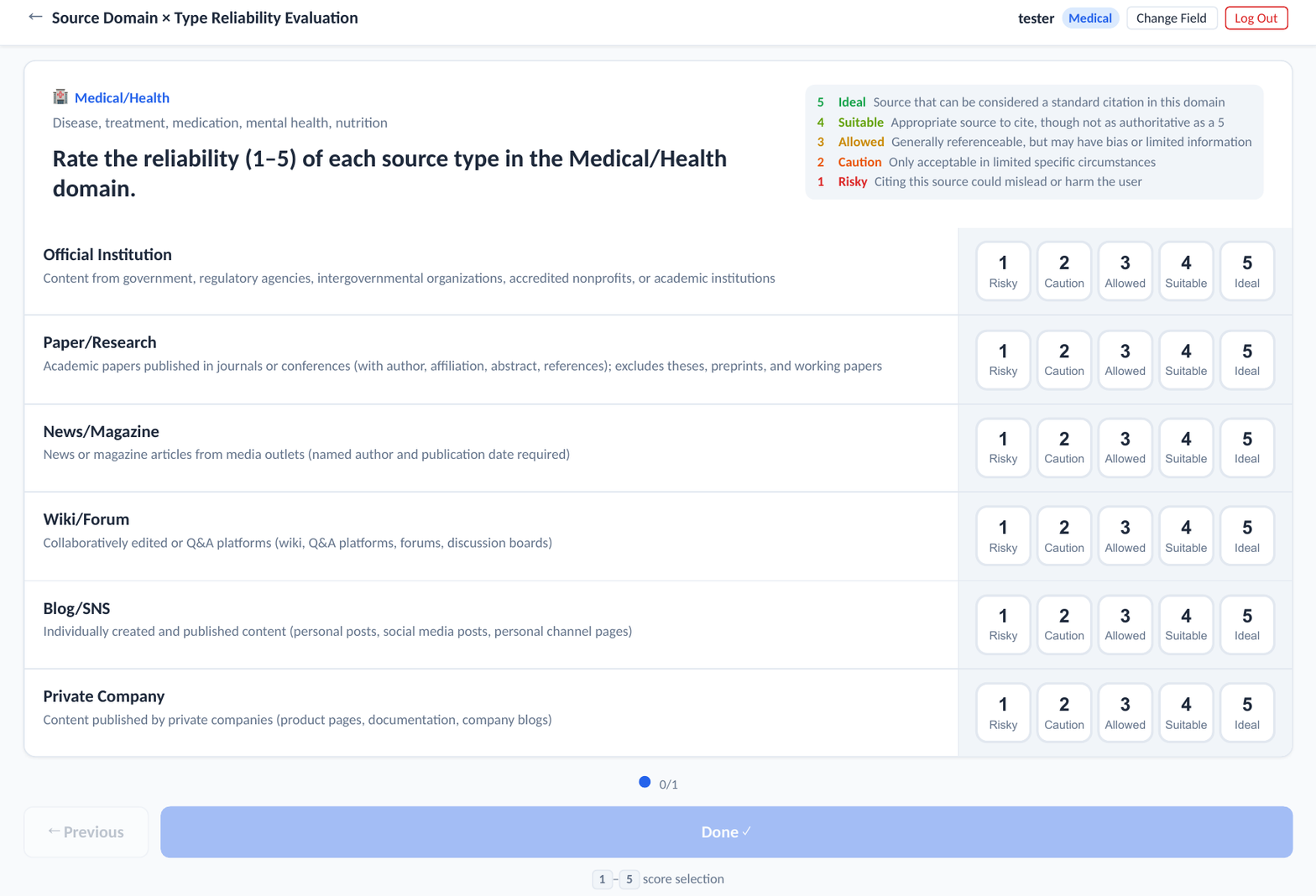}
\caption{Screenshot of the SSM expert-validation interface (\S\ref{app:axis3-detail}). On a single page, validators rate the reliability of each of the six source types (ST1--ST6) for one assigned source domain (here SD1 Medical/Health). Each of the ten domains is rated independently by $n=10$ domain experts.}
\label{fig:ui_sdst}
\end{figure}
\clearpage
\definecolor{asOne}{HTML}{7A4A18}
\definecolor{asTwo}{HTML}{A86F2A}
\definecolor{asThree}{HTML}{C4883A}
\definecolor{asFour}{HTML}{DEB57A}
\definecolor{asFive}{HTML}{F0DDB8}

\section{The \VMshort{} Effect: Detailed Empirical Results}
\label{app:results}
This appendix expands the empirical results summarized in
Section~\ref{sec:vm_effect}. Seven subsections organize the material
from the broadest aggregate view to fine-grained robustness checks:
structural composition of the source pool
(\S\ref{app:biased-pool}), aggregate failure rates across the three
evaluation dimensions (\S\ref{app:failure-profile}), the
fidelity--suitability trade-off across models
(\S\ref{app:three-failures}), how provider, scale, and reasoning
shape citation quality (\S\ref{app:search-gen-independent}),
response-level failure exposure (\S\ref{app:citations-to-users}),
robustness and lower bounds (\S\ref{app:robustness-bounds}), and
qualitative failure analysis (\S\ref{app:cases}). All metrics are
computed on the final evaluable pool of 761{,}495 citation pairs
(Appendix~\ref{app:crawling-sources}, Table~\ref{tab:b7_outcome}).

\subsection{A Structurally Biased Source Pool}
\label{app:biased-pool}
\label{app:source_composition}

\paragraph{Source pool composition.}
The pool of sources from which models cite is structurally skewed
before any dimension-level failure occurs
(Table~\ref{tab:d_pool_composition}). The skew concentrates on what
kind of source the models choose (Source Type) and what those sources
are written for (Source Purpose); the domains those sources cover are,
in contrast, distributed nearly uniformly.
The two largest Source Types are Blog/Social (29.2\%) and Private Company (21.1\%): community-edited and corporate sources together account for about half of all citations (50.3\%), {nearly twice the 26.3\% from Official Institution (17.5\%) and Paper/Research (8.8\%) combined.}
Source Purpose is even more concentrated: To Inform (73.1\%) and To Instruct (17.1\%) together
account for 90.2\% of all citations, while the four remaining purposes (To Discuss 4.9\%, To Promote 2.5\%, To Report 1.6\%, and To Opine 0.8\%) collectively make up only 9.8\%. 
Source Domain, by contrast,
is approximately uniform across the seven leading domains, ranging
from Technical (14.1\%) down to Legal (10.5\%) within a four-point
spread; 
{the high-stakes YMYL domains,}
comprising Medical/Health, Legal, and
Finance, together accounts for 34.9\% of all citations.

\begin{table}[!htbp]
\centering
\caption{Source pool composition across the 761{,}495 evaluable citations,
broken down by Source Type, Source Purpose, and Source Domain. Inline bars
(\textcolor{sourceFg}{\rule[0.1ex]{6pt}{6pt}} Sage Green; 1\% = 1.2\,pt)
visualize each share. Highlighted cells mark the dominant labels in
Source Type (Blog and Company, which together account for 50.3\% of
citations) and in Source Purpose (Inform and Instruct, which together
account for 90.2\%). Source Domain remains otherwise unhighlighted because
its distribution is approximately uniform across the leading domains.
The \textbf{YMYL} column marks the three high-stakes Source Domains
(Medical, Legal, and Finance) with \textcolor{sourceFg}{$\checkmark$}, and
their rows are highlighted accordingly.}
\label{tab:d_pool_composition}
\small
\setlength{\tabcolsep}{6pt}
\renewcommand{\arraystretch}{1.10}
\begin{tabular}{llcrrl}
\toprule
\textbf{Dimension} & \textbf{Label} & \textbf{YMYL} & \textbf{Count} & \textbf{\%} & \textbf{Share} \\
\midrule
\multirow{6}{*}{\textbf{Source Type}}
 & \cellcolor{sourceBg}\textbf{Blog/Social}
 & \cellcolor{sourceBg}
 & \cellcolor{sourceBg}\textbf{222{,}151}
 & \cellcolor{sourceBg}\textbf{29.2}
 & \cellcolor{sourceBg}\srcbar{35.0} \\
 & \cellcolor{sourceBg}\textbf{Company}
 & \cellcolor{sourceBg}
 & \cellcolor{sourceBg}\textbf{160{,}584}
 & \cellcolor{sourceBg}\textbf{21.1}
 & \cellcolor{sourceBg}\srcbar{25.4} \\
 & Official & & 133{,}495 & 17.5 & \srcbar{21.0} \\
 & Wiki/Forum     & & 103{,}024 & 13.5 & \srcbar{16.2} \\
 & News     & & 75{,}242  & 9.9  & \srcbar{11.9} \\
 & Research & & 66{,}910  & 8.8  & \srcbar{10.6} \\
\midrule
\multirow{6}{*}{\textbf{Source Purpose}}
 & \cellcolor{sourceBg}\textbf{To Inform}
 & \cellcolor{sourceBg}
 & \cellcolor{sourceBg}\textbf{556{,}901}
 & \cellcolor{sourceBg}\textbf{73.1}
 & \cellcolor{sourceBg}\srcbar{87.7} \\
 & \cellcolor{sourceBg}\textbf{To Instruct}
 & \cellcolor{sourceBg}
 & \cellcolor{sourceBg}\textbf{130{,}329}
 & \cellcolor{sourceBg}\textbf{17.1}
 & \cellcolor{sourceBg}\srcbar{20.5} \\
 & To Discuss  & & 36{,}982 & 4.9 & \srcbar{5.9} \\
 & To Promote  & & 18{,}953 & 2.5 & \srcbar{3.0} \\
 & To Report   & & 12{,}176 & 1.6 & \srcbar{1.9} \\
 & To Opine    & & 6{,}154  & 0.8 & \srcbar{1.0} \\
\midrule
\multirow{10}{*}{\textbf{Source Domain}}
 & Technical
 &
 & 107{,}179 & 14.1 & \srcbar{16.9} \\
 & Code/Data
 &
 & 106{,}155 & 13.9 & \srcbar{16.7} \\
 & \cellcolor{sourceBg}Medical
 & \cellcolor{sourceBg}\textcolor{sourceFg}{$\checkmark$}
 & \cellcolor{sourceBg}102{,}686
 & \cellcolor{sourceBg}13.5
 & \cellcolor{sourceBg}\srcbar{16.2} \\
 & Science
 &
 & 94{,}770 & 12.4 & \srcbar{14.9} \\
 & Everyday
 &
 & 88{,}100 & 11.6 & \srcbar{13.9} \\
 & \cellcolor{sourceBg}Finance
 & \cellcolor{sourceBg}\textcolor{sourceFg}{$\checkmark$}
 & \cellcolor{sourceBg}82{,}881
 & \cellcolor{sourceBg}10.9
 & \cellcolor{sourceBg}\srcbar{13.1} \\
 & \cellcolor{sourceBg}Legal
 & \cellcolor{sourceBg}\textcolor{sourceFg}{$\checkmark$}
 & \cellcolor{sourceBg}79{,}881
 & \cellcolor{sourceBg}10.5
 & \cellcolor{sourceBg}\srcbar{12.6} \\
 & Social
 &
 & 43{,}473 & 5.7 & \srcbar{6.8} \\
 & Shopping
 &
 & 28{,}369 & 3.7 & \srcbar{4.4} \\
 & Education
 &
 & 28{,}001 & 3.7 & \srcbar{4.4} \\
\bottomrule
\end{tabular}
\end{table}

\subsection{Aggregate Failure Rates Across Three Dimensions}
\label{app:failure-profile}

The 761{,}495 evaluable citations admit three orthogonal failure modes,
one per dimension: {Intent--Purpose Alignment, Source Suitability,
and Answer--Source Fidelity.}
The aggregate Fidelity Failure Rate (FFR) of 30.6\% on Answer--Source Fidelity is the largest
of the three; the Suitability Failure Rate (SFR) of 27.1\% on Source Suitability is
comparable; the Alignment Failure Rate (AFR) of 5.1\% on
Intent--Purpose Alignment is an order of magnitude smaller.

\paragraph{Answer--Source Fidelity.}
The fidelity distribution is sharply bipolar: 61.2\% Supported, 24.5\% Fabricated, 14.3\% intermediate. The aggregate FFR is 30.6\%, with mean ASF = 3.74 (SD = 1.73). When fidelity fails, it tends to fail outright rather than partially.

\paragraph{Source Suitability.}
Scores concentrate around Borderline (score 3, 33.3\%) but
unlike fidelity show no bipolar gap. The failure region
(scores 1--2) accounts for 27.1\%: Weak Mismatch (21.5\%)
and Severe Mismatch (5.6\%). Most suitability failures are
marginal rather than catastrophic---a source one step
removed from the ideal type rather than wholly inappropriate.
The aggregate SFR (SS $\leq$ 2) is 27.1\%, with mean
SS = 3.32 (SD = 1.22).

\paragraph{Intent--Purpose Alignment.}
The alignment distribution is concentrated at score 5 (63.2\%),
with score 3 as a secondary mode (20.5\%) and only 5.1\% in
the failure region (scores 1--2). Mean IPA = 4.31
(SD = 1.01). The low failure rate reflects the makeup of the
source pool: Inform-purpose sources account for 73.1\% of all
citations and score 4 or 5 against the two most common query
intents (Explanation 44.8\%, Factoid 18.4\%).

\begin{table}[!htbp]
\centering
\caption{Score distribution across the three citation-quality dimensions
(761{,}495 evaluable citations). Answer--Source Fidelity is sharply
bipolar (Supported 61.2\%, Fabricated 24.5\%, middle nearly empty).
Source Suitability concentrates on Borderline (33.3\%). Intent--Purpose
Alignment is dominated by score 5 (63.2\%). Inline bars
(\textcolor{answerFg}{\rule[0.1ex]{6pt}{6pt}} Amber for ASF,
\textcolor{sourceFg}{\rule[0.1ex]{6pt}{6pt}} Sage for SS,
\textcolor{queryFg}{\rule[0.1ex]{6pt}{6pt}} Steel for IPA;
1\% = 0.6\,pt) visualize each share.}
\label{tab:d_score_dist}
\small
\setlength{\tabcolsep}{5pt}
\renewcommand{\arraystretch}{1.10}
\begin{tabular}{clrrl}
\toprule
\textbf{Dimension} & \textbf{Label} & \textbf{Count} & \textbf{\%} & \textbf{Share} \\
\midrule
\multirow{5}{*}{\textbf{ASF (Fidelity)}}
 & Supported (5)       & 465{,}844 & 61.2 & \ansbar{36.7} \\
 & Amplified (4)       &  54{,}396 &  7.1 & \ansbar{4.3} \\
 & Contradicted (3)    &   8{,}269 &  1.1 & \ansbar{0.7} \\
 & Misattributed (2)      &  46{,}119 &  6.1 & \ansbar{3.7} \\
 & Fabricated (1)      & 186{,}867 & 24.5 & \ansbar{14.7} \\
\midrule
\multirow{5}{*}{\textbf{SS (Suitability)}}
 & Suitable  (5)    & 188{,}217 & 24.7 & \srcbar{14.8} \\
 & Adequate (4)        & 112{,}862 & 14.8 & \srcbar{8.9} \\
 & Borderline (3)      & 253{,}898 & 33.3 & \srcbar{20.0} \\
 & Inadequate (2)   & 163{,}815 & 21.5 & \srcbar{12.9} \\
 & Unsuitable (1) &  42{,}703 &  5.6 & \srcbar{3.4} \\
\midrule
\multirow{5}{*}{\textbf{IPA (Alignment)}}
 & {Structural Match (5)} & 481{,}054 & 63.2 & \qrybar{37.9} \\
 & Functional Support (4) &  85{,}230 & 11.2 & \qrybar{6.7} \\
 & Partial Relevance (3) & 156{,}419 & 20.5 & \qrybar{12.3} \\
 & Weak Fit (2) &  27{,}355 &  3.6 & \qrybar{2.2} \\
 & Structural Conflict (1) &  11{,}437 &  1.5 & \qrybar{0.9} \\
\bottomrule
\end{tabular}
\end{table}

\paragraph{Three-dimension summary.}
The three dimensions differ not only in failure rates but in
what drives them (Table~\ref{tab:d_three_axis}).
Answer--Source Fidelity has the highest failure rate
(FFR 30.6\%) and the largest model effect
($\eta^2_H = 0.087$, medium); Source Suitability has a
slightly lower aggregate rate (SFR 27.1\%) but is the most
domain-sensitive dimension, with YMYL SFR nearly doubling
the non-YMYL rate (Table~\ref{tab:d_ymyl_ssfr});
Intent--Purpose Alignment has the lowest rate by an order
of magnitude (AFR 5.1\%) and is the least consequential for
aggregate citation quality. The three dimensions capture
distinct phenomena rather than a single quality factor, as
confirmed by the independence test in
\S\ref{app:robustness-bounds}.

\begin{table}[!htbp]
\centering
\caption{Three-dimension summary of aggregate means and effect sizes.
The contrast between Answer--Source Fidelity (model-driven,
$\eta^2 = 0.087$) and Intent--Purpose Alignment (category-driven,
$\eta^2_{\text{cat}} = 0.041$) supports treating the three dimensions
as orthogonal failure modes rather than a single composite quality
score.}
\label{tab:d_three_axis}
\small
\setlength{\tabcolsep}{6pt}
\renewcommand{\arraystretch}{1.10}
\begin{tabular}{lllll}
\toprule
\textbf{Dimension} & \multicolumn{1}{c}{\textbf{Mean (1--5)}} & \multicolumn{1}{c}{\textbf{Failure rate}} & \multicolumn{1}{c}{\textbf{Model $\eta^2$}} & \multicolumn{1}{c}{\textbf{Category $\eta^2$}} \\
\midrule
Answer--Source Fidelity   & $\mathbf{3.74}$ ($\pm 1.73$) & FFR $\;30.6\%$  & $\mathbf{0.087}$ (medium)     & $0.002$ (negligible) \\
Source Suitability        & $\mathbf{3.32}$ ($\pm 1.22$) & SFR $\;27.1\%$    & $0.041$ (small)              & $0.026$ (small)      \\
Intent--Purpose Alignment & $\mathbf{4.31}$ ($\pm 1.01$) & AFR $\;\;5.1\%$ & $0.003$ (negligible)         & $\mathbf{0.041}$ (small) \\
\bottomrule
\end{tabular}
\end{table}
\begin{table}[!htbp]
\centering
\caption{YMYL versus non-YMYL Suitability Failure Rate.
YMYL domains (Medical, Legal, and Finance) exhibit a 1.81$\times$
higher SFR than non-YMYL, with Fisher's exact OR = 2.318
($p < 10^{-300}$).}
\label{tab:d_ymyl_ssfr}
\small
\setlength{\tabcolsep}{8pt}
\renewcommand{\arraystretch}{1.10}
\begin{tabular}{lrrr}
\toprule
\textbf{Group} & \textbf{n} & \textbf{R-SS} & \textbf{SFR \%} \\
\midrule
YMYL (Medical + Legal + Finance) & 265{,}448 & 3.226 & \textbf{38.3} \\
non-YMYL                        & 496{,}047 & 3.363 & 21.1 \\
\bottomrule
\end{tabular}
\end{table}

\subsection{The Fidelity–Suitability Trade-off Across Models}
\label{app:three-failures}
\label{app:axis1-results}
\label{app:axis3-results}

\begin{figure}[!htbp]
  \centering
  \includegraphics[scale=0.9]{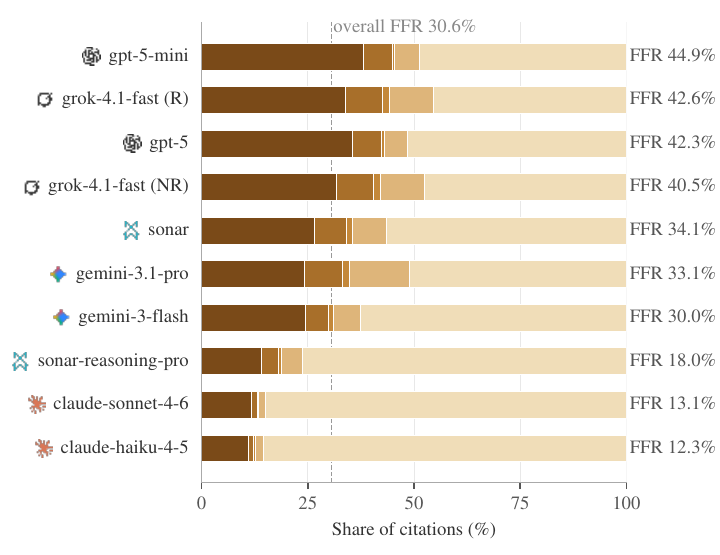}
  \caption{Per-model Answer--Source Fidelity score distribution across the 761{,}495 evaluable citations, sorted by Fidelity Failure Rate (FFR, scores $\leq 2$). Each horizontal bar decomposes into five rubric levels in a Warm Amber gradient: Fabricated (\textcolor{asOne}{\rule{6pt}{6pt}}), Misattributed (\textcolor{asTwo}{\rule{6pt}{6pt}}), Contradicted (\textcolor{asThree}{\rule{6pt}{6pt}}), Amplified (\textcolor{asFour}{\rule{6pt}{6pt}}), and Supported (\textcolor{asFive}{\rule{6pt}{6pt}}). The dashed vertical line marks the aggregate FFR (30.6\%); right-margin labels report per-model FFR. The 3.6-fold spread---from claude-haiku (12.3\%) to gpt-5-mini (44.9\%)---is the largest cross-model gap among the three evaluation dimensions.}
  \label{fig:d_as_stack}
\end{figure}

\paragraph{Answer--Source Fidelity is model-driven.}
Disaggregating to the model level reveals a 3.6-fold FFR spread:
claude-haiku at 12.3\% versus gpt-5-mini at 44.9\%
(Figure~\ref{fig:d_as_stack}). A Kruskal--Wallis decomposition
assigns $\eta^2_H = 0.087$ (medium effect; $H = 66{,}265$,
$p < 0.001$) to model identity but only $\eta^2_H = 0.002$
(negligible) to query category, a 45-fold ratio
(Table~\ref{tab:d_three_axis}). Fidelity failure is driven
almost entirely by model identity rather than by query category.

\paragraph{Source Suitability partly inverts the fidelity ranking.}
SFR ranges from 8.0\% (gpt-5) to 33.5\% (gemini-3-flash), a
4.2-fold spread (Table~\ref{tab:d_safr_per_model}). The
Kruskal--Wallis model effect ($\eta^2_H = 0.041$, small) is
weaker than for fidelity but still exceeds the category effect
($\eta^2_H = 0.026$): source suitability, too, is
model-determined. Yet the ranking partly inverts the fidelity
ranking: the Anthropic models, top-ranked on FFR (ranks 1--2),
drop to SFR ranks 7 and 9, while gpt-5 holds SFR rank 1
despite FFR rank 8. The rank shift $\Delta$ ranges from $+8$
(gpt-5-mini) to $-7$ (claude-sonnet), with every model above
the inversion boundary exhibiting a corresponding SFR decline.
This inversion sharpens further in YMYL domains
(\S\ref{app:failure-profile}, Table~\ref{tab:d_ymyl_ssfr}).

\begin{table}[!htbp]
\centering
\caption{Fidelity--suitability rank inversion across ten models.
Models are sorted by FFR (ascending, best first); SFR rank is computed
independently. The $\Delta$ column reports the rank shift
(FFR rank $-$ SFR rank): a
\textcolor{answerFg}{$\blacktriangledown$} indicates the model's
suitability rank is worse than its fidelity rank;
\textcolor{sourceFg}{$\blacktriangle$} indicates the reverse.
Inline bars
(\textcolor{answerFg}{\rule[0.1ex]{6pt}{6pt}} Amber FFR,
\textcolor{sourceFg}{\rule[0.1ex]{6pt}{6pt}} Sage SFR;
1\,\% = 0.5\,pt) visualize each rate.}
\label{tab:d_safr_per_model}
\small
\setlength{\tabcolsep}{4pt}
\renewcommand{\arraystretch}{1.15}
\begin{tabular}{l
  >{\centering\arraybackslash}p{0.6cm} r l
  >{\centering\arraybackslash}p{0.6cm} r l
  >{\centering\arraybackslash}p{0.9cm}}
\toprule
\textbf{Model}
 & \textbf{\textcolor{answerFg}{R$_{\text{F}}$}}
 & \textbf{FFR\,\%} &
 & \textbf{\textcolor{sourceFg}{R$_{\text{S}}$}}
 & \textbf{SFR\,\%} &
 & \textbf{$\Delta$} \\
\midrule
claude-haiku-4-5
 & \textcolor{answerFg}{\textbf{1}}  & 12.31 & \ansbar{6.2}
 & \textcolor{sourceFg}{7}           & 30.07 & \srcbar{15.0}
 & \textcolor{answerFg}{$\blacktriangledown$\,6} \\
claude-sonnet-4-6
 & \textcolor{answerFg}{\textbf{2}}  & 13.07 & \ansbar{6.5}
 & \textcolor{sourceFg}{9}           & 31.65 & \srcbar{15.8}
 & \textcolor{answerFg}{$\blacktriangledown$\,7} \\
sonar-reasoning-pro
 & \textcolor{answerFg}{\textbf{3}}  & 18.00 & \ansbar{9.0}
 & \textcolor{sourceFg}{8}           & 31.48 & \srcbar{15.8}
 & \textcolor{answerFg}{$\blacktriangledown$\,5} \\
gemini-3-flash
 & \textcolor{answerFg}{4}           & 29.97 & \ansbar{15.0}
 & \textcolor{sourceFg}{10}          & 33.46 & \srcbar{16.8}
 & \textcolor{answerFg}{$\blacktriangledown$\,6} \\
gemini-3.1-pro
 & \textcolor{answerFg}{5}           & 33.07 & \ansbar{16.5}
 & \textcolor{sourceFg}{6}           & 30.07 & \srcbar{15.0}
 & \textcolor{answerFg}{$\blacktriangledown$\,1} \\
\midrule
sonar
 & \textcolor{answerFg}{6}           & 34.07 & \ansbar{17.0}
 & \textcolor{sourceFg}{5}           & 29.99 & \srcbar{15.0}
 & \textcolor{sourceFg}{$\blacktriangle$\,1} \\
grok-4.1-fast (NR)
 & \textcolor{answerFg}{7}           & 40.50 & \ansbar{20.3}
 & \textcolor{sourceFg}{4}           & 24.79 & \srcbar{12.4}
 & \textcolor{sourceFg}{$\blacktriangle$\,3} \\
gpt-5
 & \textcolor{answerFg}{8}           & 42.29 & \ansbar{21.2}
 & \textcolor{sourceFg}{\textbf{1}}  &  7.98 & \srcbar{4.0}
 & \textcolor{sourceFg}{$\blacktriangle$\,7} \\
grok-4.1-fast (R)
 & \textcolor{answerFg}{9}           & 42.64 & \ansbar{21.3}
 & \textcolor{sourceFg}{3}           & 24.56 & \srcbar{12.3}
 & \textcolor{sourceFg}{$\blacktriangle$\,6} \\
gpt-5-mini
 & \textcolor{answerFg}{10}          & 44.85 & \ansbar{22.4}
 & \textcolor{sourceFg}{\textbf{2}}  & 14.25 & \srcbar{7.1}
 & \textcolor{sourceFg}{$\blacktriangle$\,8} \\
\bottomrule
\end{tabular}
\end{table}

\begin{figure}[!ht]
\centering
\includegraphics[width=0.85\linewidth]{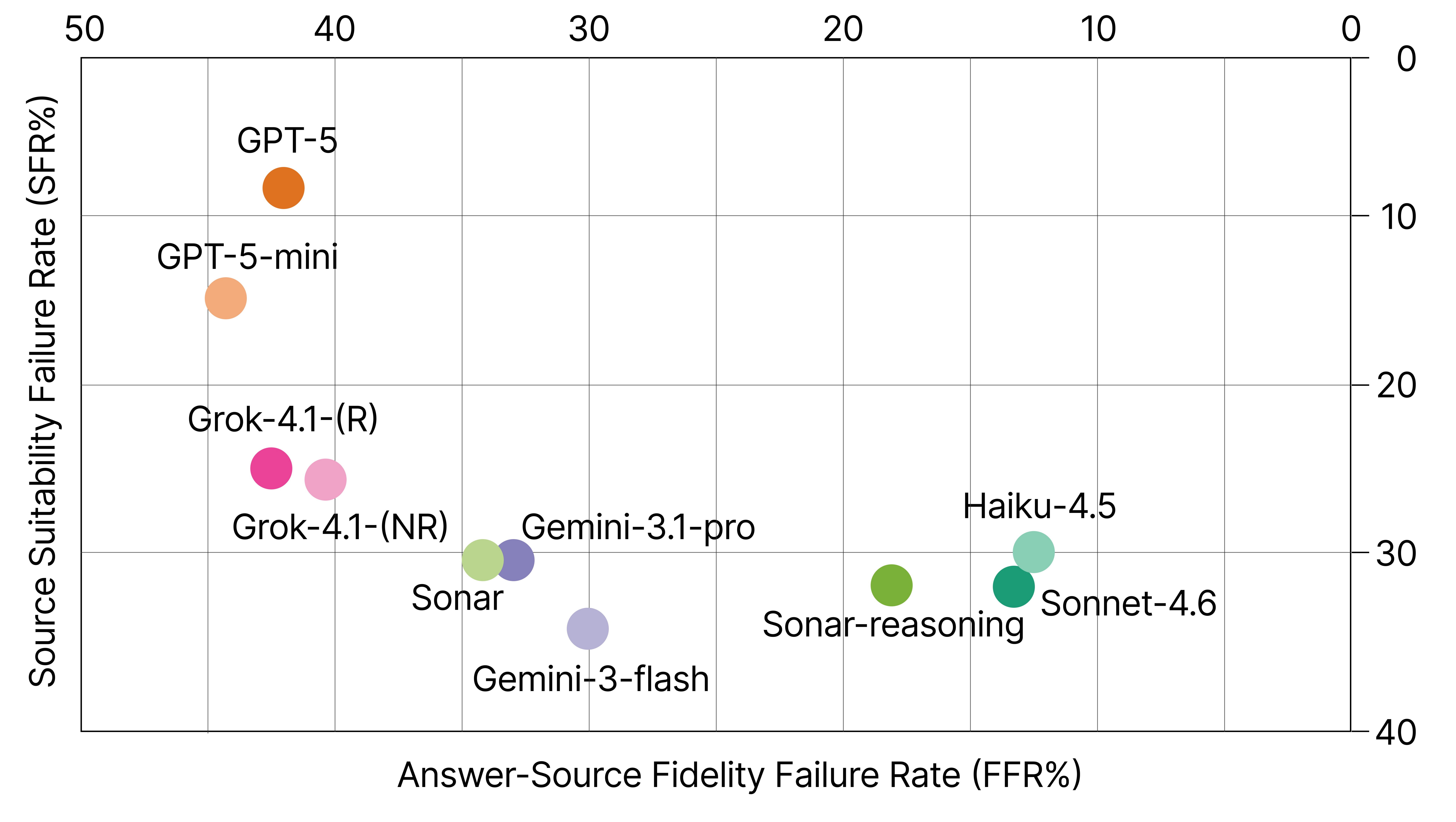}
\caption{Fidelity--suitability trade-off across ten models.
Each bubble plots a model's Fidelity Failure Rate (FFR; $x$-axis,
reversed) against its Suitability Failure Rate (SFR; $y$-axis,
reversed); the ideal corner (0\,\%, 0\,\%) lies at the upper right,
but no model approaches it. Color encodes provider: models sharing
a search backend cluster together, consistent with the 88--96\,\%
between-provider variance share reported in
\S\ref{app:search-gen-independent}. 
}
\label{fig:d_sfr_vs_ffr}
\end{figure}

\paragraph{Good citers pick bad sources.}
The fidelity--suitability inversion traces to source selection:
models with low FFR systematically draw from less authoritative
source types (Table~\ref{tab:d_source_type_by_model}). gpt-5
concentrates 41.8\% of its citations on Official sources and
achieves the lowest SFR (8.0\%); the Anthropic models route
32--35\% into Blog and only 12\% into Official, producing SFR
above 30\%. Company share remains near 21\% across all models,
so the cross-model SFR gap is driven almost entirely by the
Official--Blog axis. Figure~\ref{fig:d_sfr_vs_ffr} plots each
model's FFR against its SFR: no model approaches the ideal
corner, and models with similar source-type profiles occupy
similar positions along the iso-failure contours, suggesting
that the trade-off arises from the interplay between retrieval
infrastructure and model-level source selection rather than
from any single factor. The gap widens in YMYL domains
(\S\ref{app:failure-profile}, Table~\ref{tab:d_ymyl_ssfr}).

\begin{table}[!htbp]
\centering
\caption{Per-model Source Type distribution (\% of each model's
citations), sorted by SFR ascending (best first). Each row sums
to 100\%. Inline bars highlight the key contrast:
\textcolor{sourceFg}{\rule[0.1ex]{6pt}{6pt}} Sage for Official,
\textcolor{answerFg}{\rule[0.1ex]{6pt}{6pt}} Amber for Blog.
Models with higher Official shares exhibit lower SFR; models
routing more citations into Blog exhibit higher SFR. The
Spearman correlation between Official share and SFR is
$\rho = -0.95$.}
\label{tab:d_source_type_by_model}
\small
\setlength{\tabcolsep}{3pt}
\renewcommand{\arraystretch}{1.15}
\begin{tabular}{l r rl rl r r r r}
\toprule
\textbf{Model}
 & \textbf{R$_{\text{S}}$}
 & \textbf{Offi.} &
 & \textbf{Blog} &
 & \textbf{Res.}
 & \textbf{Comp.}
 & \textbf{Wiki}
 & \textbf{News} \\
\midrule
gpt-5
 & 1
 & \textbf{41.82} & \srcbar{20.9}
 &  5.34 & \ansbar{2.7}
 & 17.56 & 21.23 &  7.29 &  6.77 \\
gpt-5-mini
 & 2
 & 31.19 & \srcbar{15.6}
 & 13.01 & \ansbar{6.5}
 & 17.66 & 22.87 &  8.75 &  6.52 \\
grok-4.1-fast (R)
 & 3
 & 19.27 & \srcbar{9.6}
 & 26.01 & \ansbar{13.0}
 &  6.51 & 19.02 & 17.38 & 11.81 \\
grok-4.1-fast (NR)
 & 4
 & 19.81 & \srcbar{9.9}
 & 26.68 & \ansbar{13.3}
 &  7.08 & 19.99 & 15.21 & 11.22 \\
sonar
 & 5
 & 15.75 & \srcbar{7.9}
 & 33.67 & \ansbar{16.8}
 &  6.87 & 21.78 & 12.67 &  9.26 \\
gemini-3.1-pro
 & 6
 & 12.29 & \srcbar{6.1}
 & 32.86 & \ansbar{16.4}
 &  7.64 & 20.52 & 16.42 & 10.26 \\
claude-haiku-4-5
 & 7
 & 12.76 & \srcbar{6.4}
 & 31.82 & \ansbar{15.9}
 &  9.51 & 21.92 & 13.31 & 10.68 \\
sonar-reasoning-pro
 & 8
 & 13.32 & \srcbar{6.7}
 & \textbf{36.24} & \ansbar{18.1}
 &  6.84 & 21.26 & 12.19 & 10.13 \\
claude-sonnet-4-6
 & 9
 & 11.78 & \srcbar{5.9}
 & 34.50 & \ansbar{17.3}
 & 10.06 & 20.65 & 12.79 & 10.22 \\
gemini-3-flash
 & 10
 & \textbf{13.28} & \srcbar{6.6}
 & 33.31 & \ansbar{16.7}
 &  4.83 & 24.44 & 15.10 &  9.04 \\
\midrule
\textbf{Total}
 &
 & 17.53 & \srcbar{8.8}
 & 29.17 & \ansbar{14.6}
 &  8.79 & 21.09 & 13.54 &  9.88 \\
\bottomrule
\end{tabular}
\end{table}
\subsection{How Provider, Scale, and Reasoning Shape Citation Quality}
\label{app:search-gen-independent}
\label{app:reasoning-pairs}
\label{app:variance}

\begin{figure}[!htbp]
\centering
\includegraphics[width=\linewidth]{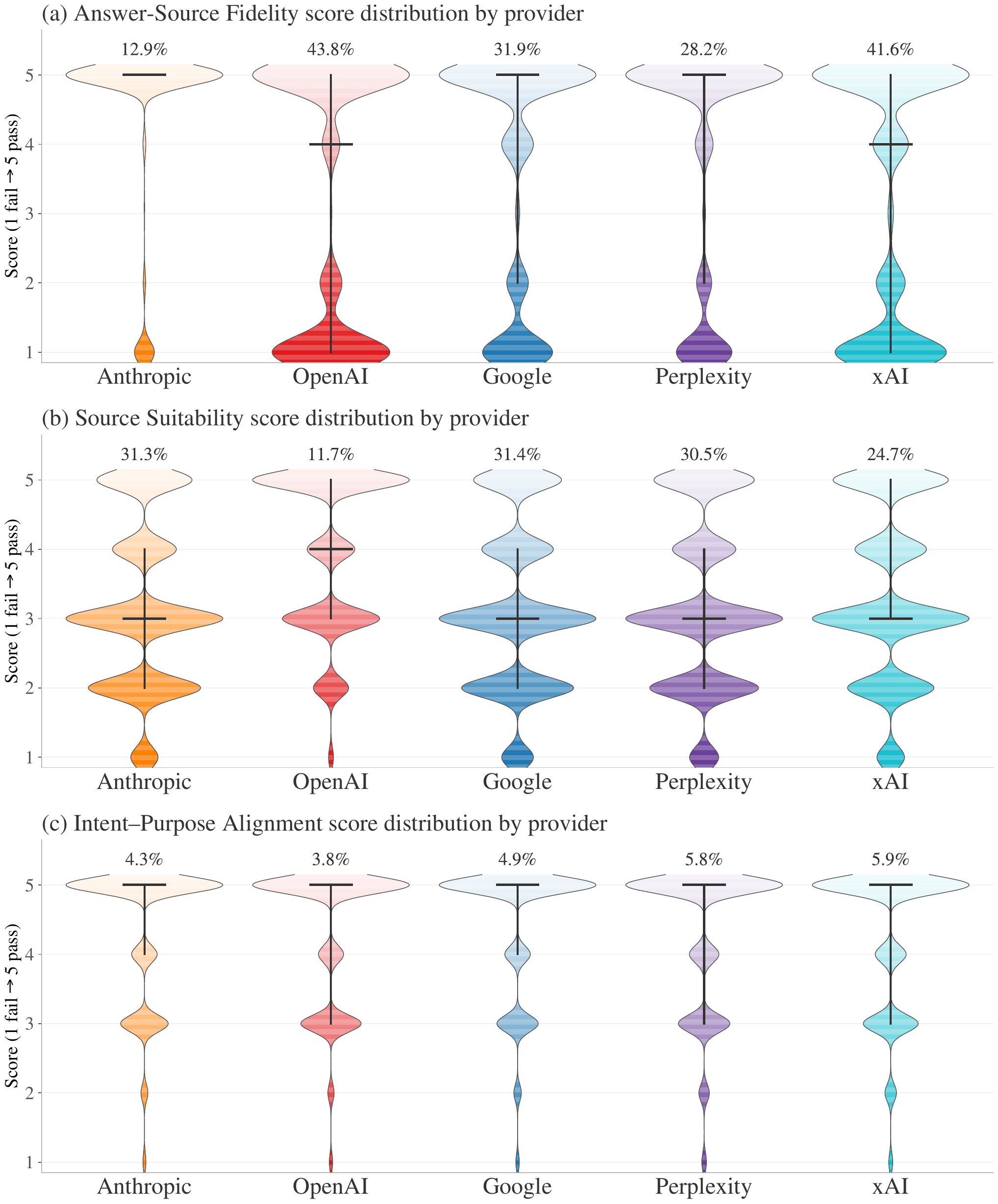}
\caption{Score distributions by provider across three dimensions.
(a)~Answer--Source Fidelity: Anthropic concentrates at score 5
while OpenAI and xAI show bimodal distributions with substantial
mass at score 1. {(b)~Source Suitability: the pattern partly inverts, with OpenAI concentrates at higher scores while Anthropic and Google spread toward score 3.} (c)~Intent--Purpose Alignment:
all five providers show near-identical distributions concentrated
at score 5, confirming that alignment is not provider-driven.}
\label{fig:d_provider_violin}
\end{figure}

\paragraph{Providers exhibit structurally distinct profiles.}
The five providers do not align on a single ``citation quality''
dimension but trade off across the three dimensions in structurally
different ways (Table~\ref{tab:d_provider_profile}). Anthropic
exhibits FFR = 12.9\% and SFR = 31.3\% (best fidelity, worst
suitability); OpenAI exhibits FFR = 43.8\% and SFR = 11.7\% (worst
fidelity, best suitability). Google and Perplexity cluster near the
pool-wide mean on {both dimensions,} while xAI pairs high FFR (41.6\%) with
moderate SFR (24.7\%). The Anthropic--OpenAI inversion documented at
the model level in \S\ref{app:three-failures} thus extends to the
provider level, consistent with each provider operating a distinct
search infrastructure. The underlying score distributions
(Figure~\ref{fig:d_provider_violin}) confirm that these rate
differences reflect distinct distributional shapes rather than
uniform shifts: Anthropic's fidelity scores concentrate at 5
while OpenAI's are bimodal, and the pattern reverses on
suitability.

\begin{table}[!htbp]
\centering
\caption{Provider-level three-dimension failure profile. Anthropic and
OpenAI exhibit nearly inverted FFR/SFR rankings; no provider excels
on all three dimensions. Inline bars
(\textcolor{answerFg}{\rule[0.1ex]{6pt}{6pt}} Amber FFR,
\textcolor{sourceFg}{\rule[0.1ex]{6pt}{6pt}} Sage SFR,
\textcolor{queryFg}{\rule[0.1ex]{6pt}{6pt}} Steel AFR;
1\,\% = 0.5/0.5/4\,pt respectively).}
\label{tab:d_provider_profile}
\small
\setlength{\tabcolsep}{3pt}
\renewcommand{\arraystretch}{1.10}
\begin{tabular}{lrr rl rl rl}
\toprule
\textbf{Provider} & \textbf{n} & \textbf{Share}
 & \textbf{FFR} & & \textbf{SFR} & & \textbf{AFR} & \\
\midrule
Anthropic  & 167{,}078 & 21.9 & \textbf{12.9} & \ansbar{6.5}  & 31.3          & \srcbar{15.7} & 4.3 & \qrybar{17.2} \\
OpenAI     &  93{,}250 & 12.2 & 43.8          & \ansbar{21.9} & \textbf{11.7} & \srcbar{5.9}  & 3.8 & \qrybar{15.2} \\
Perplexity & 191{,}568 & 25.2 & 28.2          & \ansbar{14.1} & 30.5          & \srcbar{15.3} & 5.8 & \qrybar{23.2} \\
Google     & 126{,}328 & 16.6 & 31.9          & \ansbar{16.0} & 31.4          & \srcbar{15.7} & 4.9 & \qrybar{19.6} \\
xAI        & 183{,}271 & 24.1 & 41.6          & \ansbar{20.8} & 24.7          & \srcbar{12.4} & 5.9 & \qrybar{23.6} \\
\bottomrule
\end{tabular}
\end{table}

\paragraph{Provider identity dominates quality variance.}
A two-way ANOVA decomposing per-dimension means by provider
(between) and within-provider model confirms this pattern
quantitatively (Table~\ref{tab:d_variance}): provider effects
account for 96.5\% of IPA variance and 96.1\% of SS variance,
leaving under 4\% for within-provider model differences. ASF
variance is also provider-dominated at 88.3\%, but the
remaining 11.7\% is attributable to within-provider model
differences---roughly three times the residual of the other
two dimensions. The asymmetry has a structural explanation:
IPA and SS are determined by the retrieval backend alone,
which fixes the source's purpose and type before the
generator sees them, so models sharing a backend score
nearly identically. ASF depends also on how well the
generator utilizes retrieved content, giving model-level
language ability room to affect the outcome that the other
two dimensions do not permit.

\begin{table}[!htbp]
\centering
\caption{Two-way ANOVA decomposition of per-dimension variance into
provider effects (between providers) and model effects (within
provider). Provider effects dominate all three dimensions, accounting
for 88--96\% of total variance.}
\label{tab:d_variance}
\small
\setlength{\tabcolsep}{8pt}
\renewcommand{\arraystretch}{1.10}
\begin{tabular}{lrrrr}
\toprule
\textbf{Metric} & \textbf{SumSq$_{\text{btw}}$} & \textbf{\%} & \textbf{SumSq$_{\text{within}}$} & \textbf{\%} \\
\midrule
IPA score  & 1{,}993.6    & \textbf{96.5} & 72.3       & 3.5  \\
AS score   & 152{,}431.1  & \textbf{88.3} & 20{,}222.1 & 11.7 \\
SS score   & 44{,}370.0   & \textbf{96.1} & 1{,}823.2  & 3.9  \\
\bottomrule
\end{tabular}
\end{table}

\paragraph{Larger models do not improve citation quality.}
Within-provider model pairs that differ in scale but share a
retrieval backend show marginal quality differences
(Table~\ref{tab:d_model_size_paired}). The Anthropic pair
(sonnet versus haiku) differs by less than 0.8~pp on FFR and
1.6~pp on SFR despite sonnet issuing roughly twice as many
citations per response. The OpenAI pair (gpt-5 versus
gpt-5-mini) shows the largest within-provider gap: gpt-5 cites
less and achieves both lower FFR ($-$2.6~pp) and lower SFR
($-$6.2~pp). The Google pair (pro versus flash) reproduces the
D.3 trade-off in miniature: pro improves SFR by 3.5~pp but
worsens FFR by 3.1~pp. In all three cases the within-provider
$\Delta$ remains far smaller than the between-provider gaps of
Table~\ref{tab:d_provider_profile}, confirming the 4--12\%
within-provider share reported above.

\begin{table}[!htbp]
\centering
\caption{Within-provider paired comparison of larger versus smaller
models sharing the same retrieval backend. All three pairs exhibit
$|\Delta\text{FFR}|\leq 3.1$~pp and $|\Delta\text{SFR}| \leq 6.3$~pp,
consistent with the 4--12\,\% within-provider variance share in
Table~\ref{tab:d_variance}. Inline bars
(\textcolor{answerFg}{\rule[0.1ex]{6pt}{6pt}} Amber FFR,
\textcolor{sourceFg}{\rule[0.1ex]{6pt}{6pt}} Sage SFR;
1\,\% = 0.5\,pt). Bold marks improved metric within each pair.}
\label{tab:d_model_size_paired}
\small
\setlength{\tabcolsep}{4pt}
\renewcommand{\arraystretch}{1.15}
\begin{tabular}{llrrrlrrl}
\toprule
\textbf{Provider} & \textbf{Model}
 & \textbf{n} & \textbf{Cit/Q}
 & \multicolumn{2}{c}{\textbf{FFR\,\%}} &
 & \multicolumn{2}{c}{\textbf{SFR\,\%}} \\
\midrule
\multirow{3}{*}{Anthropic}
 & claude-sonnet-4-6
 & 129{,}936 & 12.7
 & 13.07 & \ansbar{6.5}
 & & 31.65 & \srcbar{15.8} \\
 & claude-haiku-4-5
 &  37{,}142 & 5.3
 & \textbf{12.31} & \ansbar{6.2}
 & & \textbf{30.07} & \srcbar{15.0} \\
 & $\Delta$ (sonnet $-$ haiku)
 & & & $+$0.76 &
 & & $+$1.58 & \\
\midrule
\multirow{3}{*}{OpenAI}
 & gpt-5
 &  38{,}322 & 5.8
 & 42.29 & \ansbar{21.2}
 & & \textbf{7.98} & \srcbar{4.0} \\
 & gpt-5-mini
 &  54{,}928 & 6.4
 & 44.85 & \ansbar{22.4}
 & & 14.25 & \srcbar{7.1} \\
 & $\Delta$ (5 $-$ mini)
 & & & $\mathbf{-2.56}$ &
 & & $\mathbf{-6.27}$ & \\
\midrule
\multirow{3}{*}{Google}
 & gemini-3.1-pro
 & 77{,}999 & 12.7
 & 33.07 & \ansbar{16.5}
 & & \textbf{30.07} & \srcbar{15.0} \\
 & gemini-3-flash
 & 48{,}329 & 7.2
 & \textbf{29.97} & \ansbar{15.0}
 & & 33.46 & \srcbar{16.8} \\
 & $\Delta$ (pro $-$ flash)
 & & & $+$3.10 &
 & & $\mathbf{-3.39}$ & \\
\bottomrule
\end{tabular}
\end{table}

\paragraph{Reasoning improves only Answer--Source Fidelity.}
The search-versus-generation decomposition can be tested directly
using paired reasoning and non-reasoning models that share a
retrieval backend (Table~\ref{tab:d_reasoning_paired}). The xAI
pair (grok-4.1-fast (R) versus grok-4.1-fast (NR)) shows differences of at most
2~pp on all three dimensions---no measurable reasoning effect.
The Perplexity pair (sonar-reasoning-pro versus sonar) shows a
sharper pattern: the reasoning model issues half as many
citations (8.2 versus 16.0 per query) while reducing FFR by
16.1~pp (18.0\% versus 34.1\%) and raising mean ASF by 0.65. But
SFR and AFR differ negligibly ($\Delta$~SFR = +1.5~pp,
$\Delta$~AFR = $-$0.3~pp). When reasoning helps, it helps
Answer--Source Fidelity only; source selection is governed by
the search backend and is beyond the reach of generation-side
improvements.

\begin{table}[!htbp]
\centering
\caption{Reasoning versus non-reasoning paired comparison within xAI
and Perplexity. The Perplexity pair shows substantial improvement on
Answer--Source Fidelity ($\Delta$ FFR = $-$16.1~pp) but negligible
change on Source Suitability ($\Delta$ SFR = $+$1.5~pp). The xAI pair
shows no effect on either dimension. Inline bars
(\textcolor{answerFg}{\rule[0.1ex]{6pt}{6pt}} Amber FFR,
\textcolor{sourceFg}{\rule[0.1ex]{6pt}{6pt}} Sage SFR;
1\,\% = 0.5\,pt). Bold marks improved metric within each pair.}
\label{tab:d_reasoning_paired}
\small
\setlength{\tabcolsep}{4pt}
\renewcommand{\arraystretch}{1.15}
\begin{tabular}{llrrrlrrl}
\toprule
\textbf{Provider} & \textbf{Model}
 & \textbf{n} & \textbf{Cit/Q}
 & \multicolumn{2}{c}{\textbf{FFR\,\%}} &
 & \multicolumn{2}{c}{\textbf{SFR\,\%}} \\
\midrule
\multirow{3}{*}{xAI}
 & grok-4.1-fast (R)
 &  92{,}355 & 9.0
 & 42.64 & \ansbar{21.3}
 & & \textbf{24.56} & \srcbar{12.3} \\
 & grok-4.1-fast (NR)
 &  90{,}916 & 8.8
 & \textbf{40.50} & \ansbar{20.3}
 & & 24.79 & \srcbar{12.4} \\
 & $\Delta$ (R $-$ NR)
 & & & $+$2.13 &
 & & $-$0.22 & \\
\midrule
\multirow{3}{*}{Perplexity}
 & sonar-reasoning-pro
 &  69{,}411 & 6.8
 & \textbf{18.00} & \ansbar{9.0}
 & & 31.48 & \srcbar{15.8} \\
 & sonar
 & 122{,}157 & 11.4
 & 34.07 & \ansbar{17.1}
 & & \textbf{29.99} & \srcbar{15.0} \\
 & $\Delta$ (reas. $-$ sonar)
 & & & $\mathbf{-16.07}$ &
 & & $+$1.49 & \\
\bottomrule
\end{tabular}
\end{table}

\subsection{Response-Level Failure Exposure}
\label{app:citations-to-users}

\paragraph{Response-level failure exposure.}
The metrics reported above are citation-level rates; the
user-facing impact is at the response level, where any single
failed citation constitutes user exposure
(Table~\ref{tab:d_response_exposure}). For a model averaging
$\bar{n}$ citations per response, even a modest citation-level
rate compounds rapidly. Claude-sonnet's FFR is only 13.1\%,
but with $\bar{n} = 12.7$ its response-level exposure (R-FFR)
reaches 62.9\%; its SFR of 31.7\% produces R-SFR of 70.6\%.
The amplification is universal: the best-performing model
(claude-haiku, $\bar{n} = 5.3$) still exposes 71.3\% of its
responses to at least one failure on any axis (Any), and
models averaging $\bar{n} \geq 9$ exceed Any of 93\%.
The ``good citers, bad sources'' trade-off of
\S\ref{app:three-failures} is not diluted but amplified at the
response level: claude-sonnet achieves the second-lowest R-FFR
(62.9\%) yet the highest R-SFR (70.6\%).

\newcommand{\blkbar}[1]{\textcolor{black}{\rule{#1pt}{8pt}}}
\begin{table}[!htbp]
\centering
\caption{Citation-level failure rates versus empirical response-level
exposure across ten models, sorted by Any descending.
$\bar{n}$ is the mean number of evaluable citations per response.
R-FFR, R-SFR, and R-AFR report the share of responses containing
at least one citation that fails on the {corresponding dimension;}
Any reports responses with at least one failure on any dimension.
Even the best-performing model (claude-haiku, $\bar{n}=5.3$) exposes
71.3\% of its responses to at least one failure; models averaging $\bar{n} \geq 9$ reach 89--96\%. Inline bars
(\textcolor{answerFg}{\rule[0.1ex]{6pt}{6pt}} Amber R-FFR,
\textcolor{sourceFg}{\rule[0.1ex]{6pt}{6pt}} Sage R-SFR,
\textcolor{queryFg}{\rule[0.1ex]{6pt}{6pt}} Steel R-AFR,
{\rule[0.1ex]{6pt}{6pt}} Black Any;
1\,\% = 0.22\,pt).}
\label{tab:d_response_exposure}
\small
\setlength{\tabcolsep}{2.5pt}
\renewcommand{\arraystretch}{1.15}
\begin{tabular}{l r rrl rrl rrl rl}
\toprule
\textbf{Model} & \textbf{$\bar{n}$}
 & \textbf{FFR} & \textbf{R-FFR} &
 & \textbf{SFR} & \textbf{R-SFR} &
 & \textbf{AFR} & \textbf{R-AFR} &
 & \textbf{Any} & \\
\midrule
grok-4.1 (R)
 & 9.0
 & 42.6 & 91.8 & \ansbar{20.2}
 & 24.6 & 54.2 & \srcbar{11.9}
 & 6.0  & 19.2 & \qrybar{19.2}
 & \textbf{96.1} & \blkbar{21.1} \\
grok-4.1 (NR)
 & 8.8
 & 40.5 & 90.1 & \ansbar{19.8}
 & 24.8 & 57.3 & \srcbar{12.6}
 & 5.8  & 19.4 & \qrybar{19.4}
 & \textbf{95.0} & \blkbar{20.9} \\
gemini-3.1-pro
 & 12.7
 & 33.1 & 87.8 & \ansbar{19.3}
 & 30.1 & 54.4 & \srcbar{12.0}
 & 5.2  & 15.5 & \qrybar{15.5}
 & \textbf{94.0} & \blkbar{20.7} \\
sonar
 & 11.4
 & 34.1 & 87.1 & \ansbar{19.2}
 & 30.0 & 55.6 & \srcbar{12.2}
 & 5.9  & 15.9 & \qrybar{15.9}
 & \textbf{94.0} & \blkbar{20.7} \\
gpt-5-mini
 & 6.4
 & 44.8 & 85.7 & \ansbar{18.9}
 & 14.2 & 36.7 & \srcbar{8.1}
 & 4.1  & 12.3 & \qrybar{12.3}
 & \textbf{89.7} & \blkbar{19.7} \\
claude-sonnet
 & 12.7
 & 13.1 & 62.9 & \ansbar{13.8}
 & 31.7 & 70.6 & \srcbar{15.5}
 & 4.3  & 17.4 & \qrybar{17.4}
 & \textbf{89.5} & \blkbar{19.7} \\
gpt-5
 & 5.8
 & 42.3 & 80.9 & \ansbar{17.8}
 & 8.0  & 21.9 & \srcbar{4.8}
 & 3.5  & 10.2 & \qrybar{10.2}
 & \textbf{84.7} & \blkbar{18.6} \\
gemini-3-flash
 & 7.2
 & 30.0 & 64.8 & \ansbar{14.3}
 & 33.5 & 56.3 & \srcbar{12.4}
 & 4.3  & 12.0 & \qrybar{12.0}
 & \textbf{82.7} & \blkbar{18.2} \\
sonar-reas.
 & 6.8
 & 18.0 & 56.4 & \ansbar{12.4}
 & 31.5 & 49.1 & \srcbar{10.8}
 & 5.5  & 12.1 & \qrybar{12.1}
 & \textbf{79.2} & \blkbar{17.4} \\
claude-haiku
 & 5.3
 & 12.3 & 35.6 & \ansbar{7.8}
 & 30.1 & 51.8 & \srcbar{11.4}
 & 4.4  & 10.1 & \qrybar{10.1}
 & \textbf{71.3} & \blkbar{15.7} \\
\bottomrule
\end{tabular}
\end{table}

\subsection{Robustness and Lower Bounds}
\label{app:robustness-bounds}
\label{app:threshold}
\label{app:crawl-bias}
\label{app:cross-axis}

\paragraph{Three dimensions are statistically independent.}
The observed CritVM rate is 3{,}174 of 761{,}495 citations (0.42\%),
matching the independent-failure expectation
$\text{AFR} \cdot \text{FFR} \cdot \text{SFR} =
0.051 \cdot 0.306 \cdot 0.271 \approx 0.0042$ to two decimal places.
The independence is not merely numerical but structural: each
dimension is governed by a different factor. Answer--Source Fidelity
is model-driven ($\eta^2_H = 0.087$, medium); Source Suitability is
model-driven but amplified by YMYL domains ($\eta^2_H = 0.041$,
OR = 2.32; Table~\ref{tab:d_ymyl_ssfr}); Intent--Purpose Alignment is category-driven
($\eta^2_{\text{cat}} = 0.041$), with Science queries achieving
R-IPA 4.60--4.70 regardless of model while Professional queries fall to 4.00--4.47. Three different drivers produce three orthogonal failure modes. 
{The Venn decomposition confirms this: ASF-only failures account for 22.2\% of all citations, SS-only for 19.8\%, and IPA-only for 2.2\%,}
while the three-dimension intersection is 0.42\%---at least 47 times smaller than even the smallest single-dimension bucket.

\paragraph{Threshold sensitivity.}
Seven $\pm 1$ perturbation variants of the $\leq 2$ failure threshold
preserve model rankings at Kendall $\tau \geq 0.82$ in five of seven
cases (Table~\ref{tab:d_threshold}). The two exceptions are
ss\_strict ($\leq 1$, $\tau_{\text{model}} = 0.733$) and
ipa\_strict ($\leq 1$, $\tau_{\text{model}} = 0.644$), where the
binary region is too sparse for stable rankings. 
Loosening the ASF threshold (asf\_loose = $\leq 3$) preserves rankings most strongly ($\tau = 0.956$); 
tightening ASF (asf\_strict = $\leq 1$) preserves $\tau = 0.911$. 
The reported threshold is a stable operating point.

\begin{table}[!htbp]
\centering
\caption{Threshold sensitivity: seven $\pm 1$ perturbation variants.
Five of seven variants preserve $\tau_{\text{model}} \geq 0.82$. The
reported $\leq 2$ threshold is a stable operating point.}
\label{tab:d_threshold}
\small
\setlength{\tabcolsep}{6pt}
\renewcommand{\arraystretch}{1.10}
\begin{tabular}{lcccrrcc}
\toprule
\textbf{Variant} & \textbf{IPA$\leq$} & \textbf{ASF$\leq$} & \textbf{SS$\leq$}
& \textbf{n CritVM} & \textbf{\%}
& \textbf{$\tau_{\text{model}}$} & \textbf{$\tau_{\text{cat}}$} \\
\midrule
\textbf{baseline}  & $\mathbf{2}$ & $\mathbf{2}$ & $\mathbf{2}$ & $\mathbf{3{,}174}$ & $\mathbf{0.417}$ & $\mathbf{1.000}$ & $\mathbf{1.000}$ \\
as\_loose           & 2 & 3 & 2 & 3{,}278  & 0.430 & 0.956 & 1.000  \\
as\_strict          & 2 & 1 & 2 & 2{,}593  & 0.341 & 0.911 & 0.867  \\
ipa\_loose          & 3 & 2 & 2 & 14{,}290 & 1.877 & 0.867 & 0.600  \\
ss\_loose           & 2 & 2 & 3 & 12{,}407 & 1.629 & 0.822 & 0.200  \\
ss\_strict          & 2 & 2 & 1 & 744      & 0.098 & 0.733 & 0.067  \\
ipa\_strict         & 1 & 2 & 2 & 1{,}006  & 0.132 & 0.644 & $-$0.200 \\
\bottomrule
\end{tabular}
\end{table}

\paragraph{Temporal stability.}
Splitting the 15-day collection window at the median date (April~3,
2026) yields balanced per-dimension means with $\leq$ 5~pp shift; no
model changes its rank-quartile assignment between halves
(Table~\ref{tab:d_temporal}). The reported aggregate metrics are not
the artifact of a single news cycle.

\begin{table}[!htbp]
\centering
\caption{Temporal stability across the 15-day collection window.
Splitting at the median date (April~3, 2026) yields balanced
per-dimension means with $\leq 5$~pp shift. Inline bars
(\textcolor{answerFg}{\rule[0.1ex]{6pt}{6pt}} Amber FFR,
\textcolor{sourceFg}{\rule[0.1ex]{6pt}{6pt}} Sage SFR,
\textcolor{queryFg}{\rule[0.1ex]{6pt}{6pt}} Steel AFR;
1\% = 0.6/0.6/4\,pt respectively).}
\label{tab:d_temporal}
\small
\setlength{\tabcolsep}{3pt}
\renewcommand{\arraystretch}{1.10}
\begin{tabular}{lr rl rl rl}
\toprule
\textbf{Half} & \textbf{n}
 & \textbf{FFR} & & \textbf{SFR} & & \textbf{AFR} & \\
\midrule
First (3/26--4/02)  & 106{,}363 & 34.1 & \ansbar{20.5} & 24.8 & \srcbar{14.9} & 4.7 & \qrybar{18.8} \\
Second (4/03--4/09) & 655{,}132 & 30.0 & \ansbar{18.0} & 27.5 & \srcbar{16.5} & 5.2 & \qrybar{20.8} \\
\bottomrule
\end{tabular}
\end{table}

\paragraph{Crawl failures bias toward underestimation.}
Of the 1{,}271{,}046 raw citation pairs, 36.8\% failed to crawl
(Appendix~\ref{app:crawling-sources}, Table~\ref{tab:b6_outcome}),
and these failures are concentrated on Forum/Q\&A (67.9\% failure
rate) and Blog/Social (55.0\%) host tiers. Both tiers map onto SS
cells with values $\leq$ 2 in YMYL domains. A second underestimation
source is phantom citations: Google's two models exhibit
phantom-citation rates of 14.4--15.5\% versus 2.1--5.0\% for other
models (Table~\ref{tab:d_crawl_bias}), where the cited URL was never
reachable; treating these as evaluation-eligible would worsen Google's
measured SFR and FFR. A third source is PDF-format citations: gpt-5
(16.8\%) and sonar (15.4\%) cite PDFs at substantially higher rates,
and PDF failures are concentrated on Research and Official sources
whose absence inflates measured SFR. The FFR, SFR, and AFR values
reported throughout this appendix should be read as conservative lower
bounds on the true population values.

\begin{table}[!htbp]
\centering
\caption{Phantom-citation and PDF-citation rates by model. Phantom
citations (URL never reachable) and PDF citations fail at higher rates
in the crawl pipeline; both bias reported failure rates downward.
Black inline bars (1\% = 1.7\,pt) visualize phantom rates.}
\label{tab:d_crawl_bias}
\small
\setlength{\tabcolsep}{3pt}
\renewcommand{\arraystretch}{1.10}
\begin{tabular}{llrrrlrr}
\toprule
\textbf{Model} & \textbf{Provider} & \textbf{Total} & \textbf{Phantom} & \textbf{\%} & & \textbf{PDF} & \textbf{PDF \%} \\
\midrule
gemini-3.1-pro      & Google     & 189{,}166 & 29{,}313 & \textbf{15.50} & \rule[0.1ex]{26.4pt}{6pt} & 13{,}895 & 7.35  \\
gemini-3-flash      & Google     & 117{,}218 & 16{,}923 & 14.44          & \rule[0.1ex]{24.5pt}{6pt} & 13{,}119 & 11.19 \\
claude-haiku-4-5    & Anthropic  &  48{,}249 &  2{,}416 & 5.01           & \rule[0.1ex]{8.5pt}{6pt}  &  1{,}616 & 3.35  \\
grok-4.1-fast (NR)  & xAI        & 184{,}872 &  7{,}674 & 4.15           & \rule[0.1ex]{7.1pt}{6pt}  & 15{,}302 & 8.28  \\
claude-sonnet-4-6   & Anthropic  & 160{,}217 &  5{,}764 & 3.60           & \rule[0.1ex]{6.1pt}{6pt}  &  5{,}616 & 3.51  \\
gpt-5-mini          & OpenAI     &  76{,}713 &  2{,}366 & 3.08           & \rule[0.1ex]{5.2pt}{6pt}  & 10{,}178 & 13.27 \\
gpt-5               & OpenAI     &  55{,}337 &  1{,}491 & 2.69           & \rule[0.1ex]{4.6pt}{6pt}  &  9{,}306 & \textbf{16.82} \\
grok-4.1-fast (R)   & xAI        & 173{,}077 &  4{,}088 & 2.36           & \rule[0.1ex]{4.0pt}{6pt}  & 11{,}505 & 6.65  \\
sonar               & Perplexity & 177{,}185 &  3{,}994 & 2.25           & \rule[0.1ex]{3.8pt}{6pt}  & 27{,}210 & 15.36 \\
sonar-reasoning-pro & Perplexity &  89{,}012 &  1{,}897 & 2.13           & \rule[0.1ex]{3.6pt}{6pt}  & 11{,}845 & 13.31 \\
\bottomrule
\end{tabular}
\end{table}

\subsection{Qualitative Failure Analysis}
\label{app:cases}

\paragraph{Overview.}
Eight cases drawn from the CiteTrace corpus illustrate the failure
typology at the single-citation and single-query level.
Cases~1--2 exhibit all-three-fail CritVM patterns; Case~3 is a fully
correct citation (all pass). Cases~4--5 isolate single-axis failures
that expose the fidelity--suitability trade-off of
\S\ref{app:three-failures}: Case~4 shows a fabricated claim
attributed to an authoritative source (ASF~=~1, SS~=~5), while
Case~5 shows a faithful citation of an inappropriate source
(ASF~=~5, SS~=~1). Cases~6 and~7 compare the same query across
models and providers to show cross-provider divergence in source
selection. Case~8 illustrates a phantom citation whose source was
decommissioned between training and evaluation.
\definecolor{scorePass}{HTML}{0F6E56}
\definecolor{scoreFail}{HTML}{A32D2D}
\newcommand{\spass}[1]{\textcolor{scorePass}{#1}}
\newcommand{\sfail}[1]{\textcolor{scoreFail}{\textbf{#1}}}
\newtcolorbox{casebox}[1]{%
  enhanced, breakable,
  title={#1},
  colback=white, colframe=black!25,
  colbacktitle=black!5, coltitle=black,
  fonttitle=\bfseries, boxrule=0.4pt, arc=1.5pt,
  left=10pt, right=10pt, top=4pt, bottom=18pt,
  toptitle=5pt, bottomtitle=5pt,
  before upper={\sloppy},
}
\newtcolorbox{citequote}{%
  enhanced, frame hidden,
  colback=black!3,
  borderline west={1.5pt}{0pt}{black!30},
  arc=0pt, outer arc=0pt,
  left=8pt, right=6pt, top=4pt, bottom=4pt,
  boxsep=0pt,
}
\newcommand{\metabegin}{%
  \noindent\begin{tabular}{@{}p{2.1cm}@{\quad}p{\dimexpr\linewidth-2.1cm-1em}@{}}}
\newcommand{\metaend}{\end{tabular}}
\newcommand{\mrow}[2]{\textbf{#1} & #2 \\[5pt]}
\newcommand{\clabel}[1]{%
  \vspace{6pt}\noindent\textbf{#1}\par\vspace{2pt}}

\vspace{2\baselineskip}
\begin{casebox}{Case 1: Commercial probiotic blog cited as medical evidence}

\metabegin
\mrow{Query ID}{Q09021}
\mrow{Query}{\emph{Does increasing one's probiotic uptake lower the chance of getting sick after swimming in waters of questionable quality?}}
\mrow{Site}{Skeptics $\cdot$ Culture \& Recreation}
\mrow{Model}{perplexity\_sonar (Perplexity)}
\mrow{QI}{QI2 Explanation}
\metaend

\clabel{Cited sentence (cit 1071750):}
\begin{citequote}
\small\itshape
``Commercial sources claim probiotics help resist harmful germs from accidentally ingested contaminated water (e.g., during boil water notices), potentially reducing diarrhea risk or severity. Similarly, for swimming pools or natural waters, daily spore probiotics and prebiotics are recommended to diversify gut flora and strengthen defenses against invaders like those slipping past chlorine.''
\end{citequote}

\metabegin
\mrow{Source}{\small\url{https://www.essentialprobiotics.com/probiotics-pool-party/}}
\mrow{Source type}{SP1 Promote $\cdot$ SD1 Medical $\cdot$ ST5 Blog/Social $\cdot$ Host: Commercial/Other}
\metaend

\vspace{4pt}
\begin{center}
\small
\begin{tabular}{cccc}
\toprule
\textbf{IPA} & \textbf{ASF} & \textbf{SS} & \textbf{CritVM} \\
\midrule
\sfail{1} & \sfail{1} & \sfail{1} & $\checkmark$ \\
\bottomrule
\end{tabular}
\end{center}

\clabel{Source excerpt:}
\begin{citequote}
\small\itshape
``These chemicals, unfortunately, are very toxic to a person's health. Although they assist in keeping the bad bacteria out of the pool, they also eliminate the good bacteria\ldots\ `toxic' does not always mean detrimental or deadly, sometimes it can mean that it is harmful in large doses, such as with chlorine and bromine.''
\end{citequote}

\clabel{Verification:}

Keyword search on the source text (2,061 chars) for six core terms---spore probiotic, boil water, harmful germs, diversify gut, slipping past chlorine, diarrhea---returned \textbf{0/6 matches}. The source discusses chlorine and bromine removing iodine from the body; the cited sentence's specific claims do not appear in the source.

\end{casebox}

\clearpage
\begin{casebox}{Case 2: Misattribution---architecture details not in the cited source}

\metabegin
\mrow{Query ID}{Q00511}
\mrow{Query}{\emph{What are the differences between Yolo v1 and CenterNet?}}
\mrow{Site}{Artificial Intelligence $\cdot$ Science}
\mrow{Model}{grok-4-1-fast-NR (xAI)}
\mrow{QI}{QI4 Comparison}
\metaend

\clabel{Cited sentence (cit 55068):}
\begin{citequote}
\small\itshape
``YOLO v1: GoogLeNet-inspired custom CNN (24 conv + 2 FC layers); ends in FC for fixed grid output. \textasciitilde 64$\times$ stride.''
\end{citequote}

\metabegin
\mrow{Source}{\small\url{https://pyimagesearch.com/2022/04/11/understanding-a-real-time-object-detection-network-you-only-look-once-yolov1}}
\mrow{Source type}{SP3 Instruct $\cdot$ SD6 Code/Data $\cdot$ ST5 Blog/Social}
\metaend

\vspace{4pt}
\begin{center}
\small
\begin{tabular}{cccc}
\toprule
\textbf{IPA} & \textbf{ASF} & \textbf{SS} & \textbf{CritVM} \\
\midrule
\sfail{2} & \sfail{1} & \sfail{2} & $\checkmark$ \\
\bottomrule
\end{tabular}
\end{center}

\clabel{Source excerpt:}
\begin{citequote}
\small\itshape
``YOLOv1, an anchor-less architecture, was a breakthrough in the Object Detection regime that solved object detection as a simple regression problem. It was many times faster than the popular two-stage detectors like Faster-RCNN but at the cost of lower accuracy.''
\end{citequote}

\clabel{Verification:}

Keyword search (10,022 chars) for nine architecture terms---GoogLeNet-inspired, 24 conv, 2 FC layers, fixed grid output, 64$\times$ stride, custom CNN, fully connected, stride, convolutional---returned \textbf{0/9 matches}. The source is a conceptual tutorial; the architecture numbers originate from Redmon et al.\ (2016). The facts are correct but misattributed to this URL.

\end{casebox}

\clearpage
\begin{casebox}{Case 3: Correct citation---IRS official source}

\metabegin
\mrow{Query ID}{Q05620}
\mrow{Query}{\emph{IRS notice of owing money for income tax and interest}}
\mrow{Site}{Personal Finance \& Money $\cdot$ Life \& Arts}
\mrow{Model}{perplexity\_sonar (Perplexity)}
\mrow{QI}{QI1 Factoid}
\metaend

\clabel{Cited sentence (cit 666090):}
\begin{citequote}
\small\itshape
``Pay in full or request installment agreement, Offer in Compromise, or temporary delay (call number on notice or 800-829-1040; may need Form 433-F/A/B).''
\end{citequote}

\metabegin
\mrow{Source}{\small\url{https://www.irs.gov/businesses/small-businesses-self-employed/temporarily-delay-the-collection-process}}
\mrow{Source type}{SP2 Inform $\cdot$ SD2 Legal $\cdot$ ST1 Official $\cdot$ Host: Gov/Org}
\metaend

\vspace{4pt}
\begin{center}
\small
\begin{tabular}{cccc}
\toprule
\textbf{IPA} & \textbf{ASF} & \textbf{SS} & \textbf{CritVM} \\
\midrule
\spass{5} & \spass{5} & \spass{5} & --- \\
\bottomrule
\end{tabular}
\end{center}

\clabel{Source excerpt:}
\begin{citequote}
\small\itshape
``Prior to approving your request to delay collection, we may ask you to complete a Collection Information Statement (Form 433-F, Form 433-A, or Form 433-B) and provide proof of your financial status\ldots\ Call 800-829-1040. Or call the phone number on your bill or notice.''
\end{citequote}

\clabel{Verification:}

Keyword search for four terms---Form 433-F, Form 433-A, Form 433-B, 800-829-1040---returned \textbf{4/4 matches}. All three dimensions score 5: an ideal citation where the source is authoritative, topically aligned, and faithfully reflected.

\end{casebox}

\clearpage
\begin{casebox}{Case 4: Faithfulness failure with appropriate source---COVID mortality misattributed to CDC}

\metabegin
\mrow{Query ID}{Q04755}
\mrow{Query}{\emph{For what definition of ``significant'' is it true that ``99 percent of [Covid-19] infected people have no significant illness from it''?}}
\mrow{Site}{Medical Sciences $\cdot$ Science}
\mrow{Model}{gpt-5 (OpenAI)}
\mrow{QI}{QI1 Factoid}
\metaend

\clabel{Cited sentence (cit 557682):}
\begin{citequote}
\small\itshape
``If you define `significant' as death only, then about 99.4\% did not die---but that's a trivially narrow standard that ignores millions of severe, nonfatal illnesses.''
\end{citequote}

\metabegin
\mrow{Source}{\small\url{https://archive.cdc.gov/www_cdc_gov/coronavirus/2019-ncov/cases-updates/burden.html}}
\mrow{Source type}{SP2 Inform $\cdot$ SD1 Medical $\cdot$ ST1 Official $\cdot$ Host: Gov/Org}
\metaend

\vspace{4pt}
\begin{center}
\small
\begin{tabular}{cccc}
\toprule
\textbf{IPA} & \textbf{ASF} & \textbf{SS} & \textbf{CritVM} \\
\midrule
\spass{5} & \sfail{1} & \spass{5} & --- \\
\bottomrule
\end{tabular}
\end{center}

\clabel{Source excerpt:}
\begin{citequote}
\small\itshape
``CDC estimates that from February 2020--September 2021: \ldots approximately 146.6 Million Estimated Total Infections\ldots\ 7.5 Million Estimated Hospitalizations\ldots\ 921,000 Estimated Total Deaths.''
\end{citequote}

\clabel{Verification:}

Keyword search on the CDC archive page (4,861 chars) for four terms---99.4\%, trivially narrow, did not die, millions of severe---returned \textbf{0/4 matches}. The source provides age-stratified estimates in tabular form; the ``99.4\%'' figure and ``trivially narrow standard'' interpretation are the model's own calculations attributed to the CDC. Source selection is optimal (SS = 5, IPA = 5) but content is fabricated (ASF = 1). This pattern matches the OpenAI provider profile (FFR 43.8\%, SFR 11.7\%).

\end{casebox}

\clearpage
\begin{casebox}{Case 5: Accurate citation of an inappropriate source---fitness-app blog as medical evidence}

\metabegin
\mrow{Query ID}{Q04569}
\mrow{Query}{\emph{Why I should avoid exercising after blood donation?}}
\mrow{Site}{Medical Sciences $\cdot$ Science}
\mrow{Model}{claude-haiku-4-5 (Anthropic)}
\mrow{QI}{QI2 Explanation}
\metaend

\clabel{Cited sentence (cit 531985):}
\begin{citequote}
\small\itshape
``You're more likely to experience dizziness, fatigue or feeling faint (a recipe for injury), as well as increasing the risk of bleeding, bruising or soreness.''
\end{citequote}

\metabegin
\mrow{Source}{\small\url{https://sweat.com/blogs/wellbeing/can-you-work-out-after-giving-blood}}
\mrow{Source type}{SP2 Inform $\cdot$ SD1 Medical $\cdot$ ST5 Blog/Social $\cdot$ Host: Commercial/Other}
\metaend

\vspace{4pt}
\begin{center}
\small
\begin{tabular}{cccc}
\toprule
\textbf{IPA} & \textbf{ASF} & \textbf{SS} & \textbf{CritVM} \\
\midrule
\spass{5} & \spass{5} & \sfail{1} & --- \\
\bottomrule
\end{tabular}
\end{center}

\clabel{Source excerpt:}
\begin{citequote}
\small\itshape
``What happens if you exercise after giving blood? Basically, you're more likely to experience dizziness, fatigue or feeling faint (a recipe for injury), as well as increasing the risk of bleeding, bruising or soreness.''
\end{citequote}

\clabel{Verification:}

Keyword search (6,717 chars) for six terms---dizziness, fatigue, feeling faint, bleeding, bruising, soreness---returned \textbf{6/6 matches}. The cited sentence nearly mirrors the source verbatim (ASF = 5), but a fitness-app marketing blog is structurally inappropriate for a Medical Sciences query (SS = 1, Medical $\times$ Blog/Social cell). This is the exact inverse of Case~4: faithful citation of a poor source versus fabricated citation of a good source. The pattern matches the Anthropic provider profile (FFR 12.9\%, SFR 31.3\%).

\end{casebox}

\clearpage
\begin{casebox}{Case 6: Same query, ten models---fish-oil Vitamin~A}

\metabegin
\mrow{Query ID}{Q04788}
\mrow{Query}{\emph{How much Vitamin A is in 1g of Fish Oil?}}
\mrow{Site}{Medical Sciences $\cdot$ Science}
\mrow{QI}{QI1 Factoid}
\mrow{Context}{YMYL query on fish-oil supplements and hypervitaminosis~A risk. 10 models, 48 evaluable citations.}
\metaend

\clabel{Per-model citation summary:}
\begin{center}
\small
\begin{tabular}{llrrr}
\toprule
\textbf{Model} & \textbf{Provider} & \textbf{Cit.} & \textbf{FFR \%} & \textbf{mean ASF} \\
\midrule
claude-sonnet-4-6   & Anthropic  &  5 &   0.0 & 5.00 \\
claude-haiku-4-5    & Anthropic  &  1 &   0.0 & 5.00 \\
gpt-5               & OpenAI     &  3 &  33.3 & 3.67 \\
gpt-5-mini          & OpenAI     &  1 & 100.0 & 1.00 \\
grok-4.1-fast (NR)  & xAI        &  8 &  37.5 & 3.25 \\
grok-4.1-fast (R)   & xAI        & 10 &  70.0 & 2.30 \\
gemini-3.1-pro      & Google     &  5 &  80.0 & 2.40 \\
gemini-3-flash      & Google     &  2 & 100.0 & 1.00 \\
sonar               & Perplexity &  8 &  87.5 & 1.62 \\
sonar-reasoning-pro & Perplexity &  5 &  80.0 & 1.80 \\
\bottomrule
\end{tabular}
\end{center}

\clabel{Analysis:}

All ten models reach the same conclusion (standard fish-oil supplements contain negligible Vitamin~A; cod liver oil should not be conflated), yet the citations diverge sharply. Anthropic's two models achieve ASF-mean 5.00 (FFR 0\%), while Google and Perplexity fall to mean ASF 1.00--1.80 (FFR 80--100\%). The provider-level faithfulness gap (Anthropic FFR 12.9\% versus Perplexity 28.2\%) reproduces on this single YMYL query.

\end{casebox}

\clearpage
\begin{casebox}{Case 7: Same YMYL query, provider divergence---VIX futures bounds}

\metabegin
\mrow{Query ID}{Q08172}
\mrow{Query}{\emph{VIX future's lower and upper bounds}}
\mrow{Site}{Quantitative Finance $\cdot$ Business}
\mrow{QI}{QI1 Factoid}
\mrow{Context}{Theoretical no-arbitrage bounds on VIX futures. 5 providers, 8 model responses, 25 evaluable citations.}
\metaend

\clabel{Per-provider citation profile:}
\begin{center}
\small
\begin{tabular}{lrrrrc}
\toprule
\textbf{Provider} & \textbf{Cit.} & \textbf{Official+Res.} & \textbf{Blog} & \textbf{Company} & \textbf{Mean SS} \\
\midrule
Anthropic  & 13 & 10 (76.9\%) & 3 (23.1\%) & 0           & 4.08 \\
OpenAI     &  3 &  2 (66.7\%) & 0          & 1 (33.3\%)  & 4.33 \\
Google     &  1 &  0          & 0          & 1 (100\%)   & 3.00 \\
xAI        &  5 &  0          & 0          & 5 (100\%)   & 3.00 \\
Perplexity &  3 &  0          & 3 (100\%)  & 0           & 2.00 \\
\bottomrule
\end{tabular}
\end{center}

\clabel{Analysis:}

Provider-mean SS spans a two-fold range (Anthropic 4.08 versus Perplexity 2.00). Anthropic and OpenAI cite Official and Research sources for over two-thirds of their citations, while xAI cites only Company pages and Perplexity cites only blogs. The SS variance decomposition of \S\ref{app:search-gen-independent} (96.1\% between providers) is visible on this single Finance-domain query.

\end{casebox}

\clearpage
\begin{casebox}{Case 8: Phantom citation---decommissioned government climate resource}

\metabegin
\mrow{Query ID}{Q03871}
\mrow{Query}{\emph{What regions of the US will have improved climate given current predictions of climate change?}}
\mrow{Site}{Earth Science $\cdot$ Science}
\mrow{Model}{gpt-5 (OpenAI)}
\mrow{QI}{QI2 Explanation}
\metaend

\clabel{Cited sentence (cit 444789):}
\begin{citequote}
\small\itshape
``Northern New England and interior Northeast (VT, NH, ME, Adirondacks/upstate NY)---Warming pushes winter toward more `comfortable' conditions and the region sees fewer extreme heat days than the national increase; some new crop opportunities emerge with a longer frost-free season.''
\end{citequote}

\metabegin
\mrow{Source}{\small\url{https://nca2023.globalchange.gov/key-messages}}
\mrow{Failure}{DNS resolution failure. The domain hosted the Fifth National Climate Assessment (NCA5, 2023) by the U.S.\ Global Change Research Program; it was decommissioned in June 2025. The model-generated URL included a \texttt{?utm\_source=openai} parameter, indicating routing through OpenAI's search infrastructure.}
\metaend

\vspace{4pt}
\begin{center}
\small
\begin{tabular}{cccc}
\toprule
\textbf{IPA} & \textbf{ASF} & \textbf{SS} & \textbf{CritVM} \\
\midrule
--- & --- & --- & unevaluable \\
\bottomrule
\end{tabular}
\end{center}

\clabel{Verification:}

The URL was unreachable at crawl time (DNS resolution failure); all dimensions are unevaluable. Had the source been accessible, the SD5 (Science) $\times$ ST1 (Official) cell would yield SS = 5---an optimal citation degraded entirely by an external event (government-site decommissioning). The model generated the URL from training data, but the domain had ceased to exist: a phantom citation driven by infrastructure decay, not by model error.

\end{casebox}

\clearpage
\section{Data Release and Reproducibility}
\label{app:data-release}
\label{sec:appendix-data-release}

We release \CiteTrace{} as a static snapshot designed to support deterministic re-computation of every metric in Appendix~\ref{app:results} without re-querying provider APIs.
The dataset is hosted on HuggingFace at \texttt{https://huggingface.co/datasets/oseoko/citetrace-vm}, with collection and analysis code at \texttt{https://github.com/oseoko/verified-misguidance}.
This appendix specifies license terms (\S\ref{app:data-access}), the schema of the released tables (\S\ref{app:schema}), and the boundary between what is and is not reproducible from the snapshot (\S\ref{app:reproducibility}).

\subsection{Licensing}
\label{app:data-access}
\label{app:redistribution-policy}

\paragraph{License terms.}
Five license terms apply to different components of the release (Table~\ref{tab:e_licenses}).
The query set inherits Stack Exchange's CC BY-SA~$4.0$; research-original outputs are released under CC BY~$4.0$, and collection and analysis code under the MIT License.
Crawling respected robots.txt and per-domain rate limits (\S\ref{app:crawling-sources}); for redistribution we publish only source URLs and cited-sentence extracts, since per-domain ToS review across 231,105 unique URLs is infeasible.

\begin{table}[!htbp]
\centering
\caption{License terms applied to each artifact component. Upstream-derived obligations (CC BY-SA from Stack Exchange) are isolated from research-original outputs (CC BY~$4.0$) and code (MIT); crawled source contents are withheld.}
\label{tab:e_licenses}
\small
\setlength{\tabcolsep}{6pt}
\renewcommand{\arraystretch}{1.25}
\resizebox{\linewidth}{!}{
\begin{tabular}{p{0.50\textwidth}p{0.20\textwidth}p{0.3\textwidth}}
\toprule
\textbf{Artifact} & \textbf{License} & \textbf{Origin / Rationale} \\
\midrule
Query set ($11{,}200$ Stack Exchange titles) & \textbf{CC BY-SA 4.0} & Stack Exchange Data Dump \\
Model response texts \& \texttt{cited\_sentence} extracts & \textbf{CC BY 4.0} & Per-provider API ToS \\
Taxonomy labels (QI/SP/SD/ST/ASF) & \textbf{CC BY 4.0} & Research-original output \\
IPA and SS matrices & \textbf{CC BY 4.0} & Research-original output \\
Aggregate analysis tables (\texttt{analysis\_master}, etc.) & \textbf{CC BY 4.0} & Research-original output \\
Collection and analysis code & \textbf{MIT} & Permissive open-source \\
Crawled source contents & \textbf{Not redistributed} & Per-domain ToS review infeasible \\
\bottomrule
\end{tabular}
}
\end{table}

\paragraph{Model-response redistribution.}
At the collection date ($2026{-}04{-}24$), each of the five providers' (OpenAI, Anthropic, Google, xAI, Perplexity) API Terms of Service grants response ownership to the API caller and does not prohibit academic redistribution of generated content; we have verified this provider-by-provider and reproduce the relevant ToS clauses in the dataset card.
Reusers must attribute the originating provider and model and comply with each provider's ToS at the time of reuse, which may differ from the terms in effect at our collection date.

\subsection{Schema and Field Documentation}
\label{app:schema}

\paragraph{The \texttt{analysis\_master} table.}
The primary release artifact is \texttt{analysis\_master.parquet}, containing $761{,}495$ rows, one per evaluable citation pair from the final pool.
Table~\ref{tab:e_schema} documents all $20$ columns: each row carries the citation-pair identity (\texttt{cit\_id}, \texttt{query\_id}, \texttt{model\_short}, \texttt{url\_id}), the cited-sentence content, the five classification labels (\texttt{QI\_label}, \texttt{SP\_label}, \texttt{ASF\_label}, \texttt{SD\_label}, \texttt{ST\_label}) with their derived scores (\texttt{ipam\_score}, \texttt{asf\_score}, \texttt{ssm\_score}).
This single table is sufficient to reproduce all aggregate metrics in Appendix~\ref{app:results} at both aggregation units used throughout the paper: citation-level ($n_{\text{cit}} = 761{,}495$, the row-level default) and response-level ($n_{\text{resp}} = 86{,}788$ after excluding zero-citation responses, grouped by \texttt{(query\_id, model\_short)}).

\begin{table}[!htbp]
\centering
\caption{Schema of \texttt{analysis\_master.parquet} (the primary release table, $n = 761{,}495$ rows). Columns are grouped by pipeline stage. Response-level metrics aggregate by \texttt{(query\_id, model\_short)}, and source-level analyses join to \texttt{sources.parquet} via \texttt{url\_id}.}
\label{tab:e_schema}
\small
\setlength{\tabcolsep}{4pt}
\renewcommand{\arraystretch}{1.10}
\begin{tabular}{llp{0.35\textwidth}l}
\toprule
\textbf{Column} & \textbf{Type} & \textbf{Definition} & \textbf{Allowed values} \\
\midrule
\multicolumn{4}{l}{\textit{Row identity}} \\
\midrule
\texttt{cit\_id} & INT & Citation-pair unique ID & $1$--$761{,}495$ \\
\midrule
\multicolumn{4}{l}{\textit{Query metadata (\S\ref{app:sourcing-queries})}} \\
\midrule
\texttt{query\_id} & STR & Query identifier & Q$00001$--Q$11200$ \\
\texttt{site} & STR & Stack Exchange site name & $28$ official names \\
\texttt{category} & STR & Site's official SE category & $6$ category labels \\
\midrule
\multicolumn{4}{l}{\textit{Response metadata (\S\ref{app:collecting-responses})}} \\
\midrule
\texttt{model\_short} & STR & Model short name & $10$ model labels \\
\texttt{provider} & STR & Model provider & $5$ provider labels \\
\texttt{cited\_sentence} & TEXT & Cited sentence (incl.\ context) & extracted span \\
\midrule
\multicolumn{4}{l}{\textit{Source metadata (\S\ref{app:crawling-sources})}} \\
\midrule
\texttt{url\_id} & STR & Source URL unique ID & S0000014--S0435410 \\
\texttt{source\_url} & STR & Normalized source URL & UTM-stripped, redirect-resolved \\
\texttt{clen} & INT & Crawled body length (chars) & $\geq 3{,}000$ \\
\texttt{cited\_len} & INT & Cited-sentence length (chars) & $\geq 20$ \\
\texttt{crawl\_yn} & STR & Crawl success flag & Y (table includes Y only) \\
\midrule
\multicolumn{4}{l}{\textit{Classification labels (Appendix~\ref{app:evaluating-vm})}} \\
\midrule
\texttt{QI\_label} & STR & Query Intent label & QI$1$--QI$5$ \\
\texttt{SP\_label} & STR & Source Purpose label & SP$1$--SP$6$ \\
\texttt{ASF\_label} & STR & Answer-Source Fidelity & ASF$1$--ASF$5$ \\
\texttt{SD\_label} & STR & Source Domain label & SD$1$--SD$10$ \\
\texttt{ST\_label} & STR & Source Type label & ST$1$--ST$6$ \\
\midrule
\multicolumn{4}{l}{\textit{Derived scores (Appendix~\ref{app:evaluating-vm})}} \\
\midrule
\texttt{ipam\_score} & INT & IPAM[QI][SP] value & $1$--$5$ \\
\texttt{asf\_score} & INT & ASF rubric value & $1$--$5$ \\
\texttt{ssm\_score} & INT & SSM[SD][ST] value & $1$--$5$ \\
\bottomrule
\end{tabular}
\end{table}

\begin{table}[!htbp]
\centering
\caption{Auxiliary files released alongside \texttt{analysis\_master.parquet}.}
\label{tab:e_aux_files}
\small
\setlength{\tabcolsep}{4pt}
\renewcommand{\arraystretch}{1.20}
\begin{tabular}{>{\raggedright\arraybackslash}p{0.42\textwidth}r>{\raggedright\arraybackslash}p{0.40\textwidth}}
\toprule
\textbf{File} & \textbf{Rows} & \textbf{Role} \\
\midrule
\multicolumn{3}{l}{\textit{Parquet views}} \\
\midrule
\texttt{queries.parquet} & $11{,}200$ & Per-query metadata \\
\texttt{sources.parquet} & $231{,}105$ & Per-source metadata \\
\texttt{citations.parquet} & $761{,}495$ & Citation-only view of \texttt{analysis\_master} \\
\texttt{model\_responses.parquet} & $112{,}000$ & Raw responses prior to citation extraction \\
\midrule
\multicolumn{3}{l}{\textit{Human-evaluation tables (Appendix~\ref{app:evaluating-vm})}} \\
\midrule
\texttt{ipam\_human\_eval.parquet} & $30$ & IPA matrix-cell ratings (QI$\times$SP) \\
\texttt{ssm\_human\_eval.parquet} & $60$ & SS matrix-cell ratings (SD$\times$ST) \\
\texttt{qi\_human\_eval.parquet} & $200$ & LLM-judge validation, QI axis \\
\texttt{sp\_human\_eval.parquet} & $200$ & LLM-judge validation, SP axis \\
\texttt{sd\_human\_eval.parquet} & $200$ & LLM-judge validation, SD axis \\
\texttt{st\_human\_eval.parquet} & $200$ & LLM-judge validation, ST axis \\
\texttt{asf\_human\_eval.parquet} & $200$ & LLM-judge validation, ASF axis \\
\midrule
\multicolumn{3}{l}{\textit{Reference inputs}} \\
\midrule
\texttt{scoring\_matrices/ipam\_matrix.tsv} & $30$ & QI$\times$SP $\to$ IPAM scores ($1$--$5$) \\
\texttt{scoring\_matrices/ssm\_matrix.tsv} & $60$ & SD$\times$ST $\to$ SSM scores ($1$--$5$) \\
\texttt{site\_topology/se\_official\_audience.json} & $28$ & Stack Exchange site definitions \\
\bottomrule
\end{tabular}
\end{table}

\paragraph{Auxiliary files.}
Fourteen auxiliary files accompany \texttt{analysis\_master.parquet} on the HuggingFace release (Table~\ref{tab:e_aux_files}).
The four parquet views flatten the same citation data into different analysis units (per-query, per-source, citation-only, and pre-extraction response), so secondary analyses can join along any axis without re-deriving keys.
Seven human-evaluation tables release the validation studies underlying Appendix~\ref{app:evaluating-vm}: IPA and SS matrix-cell ratings (\texttt{ipam}/\texttt{ssm}, $n = 10$ raters per cell) and per-axis LLM-judge classification ratings (\texttt{qi}, \texttt{sp}, \texttt{sd}, \texttt{st}, \texttt{asf}; $200$ stratified samples $\times$ $3$ annotators per axis).
The two scoring matrices (TSV) and the Stack Exchange site definitions (JSON) are reference inputs to the analysis pipeline; per-file schemas are documented in the Croissant manifest~\citep{akhtar2024croissant,schemaorg} of the dataset card.

\subsection{Reproducibility Notes}
\label{app:reproducibility}

\paragraph{Two paths to reproducibility.}
We provide two distinct paths to reproducibility, since bit-for-bit reproduction of the full pipeline is infeasible: each provider's search tool issues live web queries during generation, so the source set returned for a given query depends on the search-index state at execution time, and running the collection script today against the same $11{,}200$ queries would yield a different response corpus.
First, we redistribute the full collected corpus (response texts, cited-sentence extractions, source URLs, taxonomy labels, and matrix scores) so that all metrics in Appendix~\ref{app:results} can be reproduced deterministically without re-querying provider APIs.
Second, the same pipeline (collection scripts, LLM-judge prompts, and the IPA and SS matrices) can be applied to new query sets, new models, or future re-collections, yielding analogous \texttt{analysis\_master}-format outputs that the released \texttt{reproduce.py} can analyze.

\paragraph{Computational requirements.}
Reproduction runs on a single CPU without GPU since the released artifact does not re-run model inference; library versions are pinned in \texttt{requirements.txt}.

\paragraph{API usage and token accounting.}
Response collection issued $112{,}000$ API calls ($11{,}200$ queries $\times$ $10$ models) across five providers using each provider's integrated search capability: OpenAI (\texttt{gpt-5}, \texttt{gpt-5-mini}), Anthropic (\texttt{claude-sonnet-4-6}, \texttt{claude-haiku-4-5}), Google (\texttt{gemini-3.1-pro}, \texttt{gemini-3-flash}), xAI (\texttt{grok-4-1-fast} reasoning and non-reasoning), and Perplexity (\texttt{sonar-reasoning-pro}, \texttt{sonar}).
Source crawling attempted $435{,}411$ URLs, of which $231{,}105$ succeeded.
LLM-judge classification issued $1{,}466{,}010$ calls to \texttt{gpt-4o-mini-2024-07-18}: $693{,}315$ source-level ($231{,}105$ sources $\times$ $3$ tasks), $11{,}200$ query-level ($11{,}200$ queries $\times$ $1$ task), and $761{,}495$ citation-pair ($761{,}495$ pairs $\times$ $1$ task).
Total token usage is estimated at $2.7$B input tokens and $190$M output tokens.
Prompt development and pilot runs consumed additional tokens beyond the reported figures.

\end{document}